\documentclass[onecolumn,draft,12pt]{IEEEtran}

\usepackage[final]{graphicx}
\usepackage{bm}
\usepackage{amssymb}

\newcommand{\diagM}{{\rm diag}}

\newcommand{\CN}{\mathcal{CN}}
\newcommand{\TCN}{\mathcal{TCN}}

\newcommand{\Leps}{ {\textit{\LARGE$\varepsilon$}} }

\newcommand{\Sf}{{\mathcal{S}}}

\newcommand{\Ft}{{\openmtr{F}}}

\newcommand{\Ct}{{\rm C}}

\newcommand{\It}{{\textsf{I}}}
\newcommand{\Iv}{{{I}}}
\newcommand{\Tc}{{T_{\rm c}}}
\newcommand{\hTc}[1][ ]{{\hat{T}_{{\rm c}#1}}}

\newcommand{\Jc}{{\mathcal{J}}}
\newcommand{\Js}{{\textsf{J}}}
\newcommand{\Pc}{{\mathcal{P}}}

\newcommand{\RV}[3][ ]{{ {\bf {#2}}_{#3}^{#1}}}
\newcommand{\FV}[3][ ]{{ \tilde{\bf {#2}}_{#3}^{#1}}}
\iffalse   %use to show FINAL
\usepackage{color}
\newcommand{\CommentApril}[1]{{\color{blue}{#1}}}
\newcommand{\CommentDeleteApril}[1]{{\color{red}{#1}}}
\else
\newcommand{\CommentApril}[1]{{{#1}}}
\newcommand{\CommentDeleteApril}[1]{{{}}}
\fi

\newcommand{\uv}{{\bf u}}
\newcommand{\yv}{{\bf y}}
\newcommand{\zv}{{\bf z}}

\newcommand{\mtr}[1]{\textsf{#1}}
\newcommand{\openmtr}{\sf}

\newtheorem{Lemma}{Lemma}

\newtheorem{theorem}{Theorem}

\newtheorem{assumption}{Assumption}
\newtheorem{definition}{Definition}

\renewcommand{\(}{\left(}
\renewcommand{\)}{\right)}

\begin{document}

\title{Bounds on the capacity of OFDM underspread frequency selective fading channels}
\author{Itsik Bergel$^{\dag*}$, Sergio Benedetto$^\ddag$
\thanks{
   $^\dag$ School of Engineering, Bar-Ilan University, 52900 Ramat-Gan, Israel
   {\tt e-mail: bergeli@eng.biu.ac.il}
   }
\thanks{
   $^\ddag$ Dipartimento di Elettronica - Politecnico di Torino
   {\tt e-mail: benedetto@polito.it}
   }
\thanks{This work has been funded by PRIMO, a research project financed by MIUR, the Italian Ministry of Education and Research} %\\
\thanks{$^*$ Corresponding author.
   }
} \markboth {Submitted to The IEEE Transactions on Information
Theory} {Bounds on the capacity of  strictly underspread frequency
selective fading channels}

\maketitle

\begin{abstract}
\CommentDeleteApril{Although communication systems are often
analyzed under the assumption of some prior knowledge of the
channel, this is rarely the case in practical systems, which usually
need to estimate the channel from the received signal. }The analysis
of the channel capacity in the absence of prior channel knowledge
(\CommentApril{noncoherent} channel) has gained increasing interest
in recent years, but it is still  unknown for the general case. In
this paper we derive bounds on the capacity of the noncoherent,
\textit{underspread} complex Gaussian, orthogonal frequency division
multiplexing (OFDM), wide sense stationary channel with uncorrelated
scattering (WSSUS), under a peak power constraint or a constraint on
the second and fourth moments of the transmitted signal. These
bounds are characterized only by the system signal-to-noise ratio
(SNR) and by a newly defined quantity termed \textit{effective
coherence time}. Analysis of the effective coherence time reveals
that it can be interpreted as the length of a block in the block
fading model in which a system with the same SNR will achieve the
same capacity as in the analyzed channel. Unlike commonly used
coherence time definitions, it is shown that the effective coherence
time depends on the SNR, and is a \CommentApril{nonincreasing}
function of it.

\CommentApril{We show that for low SNR the capacity is proportional
to the effective coherence time, while for higher SNR the coherent
channel capacity can be achieved provided that the effective
coherence time is large enough.}\CommentDeleteApril{ The lower
bounds show that, for sufficiently large effective coherence times,
a system that employs a minimum Euclidean distance receiver with
channel estimation based on past received symbols can achieve
information rates that are very close to the channel capacity.
However, as the effective coherence time is typically a decreasing
function of the SNR, our results cannot give the capacity for very
high SNRs. For such SNRs we claim that the channel can no longer be
considered as underspread.}
\end{abstract}



% for peerreview papers, inserts a page break and creates the second title.
% Will be ignored for other modes.
\IEEEpeerreviewmaketitle

\section{Introduction}\label{sec. Introduction}
The analysis of communication systems is often performed under the
assumption of perfect channel knowledge. In practical systems,
however, the channel needs to be estimated from the communication
signal itself, making this assumption unrealistic.

In this paper we study the capacity of the noncoherent, underspread,
complex Gaussian, orthogonal frequency division multiplexing (OFDM),
wide sense stationary channel with uncorrelated scattering (WSSUS),
under a peak power constraint or a constraint on the second and
fourth moments of the transmitted signal (commonly termed
\textit{quadratic power constraint}). We use the term noncoherent
channel capacity to describe the capacity of the channel when
neither the transmitter nor the receiver have any prior knowledge on
the channel realization, \CommentApril{but both have exact knowledge
on the channel statistics}.

The OFDM model provides a simple representation of an underspread
frequency selective channel \cite{tse2005fwc}. In underspread
channels the \CommentApril{channel} delay spread
\CommentDeleteApril{(also known as channel memory length) }is
significantly smaller than the channel coherence time
\cite{kozek1998nonorthogonal}. Choosing the OFDM symbol length to be
significantly larger than the \CommentApril{channel delay spread},
and yet significantly smaller than the channel coherence time,
\CommentApril{results in} a useful approximation of the underspread
channel which is both accurate and convenient for the analysis.
\CommentApril{(Detailed description on the connection between the
OFDM model and the general WSSUS channel can be found in
\cite{durisi2010noncoherent,Hlawatsch2011}.)}

This paper presents novel bounds on the capacity of the noncoherent
underspread WSSUS channel that are characterized only by the system
signal-to-noise ratio (SNR) and by a newly defined quantity termed
\textit{effective coherence time}. We show that the bounds are
\CommentApril{good }for almost all of the SNR range as long as the
effective coherence time is significantly larger than the OFDM
symbol length. Analysis of the effective coherence time reveals that
it characterizes the system ability to estimate the channel, and can
be interpreted as the length of a block in the block fading model
that achieves the same capacity as the analyzed channel at the same
SNR. Surprisingly, and unlike standard coherence time definitions,
the effective coherence time is actually a function of the system
SNR. This dependence of the effective coherence time on SNR stems
from the fact that at high SNR the system is more sensitive to
changes in the channel, and hence will effectively see a shorter
coherence time.\footnote{ A simple intuition for the decrease of
coherence time with SNR is as follows. Consider an intuitive
definition of coherence time for stationary channels: The coherence
time is the maximal time interval for which the channel
auto-correlation function does not decrease below a pre-specified
threshold. The question is how to select the threshold value.
Intuitively, a more sensitive system (higher SNR) will choose a
higher threshold value, and hence will see a shorter coherence
time.} The paper also includes a detailed study of the relation
between the effective coherence time and the system SNR.

The noncoherent underspread WSSUS channel capacity was well
characterized for the high SNR limit and for the low SNR limit
(which is equivalent to the large bandwidth limit). For the low SNR
limit, the capacity of the noncoherent underspread channel was shown
to be proportional to $p_x^2 \Tc / 2$, where $p_x$ is the SNR and
$\Tc$ is the channel coherence time
\CommentApril{\cite{durisi2010noncoherent}--\nocite{Medard:02,sethuraman2005cpu}\cite{sethuraman2009low}}.
These results were obtained for different channel models with
different constraints. Medard and Gallager \cite{Medard:02} analyzed
a stationary wideband channel with quadratic power constraint, and
presented a novel definition of the channel coherence time which is
calculated from the channel correlation function. In this case, the
proportionality constant is the ratio between the transmitted signal
fourth moment and the square of its second moment (also termed
\textit{Kurtosis}). The results in that work were derived for the
wide bandwidth limit (in which the capacity depends on the number of
resolvable channel taps, see also
\cite{Telatar:00,porrat2007channel}). In this paper, the channel
bandwidth is finite, and we address the above result only in its low
SNR limit interpretation.

Sethuraman and Hajek \cite{sethuraman2005cpu} considered the block
fading model with a peak power constraint. In that case the channel
coherence time is the length of a block, and the proportionality
constant was shown to be the ratio between the peak and average
powers. They also considered a flat fading stationary channel model
(i.e., without multipath spread) and achieved equivalent results,
but using measures of the channel spectral distribution function
(with no definition of channel coherence time). \CommentApril{In a
later work together with Wang and Lapidoth \cite{sethuraman2009low},
they considered delay-separable frequency selective fading channels,
and showed that the capacity is actually proportional to $p_x^2
(\Tc-1) / 2$;  here the definition of the channel coherence time is
the identical to the one used by Medard and Gallager
\cite{Medard:02} but the term coherence time is not used. (Note that
for underspread channels the difference between $\Tc$ and $\Tc-1$ is
negligible.)}

\CommentApril{Durisi et al. used a similar channel model, but paid
more attention to the discretization of the continuous time channel
(and did not require the channel to be delay separable). As a
result, they had worked with a time-frequency channel transfer
function (which resembles in nature to the OFDM model used herein).
They also did not use the term coherence time, but their results
revealed the same behavior of the capacity in the low SNR limit,
using a channel measure which is an extension of Medard and Gallager
\cite{Medard:02} coherence time to the time-frequency channel. (In
their work they also analyzed the capacity with per frequency peak
power constraint. This power constraint is not discussed herein, as
it is quite different and leads to a significantly different
capacity behavior.) }

For the high SNR limit, the main issue is whether the channel
estimation error can decrease to 0 as the SNR grows to infinity. The
channel fading is termed \textit{regular} if it cannot be completely
predicted based on the full knowledge of its past \cite{Doob:53}. An
alternative definition uses the covariance matrix of the channel
fading, and the channel is said to be regular if this matrix is full
rank \cite{Liang:04}. For regular channels, at high enough SNR, the
capacity in the absence of channel knowledge is significantly lower
than the capacity with perfect channel knowledge, and grows only
double logarithmically with SNR \cite{Lapidoth:03,Lapidoth:05,koch2005fna}. On the other hand, for
irregular channels the capacity can continue to grow logarithmically
with SNR, (with possible degradation in the pre log constant)
\cite{Lapidoth:05,koch2005fna}.

Interestingly, no capacity expression directly depends on the common
definition of coherence time (e.g., \cite{Proakis:95}), i.e., the
inverse of the channel Doppler spread. For stationary channels, the
only coherence time definition that was shown to be proportional to
the capacity is the one given by Medard \CommentApril{and Gallager }
\cite{Medard:02}, which only applies in the low SNR limit. The
definition of effective coherence time in this paper coincides with
Medard's \CommentApril{and Gallager's }coherence time definition in
the low SNR limit, and extends it to higher SNR. Thus, the
effective coherence time can characterize the channel capacity for
all SNRs.

The paper is organized as follows: the notation and channel model
are presented in Sections \ref{sec: notation} and \ref{sec. system
model} respectively. The main results, including the definition of
the effective coherence time and the capacity bounds are given in
Section \ref{sec. main results}. A discussion of the results
including the properties of the effective coherence time, analysis
of the \CommentApril{bounding gap}, numerical examples and
comparison to known results are given in Section \ref{sec.
Discussion}. Note that the proof of each of the theorems (providing
capacity bounds) is partly based on information theory and partly on
estimation theory. For convenience, we collect those parts in
different sections. Hence, Theorem $n$ ($n=1,2,3$) is proved by
lemma $n$.a in Section \ref{sec. Channel capacity and bounds}
(information theory part) and lemma $n$.b in Section \ref{Sec.
Channel estimation} (estimation theory part). Concluding remarks are
given in Section \ref{sec: summary}.

\section{Notation}\label{sec: notation}
\CommentApril{Throughout the paper we use Roman boldface lower case
and upper case letters to denote random scalars and vectors,
respectively (e.g.,
 ${\bf x}$, ${\bf X}$). Roman italic lower case and upper case letters are
used for deterministic scalars and vectors, respectively (e.g.,
 $x$, $X$).
Deterministic matrices are represented by sans-serif letters and
random matrices by calligraphic letters (e.g., $ \textsf{X}$, $
\cal{ X}$ ). Exceptions to this rule are scalar quantities that are
commonly denoted by uppercase letters. These include the channel
parameters $N$ and $L$, the coherence time symbol ($\Tc$,
$\hTc(\cdot)$ and $\hTc[_0]$), and the capacity symbol $C$. Another
exception are operators that can result in a deterministic or random
quantity according to their operand, i.e., the spectrum operator
$\Sf_{\cdot}$ (see definition below) and the symbol set operator
$\mathcal{I}(\cdot)$ (see definition in Lemma \ref{L: peak
estimation bound}).

The identity matrix is denoted by $\It$, ${\rm 0}_N$ and ${\rm 1}_N$
denote the $N \times 1$ vectors of all zeros and all ones,
respectively. A $\otimes$ symbolizes the Kronecker product, a
$\dagger$ stands for transposition and complex conjugation, $\left\|
\mtr{A} \right\|$ denotes the absolute value of the determinant of
the matrix $\mtr{A} $, and $\diagM({X})$ is the diagonal matrix with
the elements of the vector ${X}$ on its diagonal.  The vector
stacking is defined as: ${X}_k^n \triangleq
[{X}_k^T,\ldots,{X}_n^T]^T$. The $b,d$ element of a matrix $\mtr{A}$
is denoted as $(\mtr{A})_{b,d}$ and the $b$-th element of the vector
$X_k$ is denoted as $x_{k,b}$. }

For discrete Fourier transform (DFT) analysis, $\Ft$ denotes the DFT
matrix with elements:
\begin{IEEEeqnarray}{rCl}\label{d: Fourier matrix}
    \label{e. DFT matrix}
        {\openmtr{F}}_{m,n}  = \frac{1}{{\sqrt N }}e^{ - 2j\pi mn/N},
\end{IEEEeqnarray}
$\tilde{X}=\Ft^\dagger {X}$ denotes the DFT of a vector, and each
element in this vector is termed a frequency bin. The spectrum of a
vector is marked by:
\begin{IEEEeqnarray}{rCl}
    \label{e. Spectrum}
        \Sf_{X}  = \diagM(\tilde{X})\diagM(\tilde{X})^\dagger .
\end{IEEEeqnarray}
The cross-covariance matrix of the vectors $\bf X$ and $\bf Y$ is
denoted as:
\begin{IEEEeqnarray}{rCl}
    {\openmtr C}_{{\bf X},{\bf Y}}=E\left[\left({\bf X}-E[{\bf X}]\right)\left({\bf Y}-E[{\bf Y}]\right)^\dagger \right],
\end{IEEEeqnarray}
and the auto-covariance matrix of the vector ${\bf X}$ is denoted as
${\openmtr C}_{{\bf X}}={\openmtr C}_{{\bf X},{\bf X}}$.

Finally, any matrix ordering in this paper is in the positive
definite sense, i.e., $\mtr{A}\ge\mtr{B}$ means that
$\mtr{A}-\mtr{B}$ is \CommentApril{positive semi-definite}.
\section{System model}\label{sec. system model}
\subsection{Channel model}
We use the OFDM model
\cite{tse2005fwc,Hlawatsch2011,bolckei2002cob,schafhuber2005maa}, in
which the DFT of the received signal is given by:
\begin{IEEEeqnarray}{rCl}
    \label{e. received signal model}
     {\FV{Y}{k}} &=& \sqrt{N} \diagM  ({\FV{H}{k}}) {\FV{X}{k}} + {\FV{W}{k}} \nonumber \\
                 &=& \sqrt{N} \diagM  ({\FV{X}{k}}) {\FV{H}{k}} + {\FV{W}{k}}
\end{IEEEeqnarray}
where $\FV{X}{k} \triangleq \Ft^\dagger  \RV{X}{k}$ and $\FV{H}{k}
\triangleq \Ft^\dagger  \RV{H}{k}$ are the DFT of the input vector
and channel impulse response at the $k$-th symbol respectively, and
both are vectors of length $N$. Each input vector constitutes an
OFDM symbol. The frequency domain noise $\FV{W}{k} \triangleq
\Ft^\dagger \RV{W}{k}$ is a zero mean complex Gaussian vector with
covariance matrix: $
    \CommentApril{{\openmtr{C}}_{{\bf{\tilde W}}}}  = {\It}$. The multiplication by
    $\sqrt{N}$ follows from the definition of the DFT
    matrix, (\ref{d: Fourier matrix}). The channel itself is better characterized in the time domain where
$\RV{H}{k}=[\RV{h}{k,0},\ldots,\RV{h}{k,L-1},0,\ldots,0]^T$, and $L$
is termed the \CommentApril{channel delay spread}.

This model is an appropriate representation of an underspread channel in
which we assume $\Tc \gg  N \gg
     L-1$,\footnote{\CommentApril{We use the notation $N \gg
     L-1$ in order to emphasize that in the flat fading case ($L=1$) it is sufficient to use $N=1$}} where $\Tc$ is the channel coherence time \cite{Proakis:95}.
     The assumption guarantees that the channel Doppler
     spread is small enough to be negligible, and that the channel
     impulse response can be considered constant throughout the $N$
     samples that constitute an OFDM symbol (for analysis of faster fading channels see
\cite{Medard:00}). The small \CommentApril{channel delay spread},
$L$, permits to neglect the time required for cyclic
     prefix ($L-1$ samples after each OFDM symbol).

In Section \ref{sec. main results} we define a new parameter, the
\textit{effective coherence time}, $\hTc(p_x)$, which identifies the
effective number of received samples that can be used for channel
estimation. The impact of the effective coherence time on the
channel capacity will be treated in later sections.

The mathematical analysis in the paper requires the following
assumptions:

\renewcommand{\theassumption}{\Roman{assumption}}
\begin{assumption}
    \label{assump T>N>L}
     The channel is underspread, i.e., the channel coherence time and the effective coherence time (measured in
    channel samples) are large enough so that: $\Tc,\hTc(p_x) \gg  N $.
\end{assumption}
\begin{assumption}
    \label{assump Gaussian channel}
     The channel has a proper complex Gaussian distribution (\cite{neeser1993pcr}).
\end{assumption}
\begin{assumption}
    \label{assump X iid H}
     \CommentApril{The transmitter has no knowledge on the channel
     realization.}
\end{assumption}
\begin{assumption}
    \label{assump WSSUS}
     The channel is ergodic, (wide sense) stationary, with uncorrelated scattering (WSSUS channel).
\end{assumption}
\begin{assumption}
    \label{assump kronecker}
     The auto-correlation function of the different channel taps is identical up to a multiplication by a scalar.
\end{assumption}
\begin{assumption}
    \label{assump frequency flat}
     The expectation of the channel impulse response is frequency
     flat.
\end{assumption}

\CommentApril{Assumption \ref{assump T>N>L} is actually not needed
for the analysis presented herein. We state it as our first
assumption because it is the basis for the OFDM model. If the
coherence time is not significantly larger than the symbol length,
then our model, which implicitly assumes that the channel does not
change during an OFDM symbol, will not be realistic.}

Assumption \ref{assump X iid H} stems from the fact that the system
has no feedback\CommentApril{. As a result, the transmitted messages
are statistically independent from the channel}. The requirement of
uncorrelated scattering in Assumption \ref{assump WSSUS} is needed
mostly for low SNR where we show that it is better to concentrate
the transmission power in a single frequency bin. The uncorrelated
scattering guarantees that if in one OFDM symbol we transmit in a
certain frequency bin, the best channel estimate for the next OFDM
symbol will be achieved at the same frequency bin (see derivation in
Appendix \ref{app: Minimization of the estimation error}).

Assumptions \ref{assump kronecker} and \ref{assump frequency flat}
somewhat narrow the scope of the analysis. However, we keep them to
simplify the mathematical derivation, and to reach closed-form
expressions. Assumption \ref{assump kronecker}, although not
necessarily realistic, is common to many simple channel models
(e.g., assuming that all multipath components have identical doppler
spread). \CommentApril{Channels that satisfy this assumption are
sometimes referred to as {\it delay separable channels}
\cite{sethuraman2009low}.} Assumption \ref{assump frequency flat}
basically states that at most one channel tap has an expectation
different from zero, and it includes the case of Rayleigh fading
(the expectation is zero) and the case of line of sight (LOS)
propagation (one channel tap has an expectation different from
zero).

Using Assumption \ref{assump Gaussian channel}, the stacking of the
channel vector has a proper complex Gaussian distribution,
$\RV[k]{H}{0}\sim \CN (E[\RV[k]{H}{0}],{\openmtr
C}_{\RV[k]{H}{0}})$\CommentApril{. Using Assumption \ref{assump
WSSUS} the amplitudes of the different channel taps (in the time
domain, $\RV{h}{k,0},\ldots,\RV{h}{k,L-1}$) are statistically
independent. Using also Assumption \ref{assump kronecker} the
covariance matrices of the channel taps are identical up to a
multiplication by a scalar. Considering $k+1$ OFDM symbols, we
denote by $\Delta_k$ the single-tap $(k+1)\times(k+1)$ Toeplitz
covariance matrix, in which each element is given by
\begin{IEEEeqnarray}{rCl}
({{\Delta}_{k}})_{i,j}={{\openmtr
C}_{\RV{h}{i,0},\RV{h}{j,0}}}/{{\openmtr C}_{\RV{h}{0,0}}}.
\end{IEEEeqnarray}
We also define the $k \times 1$ vector ${{D}_{k}}$ in which the
$i$-th element is given by
\begin{IEEEeqnarray}{rCl}
{d}_{k,i}=({{\Delta}_{k}})_{i,k},\quad i=0,\ldots,k-1.
\end{IEEEeqnarray}
 This vector
is the correlation between the channel tap value at the $k$-th
symbol and its value at all previous symbols.
 Considering all channel taps, the covariance
matrix of the stacked channel vector $\RV[k]{H}{0}$ is:}
\begin{IEEEeqnarray}{rCl}
\label{d: kron channel matrix}
    {\openmtr C}_{\RV[k]{H}{0}}={{{\Delta}_{k}}} \otimes
    {\openmtr{C}}_{\RV{H}{0}}=
    \left( \begin{IEEEeqnarraybox}[][c]{c?c?c?c}
    ({{\Delta}_{k}})_{0,0} {\openmtr{C}}_{\RV{H}{0}} & ({{\Delta}_{k}})_{0,1} {\openmtr{C}}_{\RV{H}{0}} & \ldots & ({{\Delta}_{k}})_{0,k} {\openmtr{C}}_{\RV{H}{0}} \\
    ({{\Delta}_{k}})_{1,0} {\openmtr{C}}_{\RV{H}{0}} & ({{\Delta}_{k}})_{1,1} {\openmtr{C}}_{\RV{H}{0}} & \ldots & ({{\Delta}_{k}})_{1,k} {\openmtr{C}}_{\RV{H}{0}} \\
    \vdots & \vdots & \ddots & \vdots \\
    ({{\Delta}_{k}})_{k,0} {\openmtr{C}}_{\RV{H}{0}} & ({{\Delta}_{k}})_{k,1} {\openmtr{C}}_{\RV{H}{0}} & \ldots & ({{\Delta}_{k}})_{k,k} {\openmtr{C}}_{\RV{H}{0}}
    \end{IEEEeqnarraybox}
    \right),
\end{IEEEeqnarray}
and
${\openmtr{C}}_{\RV{H}{0}}$ %LM%
is an $N \times N$ diagonal matrix\CommentApril{, in which each
diagonal element represents the power of one channel tap }(note that
due to Assumption \ref{assump WSSUS}, \CommentApril{we can write }
${\rm{C}}_{\RV{H}{k}}={\openmtr{C}}_{\RV{H}{0}}$ for any $k$).
 \CommentApril{In the following we mostly consider the frequency domain channel, $\FV{H}{k}=\Ft^\dagger\RV{H}{k}$. Its covariance matrix is given by ${\openmtr C}_{\FV{H}{k}}=\Ft^\dagger {\openmtr C}_{\RV{H}{k}} \Ft$}. For normalization purposes we will use
${\rm Tr}[{\openmtr{C}}_{\RV{H}{0}}]=1$ (and hence all elements on
the diagonal of ${\openmtr{C}}_{\FV{H}{0}}$ equal $1/N$).

\CommentApril{Another quantity which is useful in the
characterization of the channel is the channel conditional
covariance matrix given the past transmitted and received symbols:
\begin{IEEEeqnarray}{rCl}
\tilde{\mathcal {\bm \Leps}}_k  &=&
E\left[(\FV{H}{k}-E[\FV{H}{k}])(\FV{H}{k}-E[\FV{H}{k}])^\dagger|\FV[k-1]{X}{0}
,\FV[k-1]{Y}{0}\right].
\end{IEEEeqnarray}
This conditional covariance matrix is in general a random quantity
since it depends on the random vectors $\FV[k-1]{X}{0}$ (as the channel
and the noise are jointly Gaussian, it does not
depend on $\FV[k-1]{Y}{0}$). In some cases (e.g., constant amplitude
modulations) the resulting conditional covariance matrix is
deterministic (does depend on $\FV[k-1]{X}{0}$), and hence will be
denoted by $\tilde{\mathcal { \Leps}}_k$. We will also use the limit
as both time and SNR tend to infinity $\Leps_\infty^\infty = \lim_{k
\to \infty} \lim_{p_x \to \infty}\tilde{\mathcal { \Leps}}_k$.
Further details on the conditional channel distribution given the
past transmitted and received symbols can be found in Section
\ref{Sec. Channel estimation}. }

\subsection{Power constraints}
The input signal has constraints both on its average power and on
its ``peakiness". Defining the OFDM symbol power and power matrix
as:
\begin{IEEEeqnarray}{rCl}\label{d: power variables}
\RV{p}{k} =\FV[\dagger]{X}{k}\FV{X}{k}, \quad \quad \Pc_{k}=\diagM
([\RV{p}{0},\ldots,\RV{p}{k}]),
\end{IEEEeqnarray}
respectively, we consider two types of constraints preventing
the use of very high peak powers. The first, {\em peak constraint},  limits the
peak power of each OFDM symbol:
%LM%
\begin{IEEEeqnarray}{rCl}
    \label{e. Peak constraint}
        \RV{p}{k} \le N p_x , \quad  \forall k ,
\end{IEEEeqnarray}
\CommentApril{with probability 1 (The peak power constraint does not
involve statistical averaging, and therefore it is more relevant in
practical systems)}.

The second constraint, {\em quadratic constraint}
\cite{Medard:02,Telatar:00}, limits the first and second moments of
the OFDM symbol power:
\begin{IEEEeqnarray}{rCl}
    \label{e. frequency framed Quadratic constraint}
    E[\RV{p}{k} ]\le N p_x, \quad E[\RV[2]{p}{k} ]\le \alpha N^2
    p_x^2 , \quad  \forall k ,
\end{IEEEeqnarray}
where %LM%
$\alpha \ge 1$ %LM%
is a positive constant. This constraint limits the probability of
very high peak powers, but still permits a relatively simple analysis (for example
by allowing Gaussian signaling).

As the noise variance is normalized to 1, in the following we will
mostly refer to $p_x$ as the system SNR.
\section{Main results}\label{sec. main results}
In this section we introduce without proof upper and lower bounds on
the capacity of the noncoherent channel described in the previous
section. These bounds are characterized only by the system SNR and
the {\textit{effective coherence time}}:
\begin{definition}\label{def: effective coherence time}
The {\textit{effective coherence time}} is given by:
\begin{IEEEeqnarray}{rCl}\label{d: effective coherence time}
    \hTc(p_x)
    &=&
        2N \lim \limits_{k \to \infty }
         {{D}_{k}^\dagger}
        \left( N p_x \left[{{\Delta}_{k-1}} -
        {{D}_{k}} {{D}_{k}^\dagger}  \right]
        + \It
        \right)^{ - 1}
        {{D}_{k}}
        +N.
\end{IEEEeqnarray}
\end{definition}

As it will be shown, the effective coherence time characterizes the
"effective number of channel samples usable for channel estimation".
Although the previous statement is not precise now, it will be
discussed further in Subsection \ref{sub. Effective coherence time}.

\begin{theorem}\label{T: upper bound}
    The capacity of a channel with a peak power constraint is upper bounded by:
\begin{IEEEeqnarray}{rCl}
        C \le \min\left({\rm UB}_{\rm low}^{({\rm pk})} (p_x),{\rm UB}_{\rm coh}
        (p_x)\right),
\end{IEEEeqnarray}
and the capacity of a channel with a quadratic power constraint is
upper bounded by:
\begin{IEEEeqnarray}{rCl}
        C \le \min\left({\rm UB}_{\rm low}^{({\rm qd})} (p_x),{\rm UB}_{\rm coh}
        (p_x)\right)
\end{IEEEeqnarray}
where:
\begin{IEEEeqnarray}{rCl}\label{e: T1 coh UB}
        {\rm UB}_{\rm coh} (p_x) = E \left[ {\log \left( 1+ N p_x |{\FV{h}{0,0}}|^2  \right)}\right]
\end{IEEEeqnarray}
\begin{IEEEeqnarray}{rCl}\label{e: T1 UB pk}
        {\rm UB}_{\rm low}^{({\rm pk})} (p_x)
        &=& N p_x \left|E \left[  \FV{h}{0,0} \right]
        \right|^2 +p_x
        -\frac{1}{N} \log \left( 1+\frac{N p_x }{1+\frac{p_x}{2}(\hTc(p_x)-N)}
        \right)
\end{IEEEeqnarray}
\begin{IEEEeqnarray}{rCl}\label{e: T1 final_UB_low_qd}
        {\rm UB}_{\rm low}^{({\rm qd})} (p_x) &=&
        N p_x \left|E \left[  \FV{h}{0,0}\right]\right|^2
        +
        \alpha \frac{p_x^2(\hTc(p_x)-N)}{2+p_x(\hTc(p_x)-N)}
        +\frac{1}{2} \alpha N p_x^2 .
\end{IEEEeqnarray}
\end{theorem}
{\it Proof of Theorem \ref{T: upper bound}:} see Result 1, Lemma
\ref{L: ub low pk} and Lemma \ref{L: peak estimation
bound}.\footnote{The proof of each theorem is divided into two
lemmas. In the first lemma we use information-theoretical arguments
to prove the theorem assuming a knowledge of the conditional channel
distribution (defined by the conditional channel covariance matrix),
while in the second lemma we bound the conditional channel
covariance matrix.}

The term ${\rm UB}_{\rm coh} (p_x)$ in the bound is the capacity of
the coherent channel\CommentApril{ with only average power
constraint}. We will show in Subsection \ref{sub: Bound tightness}
that it is \CommentApril{useful} in the medium-to-high SNR regime.
\CommentApril{For} low SNR the coherent channel capacity is not
achievable and tighter bounds are described by ${\rm UB}_{\rm
low}^{({\rm pk})} (p_x)$ and ${\rm UB}_{\rm low}^{({\rm qd})}
(p_x)$. For low enough SNR, using
$\log(1+x)=x+\frac{1}{2}x^2+o(x^2)$ and $ 1/(1+x)=1-x+o(x)$, these
bounds can be approximated as:
\begin{IEEEeqnarray}{rCl}\label{e: low snr approx UB peak}
        {\rm UB}_{\rm low}^{({\rm pk})} (p_x)
        &\approx& N p_x
        \left|E \left[  \FV{h}{0,0} \right]
        \right|^2
        + \frac{p_x^2}{2}\CommentApril{\hTc(p_x)},
\end{IEEEeqnarray}
\begin{IEEEeqnarray}{rCl}\label{e: low snr approx UB quad}
        {\rm UB}_{\rm low}^{({\rm qd})} (p_x)
        &\approx&
        N p_x \left|E \left[  \FV{h}{0,0}\right]\right|^2
        +
        \alpha \frac{p_x^2}{2} \hTc(p_x).
\end{IEEEeqnarray}
If the channel has a constant term ($|E [ \FV{h}{0,0}]|$) different
from zero,  this is the dominant term. Otherwise, the bounds are
proportional to $p_x^2\hTc(p_x)$, but the quadratic power constraint
allows a capacity which is $\alpha$ times larger.

For the lower bounds, we choose specific input distributions
(modulations). For low SNR it is important to use a modulation that
maximizes the estimation accuracy subject to the power constraints.
Such a maximization is achieved by constant amplitude modulation,
and
 therefore we derive a
lower bound using Quadrature Phase Shift Keying (QPSK). As it was
noted in previous works (e.g.,
\cite{sethuraman2005cpu,porrat2007channel}), in the low SNR regime
it is better to use only part of the time and/or part of the
frequency band. Using only part of the available degrees of freedom
reduces the number of parameters that need to be estimated and hence
results in better estimation. In the following bounds we allow the
use of only \CommentApril{$r$} out of the $N$ frequency bins, and
allow transmitting in $1/\beta$ of the time. In the time domain the
signal is transmitted in long blocks (significantly larger than the
effective coherence time), and the silent periods between
\CommentApril{blocks results in a }duty cycle of $1/\beta$. The
resulting bound is:
\begin{theorem}\label{T: QPSK lower bound}
The capacity of a channel with a peak power constraint is lower
bounded by:
\begin{equation}
        C \ge \max_{r\le N} {\rm LB}_{\rm QPSK} (p_x,r,1)
\end{equation}
and the capacity of a channel with a quadratic power constraint is
lower bounded by:
\begin{equation}
        C \ge \max_{r\le N} \max_{1\le \beta \le \alpha} {\rm LB}_{\rm QPSK} (p_x,r,\beta)
\end{equation}
where
\begin{equation}\label{e: QPSK LB}
{\rm LB}_{\rm QPSK}(p_x,r,\beta)=\left\{
\begin{array}{ll}
    \frac{r}{N\beta}
    E\left[ C_{{\rm QPSK}}\(\frac{ \left|\sqrt{N}E \left[  \FV{h}{0,0}\right]+\sqrt{
    1-\frac{1}{1+\frac{\beta}{r}\frac{p_x}{2}\left(\hTc\left(\frac{\beta}{r} p_x\right)-N\right)}}\uv \right|^2}
    {\frac{r}{N}\frac{1}{\beta p_x}+
        \frac{1}{1+\frac{\beta}{r}\frac{p_x}{2}\left(\hTc\left(\frac{\beta}{r} p_x\right)-N\right)}}\) \right]
& r<N \\
\frac{1}{\beta}
    E\left[ C_{{\rm QPSK}}\(\frac{ \left|\sqrt{N}E \left[  \FV{h}{0,0}\right]+\sqrt{1-\frac{1}{1+\frac{\beta}{L} \frac{p_x}{2}\left(\hTc\left(
\beta p_x\right)-N\right)}}\uv \right|^2}
    {\frac{1}{\beta p_x}+ \frac{1}{1+\frac{\beta}{L} \frac{p_x}{2}\left(\hTc\left(
\beta p_x\right)-N\right)}}\) \right] & r=N
\end{array} \right.
    , \IEEEeqnarraynumspace
    \end{equation}
%\CommentDeleteApril{$C_{{\rm QPSK}}(\rho)=log4-\sqrt{\frac{2}{\pi}}\int_{-\infty}^{\infty}e^{-y^2/2}\log \left( 1+ e^{\rho^2-2y\rho}\right) dy$}
\CommentApril{\begin{IEEEeqnarray}{rCl}\label{d: C_QPSK}
C_{{\rm
QPSK}}(\rho)=\log4-\sqrt{\frac{2}{\pi}}\int_{-\infty}^{\infty}
e^{-y^2/2} \log \left( 1+ e^{-2(\rho+\sqrt{\rho}y)}\right) dy
\end{IEEEeqnarray}}
is the achievable mutual information for
QPSK transmission over AWGN channel with SNR equal to $\rho$
(\cite{baccarelli2000ssb}) and ${\bf u} \sim {\cal CN}(0,1)$ is a
proper complex Gaussian random variable.
\end{theorem}

{\it Proof of Theorem \ref{T: QPSK lower bound}:} see Lemma \ref{L:
QPSK LB} and Lemma \ref{L: QPSK estimaiton error}.

Note that if the channel has zero mean  ($E [ \FV{H}{0}]=0$) then
the bound in (\ref{e: QPSK LB}) simplifies to:
\begin{IEEEeqnarray*}{rCl}
{\rm LB}_{\rm QPSK}(p_x,r,\beta)=\left\{
\begin{array}{ll}
    \frac{r}{N\beta}
    E\left[ C_{{\rm QPSK}}\(\frac{
\frac{N\beta^2}{r^2}\frac{p_x^2}{2}\left(\hTc\left(\frac{\beta}{r}
p_x\right)-N\right)\left|\uv \right|^2}
    {1+\frac{N}{r}{\beta p_x}+\frac{\beta}{r}\frac{p_x}{2}\left(\hTc\left(\frac{\beta}{r}
    p_x\right)-N\right)
        }\) \right]
& r<N \\
\frac{1}{\beta}
    E\left[ C_{{\rm QPSK}}\(\frac{ \frac{\beta^2}{L} \frac{p_x^2}{2}\left(\hTc\left(
\beta p_x\right)-N\right)\left|\uv \right|^2}
    {1+\beta p_x+ \frac{\beta}{L} \frac{p_x}{2}\left(\hTc\left(
\beta p_x\right)-N\right)}\) \right] & r=N
\end{array} \right.
    .
\end{IEEEeqnarray*}
This bound is especially interesting in the low SNR limit. In that
case, the \CommentApril{best} bound uses only a single frequency
bin, and the minimum allowed duty cycle. Using $C_{\rm QPSK}(x)= x
+o(x)$ and $1/(1+x)=1+o(1)$, the low SNR approximation of the above
bound for the peak power constraint is:
\begin{equation}\label{e: low snr approx LB peak}
        {\rm LB}_{\rm QPSK} (p_x,1,1)
        \approx
     N p_x\left|E \left[  \FV{h}{0,0}\right]\right|^2+
    \frac{p_x^2}{2}\left(\hTc\left(p_x\right) - N\right),
\end{equation}
and the low SNR approximation for the quadratic power constraint
is:
\begin{equation}\label{e: low snr approx LB quad}
        {\rm LB}_{\rm QPSK} (p_x,1,\alpha)
        \approx
        N p_x \left|E \left[  \FV{h}{0,0}\right]\right|^2+
    \alpha \frac{p_x^2}{2}\left(\hTc\left(\alpha p_x\right)-N\right).
\end{equation}
Comparing with  (\ref{e: low snr approx UB peak}) and (\ref{e: low snr
approx UB quad}) we see that the bounds are tight for the peak power
\CommentApril{constraint} \CommentApril{as long as
$\hTc\left(p_x\right) \gg N$,\footnote{\CommentApril{Much of the
work on OFDM is concerned with the peak-to-average power
 ratio (PAPR), while in this work we only consider the entire symbol
 energy. Yet, we note that in the low SNR limit the lower bound uses a
 single frequency bin, and hence the PAPR is 1.}}}
 while for the quadratic power constraint the bound tightness
depends on the properties of the effective coherence time.
\CommentApril{When the limit
$\hTc[_0]=\lim_{p_x \to  0} \hTc(p_x)$ exists, the bounds are
tight up to a factor of $(\hTc[_0]-N)/\hTc[_0]$. We will further
discuss properties of the effective coherence time in Subsection
\ref{sub. Effective coherence time}}.

For higher SNR we expect the coherent channel bound to be
achievable. In the coherent channel the capacity is achieved using a
wideband Gaussian modulation, and we expect it to be close to
optimal also for the noncoherent channel, in the appropriate SNR
regime. Several simulations performed using Gaussian modulation
showed convergence to the upper bound, and in Subsection \ref{sub.
Effective coherence time} we even discuss a simulation-based
approximation that holds for large enough effective coherence times
and can be used to approximately lower bound the channel capacity
using Gaussian modulation.

However, so far we have not been able to prove a useful lower bound
using Gaussian modulation. Instead, we present here a bound which is
based on truncated complex Gaussian modulation. This modulation is
defined in Section \ref{sec. Channel capacity and bounds}. In
essence, it uses a proper complex Gaussian distribution
with the condition of a minimal and maximal power for the signal
transmitted in each active bin. The presented bound will show that
indeed the coherent channel capacity is achievable for high SNR and
large enough effective coherence times.

For this bound we also require a zero mean channel ($E [
\FV{h}{0,m}]=0$). An alternative (simpler) bound that does not
require this assumption is presented in Appendix \ref{app:
aleternative TG bound}. This alternative bound is in general less
tight, and therefore we prefer to assume a zero mean channel and
focus on the following bound:

\begin{theorem}\label{T: TG lower bound}
The capacity of a channel with zero mean ($E [  \FV{h}{0,m}]=0$) and
a quadratic power constraint  is lower bounded by:
\begin{equation}\label{e: TG bound maximization quad}
        C \ge \sup_{\eta>0}\max_{r\le N} \max_{1\le\beta\le \frac{\alpha r(1+\eta)^2}{1+r(1+\eta)^2}} {\rm LB}_{\rm TG}^{({\rm qd})}
        (p_x,r,\beta,\eta)-\frac{r}{N\beta}\eta
\end{equation}
and the capacity of a channel with zero mean ($E [ \FV{h}{0,m}]=0$)
and a peak power constraint is lower bounded by:
\begin{equation}\label{e: TG bound maximization peak}
        C \ge \sup_{\eta>0}\sup_{\xi>\eta}\max_{r\le N} {\rm LB}_{\rm TG}^{({\rm pk})}
        (p_x,r,\eta,\xi)+\frac{r}{N}\log(e^{-\eta}-e^{-\xi})
\end{equation}
where
\begin{equation}\label{e: fianl LB TG}
{\rm LB}_{\rm TG}^{({\rm qd})}(p_x,r,\beta,\eta)=\left\{
\begin{array}{ll}
    \frac{r}{N\beta}
    E\left[ \log\(1+\frac{
    \frac{N}{r}\beta p_x \left|\uv \right|^2}
    {1+
        \frac{1+\frac{N}{r}\beta p_x}{\frac{\beta p_x}{\nu_{{\rm qd}}(\eta)}\frac{1}{2 r} \left(\hTc\left(\frac{\beta p_x}{r \nu_{{\rm qd}}(\eta)}
\right)-N\right)}}\)
        \right]
& r<N \\
\frac{1}{\beta}
    E\left[ \log\(1+\frac{ \beta p_x\left|\uv \right|^2}
    {1+\frac{1+ \beta p_x}{\frac{\beta p_x}{\nu_{{\rm qd}}(\eta)}\frac{1}{2 L} \left(\hTc\left(\frac{\beta p_x}{\nu_{{\rm qd}}(\eta)}
\right)-N\right)} }\) \right]  & r=N
\end{array} \right.
    , \IEEEeqnarraynumspace
    \end{equation}
\CommentApril{
\begin{equation}\label{e: fianl LB TG peak} {\rm
LB}_{\rm TG}^{({\rm pk})}(p_x,r,\eta,\xi)=\left\{
\begin{array}{ll}
    \frac{r}{N}
    E\left[ \log\(1+\frac{
    \frac{N}{r} \frac{p_x}{\xi} \(1+\frac{\eta e^{-\eta}-\xi e^{-\xi}}{e^{-\eta}-e^{-\xi}}\) \left|\uv \right|^2}
    {1+
        \frac{1+\frac{N}{r} \frac{p_x}{\xi} \(1+\frac{\eta e^{-\eta}-\xi e^{-\xi}}{e^{-\eta}-e^{-\xi}}\)}{\frac{ p_x}{\nu_{{\rm pk}}(\eta,\xi)}\frac{1}{2 r} \left(\hTc\left(\frac{ p_x}{r \nu_{{\rm pk}}(\eta,\xi)}
\right)-N\right)}}\)
        \right]
& r<N \\
    E\left[ \log\(1+\frac{
    \frac{p_x}{\xi} \(1+\frac{\eta e^{-\eta}-\xi e^{-\xi}}{e^{-\eta}-e^{-\xi}}\) \left|\uv \right|^2}
    {1+
        \frac{1+ \frac{p_x}{\xi} \(1+\frac{\eta e^{-\eta}-\xi e^{-\xi}}{e^{-\eta}-e^{-\xi}}\)}{\frac{ p_x}{\nu_{{\rm pk}}(\eta,\xi)}\frac{1}{2 L} \left(\hTc\left(\frac{ p_x}{\nu_{{\rm pk}}(\eta,\xi)}
\right)-N\right)}}\)
        \right]
 & r=N
\end{array} \right.
    , \IEEEeqnarraynumspace
    \end{equation}}
\begin{IEEEeqnarray}{rCl}
\nu_{{\rm qd}}(\eta)=(1+\eta)\log\(1+\frac{1}{\eta}\)
\end{IEEEeqnarray}
\begin{IEEEeqnarray}{rCl}
\nu_{{\rm pk}}(\eta,\xi)
=\frac{\xi\log(1+\frac{1}{\eta})}{1-e^{\eta-\xi}}
\end{IEEEeqnarray}
and ${\bf u} \sim \CN(0,1)$ is a proper complex Gaussian random
variable.
\end{theorem}

{\it Proof of Theorem \ref{T: TG lower bound}:} see Lemma \ref{L: TG
bound} and Lemma \ref{L: TG estimaiton error}.

The bound for the quadratic power constraint is
\CommentDeleteApril{very }tight for high SNR and high effective
coherence time. Together with Theorem \ref{T: QPSK lower bound} it
yields a pair of bounds encompassing the range of SNR for which the
effective coherence time is large enough. Considering in particular
the case of $\beta=1$ and $r=N$, we can see two penalty terms. A
capacity penalty term of $\eta$ (the second term on the right hand
side of (\ref{e: TG bound maximization quad})), and an estimation
penalty term in the denominator of (\ref{e: fianl LB TG}).
\CommentApril{To show the bound tightness we can choose for example
\begin{IEEEeqnarray}{rCl}\label{e: decreasing eta function}
\eta=\frac{1}{e^{\sqrt{\hTc(p_x)}}-1}.
\end{IEEEeqnarray}
Note that $\hTc(p_x)\ge N \ge 1$ and hence we have $\eta\le 1$ and
$\nu_{{\rm qd}}(\eta)\le 2\log\(1+\frac{1}{\eta}\)=2\sqrt{\hTc(
p_x)}$. Therefore, the estimation penalty term in (\ref{e: fianl LB
TG}) satisfies:
\begin{IEEEeqnarray*}{rCl}
\frac{1+ \beta p_x}{\frac{\beta p_x}{\nu_{{\rm qd}}(\eta)}\frac{1}{2
L} \left(\hTc\left(\frac{\beta p_x}{\nu_{{\rm qd}}(\eta)}
\right)-N\right)} \le \frac{1+  p_x}{\frac{ p_x}{\nu_{{\rm
qd}}(\eta)}\frac{1}{2 L} \hTc( p_x)} \le \frac{1+ p_x}{\frac{ p_x}{4
L} \sqrt{\hTc( p_x)}},
\end{IEEEeqnarray*}
and becomes negligible for large enough $\hTc( p_x)$. On the other
hand, the capacity penalty term is proportional to $\eta$, (\ref{e:
decreasing eta function}), and is also negligible for large $\hTc(
p_x)$.
 }
Thus, for high SNR and large enough effective coherence time, the
bound can be approximated as:
\begin{equation}
\sup_{0<\eta<1}{\rm LB}_{\rm TG}(p_x,N,1,\eta) \approx
    E\left[ \log\(1+p_x\left|\uv \right|^2\) \right].
    \end{equation}
 Since in this case $\FV{h}{0,0} \sim \CN (0,\frac{1}{N})$, this approximation converges
 to the coherent channel upper bound ${\rm UB}_{\rm coh} (p_x)$
 (\ref{e: T1 coh UB}).

\CommentApril{For the peak power constraint the situation is more
complicated due to the power decrease that is required to allow for
the truncated Gaussian modulation to satisfy the peak power
constraint. This power penalty term is given by $\frac{1}{\xi}
\(1+\frac{\eta e^{-\eta}-\xi e^{-\xi}}{e^{-\eta}-e^{-\xi}}\)$ (in
the nominator of (\ref{e: fianl LB TG peak})). For high enough
effective coherence time (using for example  $\eta$ as in (\ref{e:
decreasing eta function})) the estimation will be good enough so
that $\eta$ will be negligible and also the denominator of (\ref{e:
fianl LB TG peak}) will be very close to 1. In such case, at high
SNR and $r=N$ the capacity loss compared to the coherent channel
capacity converges to:
\begin{IEEEeqnarray*}{rCl}
        \log\left(\frac{1}{\xi} \(1-\frac{\xi e^{-\xi}}{1-e^{-\xi}}\)\right) +\log\left(1-e^{-\xi}\right)
        ,
\end{IEEEeqnarray*}
which is maximized by $\xi=1.79$ at a loss of $-1.21$ nats. Thus,
the bound will be tight only for very high SNR, where such capacity
loss is negligible, if the effective coherence time $\hTc(p_x)$ is
still large enough.}

\section{Discussion}\label{sec. Discussion}
\subsection{Effective coherence time}\label{sub. Effective coherence time}
The three bounds in the previous section show that the effective
coherence time can very well characterize the noncoherent channel
capacity. Yet, its definition:
\begin{IEEEeqnarray*}{rCl}
    \hTc(p_x)
    &=&
        2N \lim \limits_{k \to \infty }
         {{D}_{k}^\dagger}
        \left( N p_x \left[{{\Delta}_{k-1}} -
        {{D}_{k}} {{D}_{k}^\dagger}  \right]
        + \It
        \right)^{ - 1}
        {{D}_{k}}
        +N.
\end{IEEEeqnarray*}
 is not intuitive and needs a
discussion on its meaning and properties. In the following we
present some interesting properties of the effective coherence time,
followed by a discussion on the reasoning and motivation for each
property. In particular, we will also explain why to use the term
\textit{effective coherence time} for this quantity.

\textit{Properties of the effective coherence time}
\begin{enumerate}
%### is this a sufficient explanation ?
\item{Using wideband constant amplitude modulation (i.e., all OFDM frequency bins are active and all have identical amplitude, $r=N$), the conditional channel covariance matrix, given all past
transmitted and received symbols, is a diagonal matrix. The diagonal
element that corresponds to any channel tap such that
$({\openmtr{C}}_{\RV{H}{0}})_{l,l}>0$ is given by:
\begin{IEEEeqnarray}{rCl}
    \label{e. inverse tap etsimation error}
    \lim \limits_{k \to \infty } ({{ \Leps}}_k)_{l,l}^{-1}
&= & ({\openmtr{C}}_{\RV{H}{0}})_{l,l}^{-1} +\frac{
p_x}{2}\left(\hTc\left(p_x \cdot
({\openmtr{C}}_{\RV{H}{0}})_{l,l}\right)-N\right)
        .
\end{IEEEeqnarray}}\label{p: wideband QPSK}
\item{Using wideband constant amplitude
modulation, each diagonal element of the conditional channel
covariance matrix in the frequency domain, given all past
transmitted and received symbols, is upper bounded by:
\begin{IEEEeqnarray}{rCl}\label{e: propery 2}
    \lim \limits_{k \to \infty }(\tilde{\mathcal { \Leps}}_k)_{m,m}
    \le\frac{1}{N}\frac{1}{1+ \frac{p_x}{2 L}\left(\hTc\left(
p_x\right)-N\right)} .
\end{IEEEeqnarray}\label{p: freq wideband QPSK}
Equality in (\ref{e: propery 2}) is achieved if the channel
energy is concentrated into a single tap (i.e., if $L=1$).}
\item{Using narrowband (i.e., only a single frequency bin is active, using all of the OFDM symbol power, $r=1$) constant amplitude modulation, the relevant diagonal element of the conditional channel covariance matrix in the frequency domain, given all past
transmitted and received symbols, is given by:
\begin{IEEEeqnarray}{rCl}
    \frac{1}{1+\frac{p_x}{2}\left(\hTc\left(p_x\right)-N\right)}.
\end{IEEEeqnarray}}\label{p: narrowband QPSK}
\item{At the low SNR limit, define:
\begin{IEEEeqnarray}{rCl}
    \label{d. Tc max}
    \hTc[_0]=\lim_{p_x \to  0} \hTc(p_x)
        =2 N \lim \limits_{k \to \infty } {{D}_{k}^\dagger}
        {{D}_{k}}+N.
\end{IEEEeqnarray}
If this limit exists, it converges to the definition of coherence
time used by Medard \CommentApril{and Gallager }\cite{Medard:02}.
}\label{p: Tc0}
\item{For any $\lambda>1$ the effective coherence time satisfies:
\begin{IEEEeqnarray}{rCl}
    \hTc\left(p_x\right)/\lambda\le\hTc\left(\lambda p_x\right)\le\hTc\left(p_x\right).
\end{IEEEeqnarray}}\label{p: non increasing}
\item{If the prediction error is not zero, i.e.:
\begin{IEEEeqnarray}{rCl}
\Leps_\infty^\infty = \lim_{k \to \infty} \lim_{p_x \to \infty}
        \tilde{\mathcal { \Leps}}_k = \lim_{k \to \infty} {\openmtr{C}}_{\FV{H}{0}} [1 -
        {{D}_{k}^\dagger}
        {\Delta}_{k-1}^{-1}
        {{D}_{k}}]>0
\end{IEEEeqnarray}
then:
\begin{IEEEeqnarray}{rCl}
    \lim_{p_x \to \infty}
    \hTc(p_x) = N.
\end{IEEEeqnarray}
}\label{p: T infty}
\end{enumerate}

Properties \ref{p: wideband QPSK}--\ref{p: narrowband QPSK} are
shown in the proof of Lemma \ref{L: QPSK estimaiton error} in
Section \ref{Sec. Channel estimation}. They reflect the relation
between the effective coherence time and the conditional channel
distribution, which is the basis for the capacity bounds. They can
also be used to argue the relationship between the coherence time
and the effective coherence time. To particularize the results in a
case where we have a direct and intuitive relation between capacity
and coherence time, we consider a block fading channel
\cite{marzetta1999cmm}. In this model, the time axis is divided into
blocks of length $\Tc/N$, and the channel realization is fixed and
independent within each block. This channel is
\CommentApril{nonstationary}, and exhibits different statistics for
each symbol (as an example, the first symbol of each block has
correlation only with future symbols, while the last symbol of each
block has correlation only with past symbols). As a reference, we
assume that the $k$-th symbol is the middle symbol of a block and
that the block length is an odd multiple of the symbol length $N$.
It is easy to show\footnote{For example by using ${D}_{k}={\rm 1}_d
 $
 and ${\Delta}_{k-1}={\rm 1}_d^{\phantom{\dagger}} {\rm 1}_d^\dagger $ in (\ref{e. inverse tap etsimation error first
 one}) for the wideband case (with $\beta=1$), and in
(\ref{e. estimation error frequency minima}) for the narrowband
case, with $d=0.5(\Tc/N-1)$.} that the conditional channel
covariance matrix for the middle symbol of a block satisfies
Properties \ref{p: wideband QPSK}--\ref{p: narrowband QPSK} if we
substitute $\hTc(p_x)$ with $\Tc$, the block length. This similarity
motivates the term \textit{effective coherence time} as an extension
of the coherence time definition to more general channel models.

Our simulations have shown that Properties \ref{p: wideband
QPSK}--\ref{p: narrowband QPSK} can also well approximate the
conditional channel covariance matrix given all past transmitted and
received symbols when using Gaussian distributed input signals, as
long as the effective coherence time is large enough. The Gaussian
modulation is of interest as it achieves the coherent channel
capacity. So far, we have not been able to provide a formal proof of
this, and this is the reason why we used truncated Gaussian
modulation instead of Gaussian modulation for the capacity lower
bound in Theorem \ref{T: TG lower bound}. We demonstrate the
accuracy of this approximation in the simulation results presented
in Section \ref{sub: Numerical evaluation}.

Property \ref{p: Tc0} is easily verified from (\ref{d: effective
coherence time}). As long as $\hTc[_0]$ exists, the effective
coherence time can be seen as an extension of the coherence time
defined in \cite{Medard:02} to higher SNR. Note that the effective
coherence time gives a good characterization of the capacity even
for channels for which $\hTc[_0]$ does not exist.

The major difference between any channel coherence time definition
and the effective coherence time is that the latter is a function of
the system SNR. This may seem \CommentDeleteApril{as }a surprising
characteristic of a coherence time, since intuitively one would
expect the coherence time to characterize the channel, regardless of
the system parameters. This dependence of the effective coherence
time on SNR comes from the fact that at high SNR the system is more
sensitive to changes in the channel, and hence will effectively see
a shorter coherence time. Alternatively, one can say that a system
with higher SNR requires better channel estimations. Thus, such a
system will be able to use only channel measurements that present
higher correlation with the present channel state, so that an
increase in the SNR will reduce the number of useful measurements
and consequently reduce the effective coherence time.
\CommentApril{Note that some engineers actually regard the SNR as a
part of the channel and not as a system parameter (in the sense that
in many cases the transmission power, just like the channel, is a
limitation given to the system designer and not part of the system
design).}

Property \ref{p: non increasing} can be seen from (\ref{d: effective
coherence time}) by considering the eigenvalue decomposition of the
matrix ${{\Delta}_{k-1}} -
        {{D}_{k}} {D}_{k}^\dagger$. A more detailed description can be found in \cite{Bergel:10}. As reflected by Property \ref{p: non increasing}, the
effective coherence time is a nonincreasing function of the SNR, and
for most channel models it decreases as the SNR increases.

Property \ref{p: T infty} follows from Property \ref{p: freq
wideband QPSK} and deals with the high SNR extreme behavior of the
effective coherence time. The prediction error is the conditional
covariance of the channel given all past channel realizations (i.e.,
if the prediction error is zero then the channel is fully
predictable). In regular channel models $\Leps_\infty^\infty $ is
larger than zero \cite{Lapidoth:03}, and represents the inherent
uncertainty in the channel estimation given the full knowledge of
its past. Property \ref{p: T infty} shows that for regular channels
the effective coherence time converges to $N$ as the SNR grows to
infinity (as we assume that the channel does not change within an
OFDM symbol, the model cannot show a coherence time smaller than
$N$).

As we limit the present analysis to large effective coherence times,
for such regular channels the presented bounds will not be
tight for very high SNR. The looseness of the bounds for very high
SNR is in agreement with the results of Lapidoth \CommentApril{and
Moser }\cite{Lapidoth:03}, which showed that the coherent channel
capacity is not achievable in regular channels in the high SNR limit
(the capacity grows only doubly logarithmically with SNR).

\CommentApril{Although not proved herein, the effective coherence
time can significantly decrease even for irregular channels. As an
example consider a Clarke's fading channel
\cite{clarke1968statistical} with a doppler spread of $300$ Hz and an
OFDM symbol length of $66.7\mu$S (which corresponds to an LTE mobile
device operating at the $2$GHz band and moving at a speed of
$160$KMH). The effective coherence time of this channel at $6$dB SNR
is $50$ symbols, and the underspread assumption is well verified.
However, at a very high SNR of $116$dB the effective coherence time
drops to only $10$ symbols (at which point one might start
questioning the underspread assumption).}

\CommentApril{
\subsection{Bound gap at finite SNR}\label{sub: Bound
tightness} We have discussed the bound tightness at the high and low
SNR limits. In this subsection we discuss the gap between the bounds
in finite SNR. We will show that in most cases, if the coherence
time is large enough, this gap is not significant, and hence the
bounds describe the channel capacity very well. For simplicity, the
discussion in this subsection is limited to the case of a channel
with zero mean ($E [ \FV{H}{0}]=0$).

We start with the peak power constraint for
$p_x\le\frac{2}{\hTc(p_x)}$. Using $\log(1+x)\ge x-0.5x^2$ the upper
bound (\ref{e: T1 UB pk}) is upper bounded by:
\begin{IEEEeqnarray}{rCl}
        {\rm UB}_{\rm low}^{({\rm pk})} (p_x)
        &\le & \frac{ \frac{p_x^2}{2} }{1+\frac{p_x}{2}(\hTc(p_x)-N)} \cdot
        \frac{ \hTc(p_x)+\frac{1}{2}(\hTc(p_x)-N)^2 \cdot p_x
        }{1+\frac{p_x}{2}(\hTc(p_x)-N)}.
\end{IEEEeqnarray}
For the lower bound we substitute in (\ref{d: C_QPSK}) the
inequality\footnote{Using:
$\log(1+e^{-x})=\log2-\frac{x}{2}+\log\cosh\(\frac{x}{2}\)$, and
$\cosh(x)\le e^{\frac{1}{2}x^2}$.} $\log(1+e^{-x})\le \log
2-\frac{x}{2}+\frac{x^2}{8}$ which \CommentApril{results in}
$C_{{\rm QPSK}}(\rho)\ge \rho-\rho^2$. Substituting the inequality
and $r=\beta=1$ in the lower bound (\ref{e: QPSK LB}),
 we get:
\begin{equation}\label{e: LB on LB QSPK peak}
{\rm LB}_{\rm QPSK}(p_x,1,1)\ge
    \frac{
    \frac{p_x^2}{2}\left(\hTc\left(p_x\right)-N\right)}
    {1+\frac{p_x}{2}\left(\hTc\left(p_x\right)-N\right)+N p_x
        } -\frac{N \frac{p_x^4}{2}\left(\hTc\left(p_x\right)-N\right)^2}
    {\left(1+\frac{p_x}{2}\left(\hTc\left(p_x\right)-N\right)+N p_x
        \right)^2}.
\end{equation}
These upper and lower bounds are almost identical for very low SNR and
high effective coherence time, and start diverging as the SNR increases. For
the SNR range of interest, the largest gap appears at the highest
SNR: $p_x=\frac{2}{\hTc(p_x)}$. Defining the factor
$\kappa=\frac{\hTc\left(p_x\right)-N}{\hTc(p_x)}$ (which is
approximately $1$ for large enough effective coherence time) we have
$N p_x=2-2\kappa$:
\begin{IEEEeqnarray}{rCl}\label{e: peak UB at 1 over Tc}
        {\rm UB}_{\rm low}^{({\rm pk})} \(\frac{2}{\hTc(p_x)}\)
        &\le &
        \frac{p_x}{1+\kappa}\frac{ 1+\kappa^2
        }{1+\kappa},
\end{IEEEeqnarray}
\begin{equation}\label{e: peak LB at 1 over Tc}
{\rm LB}_{\rm QPSK}\(\frac{1}{\hTc(p_x)},1,1\)\ge
    \frac{
    \kappa p_x}
    {3-\kappa} -\frac{4(1-\kappa) \kappa^2 p_x}
    {\left(3-\kappa
        \right)^2}.
\end{equation}
For example, if $p_x=\frac{2}{\hTc(p_x)}$ is satisfied for
$\hTc(p_x)=100 N$ then $\kappa=0.99$ and the gap between (\ref{e:
peak UB at 1 over Tc}) and (\ref{e: peak LB at 1 over Tc}) is only
2.5\%.

Next, for $\frac{2}{\hTc(p_x)} \le p_x \le \frac{1}{2} $ (and still
for the peak power constraint), we turn to a numerically verified
bound: For $\uv \sim \CN(0,1)$ and $\rho<0.5$ one can verify that:
\begin{IEEEeqnarray}{rCl}
E\left[C_{{\rm QPSK}}(\rho|\uv|^2)\right]\ge 0.97
E\left[\log(1+\rho|\uv|^2)\right].
\end{IEEEeqnarray}
Using this inequality in (\ref{e: QPSK LB}) and substituting $r=N$,
$\beta=1$ we have:
\begin{IEEEeqnarray}{rCl}
{\rm LB}_{\rm QPSK}(p_x,N,1)&=&
    E\left[ C_{{\rm QPSK}}\(\frac{ \frac{p_x^2}{2}\left(\hTc\left(
 p_x\right)-N\right)|\uv |^2}
    { \frac{p_x}{2}\left(\hTc\left(
 p_x\right)-N\right)+ L(1+p_x)}\) \right]
\nonumber \\ &=&
    E\left[ C_{{\rm QPSK}}\(\frac{ p_x|\uv |^2}
    { 1+\frac{ 2L(1+p_x)}{p_x\left(\hTc\left(
 p_x\right)-N\right)}}\) \right]
\nonumber \\ &\ge&
    0.97 E\left[ \log\(1+\frac{ p_x|\uv |^2}
    {1+\frac{2 L(1+p_x)}{p_x\left(\hTc\left(
p_x\right)-N\right)}}\) \right].
    \end{IEEEeqnarray}
Comparing this bound with the coherent channel upper bound, (\ref{e:
T1 coh UB}), and recalling that ${\FV{h}{0,0}} \sim \CN
(0,\frac{1}{N})$ one can evaluate the gap between the bounds. Again
this gap is largest at the high SNR limit, $p_x=0.5$. Considering
the example of $\hTc(0.5)= 100 N$ and $N=5L$, the gap between the
bounds is at most 4\%.

Thus, for the peak power constraint case the bounds are good if the
effective coherence time is large enough for SNRs of up to $p_x\le
0.5$. As stated in Section \ref{sec. main results}, for higher SNRs
the bounds are less tight for the peak power case (due to the power
decrease that is required to allow for the truncated Gaussian
modulation to satisfy the peak power constraint). The bounds can be
tight again only at much higher SNR where the power penalty term
(which results in an asymptotic difference of $1.2$ nats) becomes
negligible (if the effective coherence time is still large enough).

In the case of the quadratic power constraint we already stated the
bounds' tightness when the effective coherence time is large enough
for the high SNR limit and for the low SNR limit if the limit
$\hTc[_0]$ exists. We next quantify the bounding gap for finite SNR.

In the low SNR regime the bounds can be quite loose, depending on
the behavior of the effective coherence time as a function of the
SNR. Typically, the bounds are less tight at the point where the two
upper bounds intersect. If $\hTc(p_x)\gg N$ then the two upper
bounds ((\ref{e: T1 coh UB}) and (\ref{e: T1 final_UB_low_qd}))
intersect very close to $p_x=\frac{2}{(\alpha-1)\hTc(p_x)}$. Using
the same derivation as in (\ref{e: LB on LB QSPK peak}) and
substituting $r=1$ and $\beta=\alpha$, we can lower bound the QPSK
lower bound by:
\begin{IEEEeqnarray*}{rCl}\label{e: LB on LB QSPK quad}
{\rm LB}_{\rm QPSK}(p_x,1,\alpha)&\ge&
    \frac{
\alpha \frac{ p_x^2}{2}\left(\hTc\left(\alpha p_x\right)-N\right)}
    {
     1+   N\alpha
p_x+\frac{\alpha p_x}{2}\left(\hTc\left(\alpha p_x\right)-N\right)}
 -
    \frac{
 N\alpha^3 \frac{ p_x^4}{2}\left(\hTc\left(\alpha
p_x\right)-N\right)^2}
    {
\(     1+   N\alpha p_x+\frac{\alpha p_x}{2}\left(\hTc\left(\alpha
p_x\right)-N\right)\)^2}.
\end{IEEEeqnarray*}
Redefining $\kappa=\frac{\hTc\left(\alpha p_x\right)-N}{\hTc(\alpha
p_x)}$, and defining
$\upsilon=\frac{\alpha}{\alpha-1}\frac{\hTc(\alpha p_x)}{\hTc(p_x)}$
we have $\upsilon=\alpha p_x \hTc(\alpha p_x) /2$, $N \alpha
p_x=\upsilon(2-2\kappa)$ and
\begin{IEEEeqnarray*}{rCl}\label{e: LB on LB QSPK quad}
{\rm LB}_{\rm QPSK}\(\frac{2}{(\alpha-1)\hTc(p_x)},1,\alpha\)\ge
    \frac{
\upsilon\kappa p_x}
    {
     1+   \upsilon(2-\kappa)}
 -
    \frac{
2 \upsilon^3\kappa^2(2-2\kappa) p_x}
    {
\(   1+   \upsilon(2-\kappa)\)^2}
\end{IEEEeqnarray*}
while the upper bounds intersect very close to:
\begin{IEEEeqnarray}{rCl}
        {\rm UB}_{\rm low}^{({\rm qd})} \(p_x\)
        &\approx &
{\rm UB}_{\rm coh} (p_x)\approx p_x
        \end{IEEEeqnarray}

Nothing that for high enough effective coherence time $\kappa
\approx 1$, the bounding gap is mostly determined by the term
$\upsilon$. Using Property \ref{p: non increasing} from Subsection
\ref{sub. Effective coherence time}, $\upsilon$ is lower bounded by
$\upsilon\ge\frac{1}{\alpha-1}$ and hence the bounds can differ by
approximately a factor of $\alpha$. However, typically, the
effective coherence time will not change that fast. If the effective
coherence time is approximately constant at the relevant SNRs
($\hTc(\alpha p_x)\approx \hTc(p_x)$) the ratio between the bounds
will be approximately $\frac{2\alpha-1}{\alpha}$ (i.e., between
$1.5$ for $\alpha=2$ and $2$ for high $\alpha$).

For higher SNR, the bounding gap becomes smaller. Inspecting the
lower bound (\ref{e: TG bound maximization quad}), we have two
penalty terms compared to the upper bound (\ref{e: T1 coh UB}). The
first is a capacity penalty term (the second term on the right hand
side), and the second is an SNR penalty term (at the denominator of
(\ref{e: fianl LB TG}) inside the $\log$). As shown above, the
bounds are tight for high enough \CommentApril{SNR} and high enough
effective coherence time. Taking as a reference SNR of $p_x=1$
(${\rm UB}_{\rm coh}
        (p_x)>0.6$ nats), setting
$r=N$, $\beta=1$, and $\eta=0.005$ will result in a capacity penalty
term of $0.005$ nats. If the effective coherence time satisfies
$\hTc( p_x)>330L+N$ then the SNR penalty term will be less than
$0.3$ dB, and the gap between the bounds will be less than 5\%. For
higher SNRs (satisfying the same condition on the effective
coherence time) or for higher effective coherence times, the gap
will be even smaller. Note however that for high enough SNRs the
effective coherence time will typically decrease to a level that
will not allow coherent communication, and the bounds will
diverge.\footnote{The two lower bounds in this work assume a
standard coherent communication scheme, i.e., the receiver estimates
the channel based on past symbols, and then uses this estimate to
detect the next symbol.}

Numerical examples showing the bounds gap are shown in
the next subsection. }
\subsection{Numerical example}\label{sub: Numerical evaluation}

In order to visually demonstrate the bounds derived in the previous
sections, we evaluate them numerically for the auto-regressive
channel model of the first order (AR1). This channel model is
defined by a single parameter, $\gamma$,  the channel forgetting
factor, and is characterized by
${\Ct}_{\RV{H}{i},\RV{H}{j}}=\gamma^{|i-j|}\Ct_{\RV{H}{0}}$. The
effective coherence time for this channel can be calculated using
(\ref{d: effective coherence time}), and its low SNR limit is
$\hTc[_0]=N(1+\gamma^2)/(1-\gamma^2)$. Throughout this section we
assume $E[\RV{H}{k}]={\rm 0}$, equal power taps (i.e., $({\openmtr
C}_{\RV{H}{k}})_{l,l}=1/L$ for $0\le l < L$) and $N=30$, $L=5$.

\begin{figure}[t]
    \includegraphics[scale=0.9]{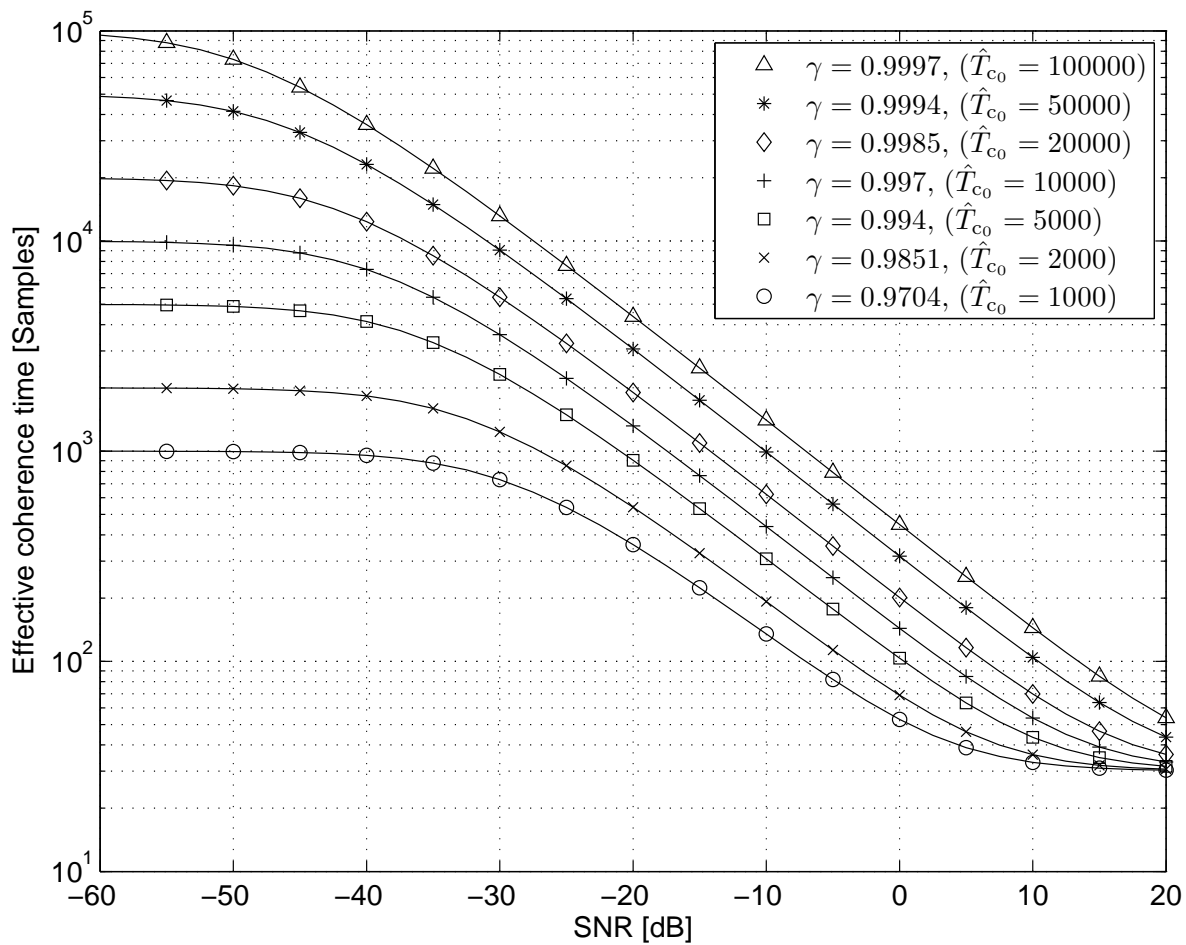}
    \caption{Effective coherence time of the AR1 channel model for various values of the channel forgetting factor.}
    \label{f:figure_Tc_Ar1}
\end{figure}

Figure \ref{f:figure_Tc_Ar1} shows the effective coherence time of
the AR1 channel for different values of the channel forgetting
factor. In all cases the effective coherence time reaches its limit,
$\hTc[_0]$, at low enough SNR (for $\hTc[_0]\ge 50,000$ this
convergence is not seen in the figure as it happens in lower SNR).
For higher SNR the effective coherence time is a decreasing function
of the SNR, until it reaches $\hTc(p_x)=N$ which is the lowest value
measurable in our model.

\begin{figure}[t]
    \includegraphics[scale=0.9]{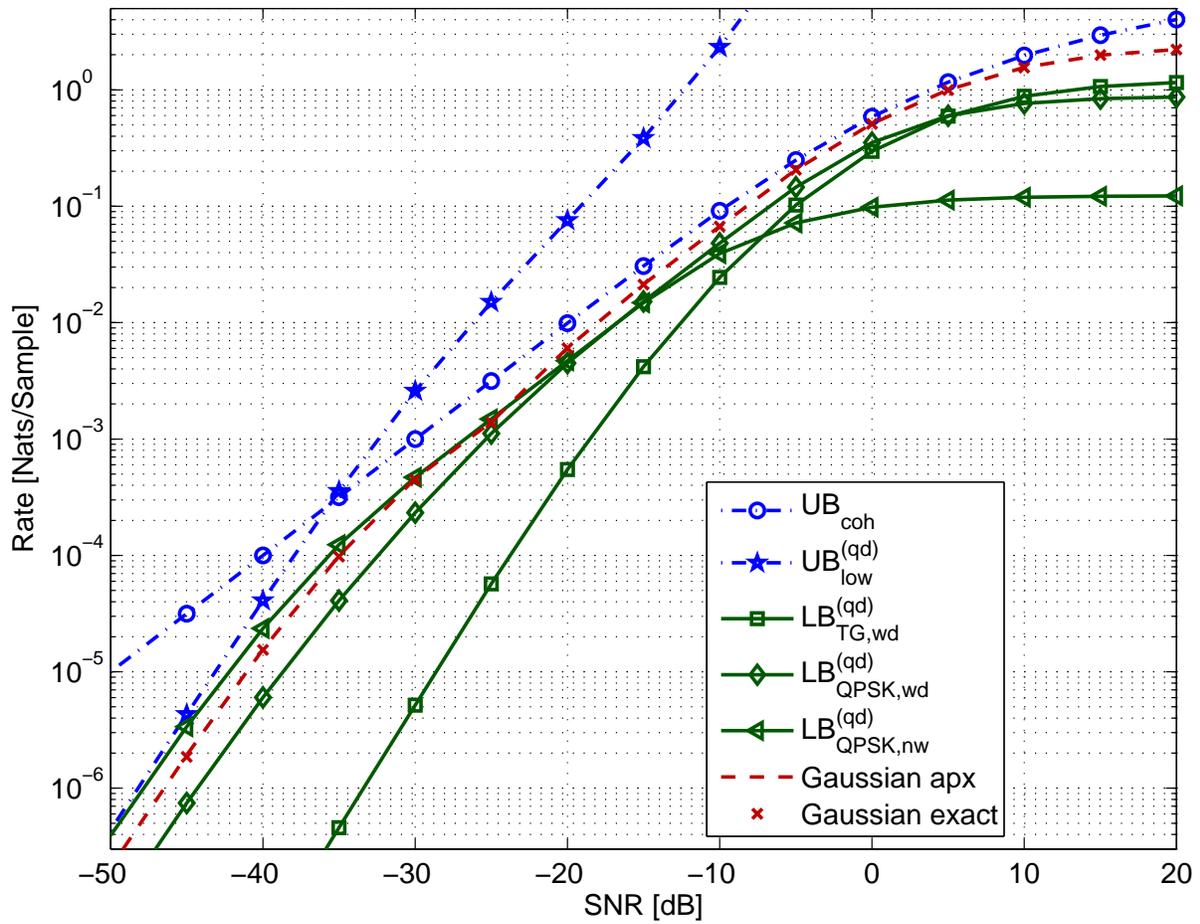}
    \caption{Upper and lower capacity bounds vs. SNR for AR1 fading channel ($\hTc[_0]=900$) and quadratic power constraint ($\alpha=10$).}
    \label{f:figure_AR1_1800}
\end{figure}

Figure \ref{f:figure_AR1_1800} shows the bounds on the capacity of
the AR1 channel with quadratic power constraint when the quadratic
constraint constant in (\ref{e. frequency framed Quadratic
constraint}) is set to $\alpha=10$. The figure shows the capacity
bounds when the channel forgetting factor is $\gamma=0.9672$
($\hTc[_0]=900$). The channel capacity is upper bounded by the low
SNR upper bound, ${\rm UB}_{\rm low}^{({\rm qd})}$ (\ref{e: T1
final_UB_low_qd}), which is effective for low SNRs, and by the
coherent channel upper bound, ${\rm UB}_{\rm coh}$ (\ref{e. known
channel bound ver 2}), which is effective in higher SNRs. In order
to demonstrate the role of the different lower bounds we draw 3
lower bound curves. Two of the bounds are based on the QPSK bound
(\ref{e: QPSK LB}). For low SNRs the tightest bound uses narrowband
QPSK signaling: ${\rm LB_{QPSK,nw}^{(qd)}}=\max_{1\le \beta \le
\alpha}{\rm LB}_{\rm QPSK}(p_x,1,\beta)$. For medium SNRs the
tightest bound uses wideband QPSK signaling: ${\rm
LB_{QPSK,wd}^{(qd)}}=\max_{1\le \beta \le \alpha}{\rm LB}_{\rm
QPSK}(p_x,N,\beta)$. For high SNRs the tightest bound uses the
truncated Gaussian signaling, (\ref{e: TG bound maximization quad}):
${\rm LB_{TG,wd}^{(qd)}}=\sup_{0<\eta<1} {\rm LB}_{\rm
TG}(p_x,N,1,\eta)$. Note that the upper bounds intersect at
$p_x=-38.5$ dB. At this point the ratio between the upper bounds and
lower bounds is $0.5$. For higher and lower SNR the bounds get
tighter, up to the point where, for high enough SNR, the effective coherence time
decreases too much and the bounds diverge.

The figure also demonstrates the achievable rates using Gaussian
signaling, and the accuracy of the approximation suggested in
Section \ref{sub. Effective coherence time}. Using Montecarlo
simulations, we evaluated the bound for proper complex Gaussian
input distribution (using (\ref{e: almost final TG}) with $\eta=0$
and the conditional channel covariance matrix from (\ref{e.
condtional channel distribution variance})). The resulting rate is
depicted in x-marks in Figure \ref{f:figure_AR1_1800} (such an
evaluation is of course feasible only for short coherence times).
The figure also depicts (in a dashed line) the resulting rate when
the conditional channel covariance matrix is approximated according
to Properties \ref{p: wideband QPSK}--\ref{p: narrowband QPSK} in
Subsection \ref{sub. Effective coherence time}  (which where derived
for the case of constant amplitude modulation, and are used here for
the approximation of the capacity in the case of Gaussian
modulation). As it can be seen, at least for the plotted case, the
suggested approximation is very accurate and the approximation error
is \CommentDeleteApril{completely }negligible. The large number of
simulations performed have supported this claim, and shown that the
approximation accuracy is even better for channels with longer
coherence times. Based on these observations we suggest that
Gaussian signaling may lead to an even tighter lower bound.

\begin{figure}[t]
    \includegraphics[scale=0.9]{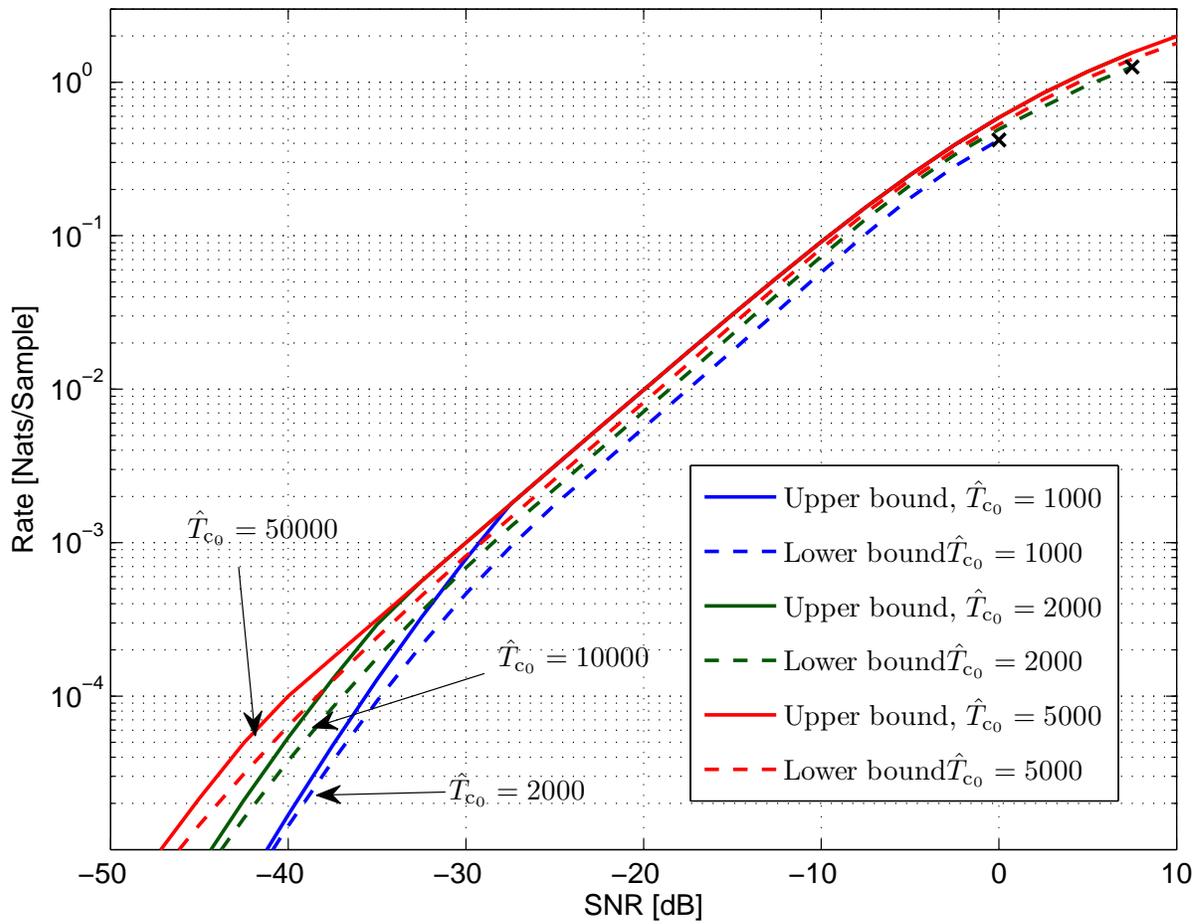}
    \caption{Upper and lower capacity bounds vs. SNR for AR1 fading channel ($\hTc[_0]=2000,10000,50000$) and quadratic power constraint ($\alpha=2$).}
    \label{f:figure_AR1_bounds}
\end{figure}

Figure \ref{f:figure_AR1_bounds} depicts the combined capacity
bounds for various values of the channel forgetting factor when the
quadratic constraint constant is $\alpha=2$. The upper bound
(depicted by solid lines) is the bound given by Theorem \ref{T:
upper bound}. The lower bound (depicted by dashed lines) is the
maximum of the bounds given by Theorems \ref{T: QPSK lower bound}
and \ref{T: TG lower bound}. The figure depicts the bounds for
forgetting factors of $\gamma=0.9851, 0.997, 0.9994$ (which
correspond to $\hTc[_0]=2,000,10,000,$ and $50,000$ respectively).

As it can be seen, the bounds are \CommentApril{good} in most of the
range. The bounds are \CommentApril{least} tight at SNR values in
which the two upper bounds intersect. In these SNR the ratio between
the \CommentApril{upper and lower bounds is $1.89$, $1.64$ and
$1.59$} for $\hTc[_0]=2,000$, $10,000$ and $50,000$ respectively,
very close to the ratio predicted in Section \ref{sub: Bound
tightness} for the case of slowly changing $\hTc(p_x)$. For higher
and lower SNR the bounds are much tighter. In particular, for high
SNR all lower bounds are close to the coherent channel upper bound,
but the lower bound is tighter for channels with higher effective
coherence time.

The x-marks in the top right end of the lower bounds (for
$\hTc[_0]=2,000$ and $\hTc[_0]=10,000$) mark the point in which the
effective coherence time dropped below $\hTc(p_x)=2N=60$. For higher
SNR the lower bounds will not be tight. More importantly, for
higher SNR our model will not represent well the physical channel,
because the assumption that the channel impulse response does not
change during one OFDM symbol can no longer be justified. For
longer coherence times, this SNR threshold is higher and hence not
shown in the figure. For any SNR below the x-marks, the bounds
derived in the previous sections are \CommentApril{good}, and
describe the channel capacity with high accuracy.

\subsection{Comparison to known results}\label{sub: Comparison to known bounds}
\CommentApril{\subsubsection{Closest results} The bounds which are
most similar to the results presented here were derived by
Sethuraman et al. \cite{sethuraman2009low} and by Durisi et al.
\cite{durisi2010noncoherent}. Sethuraman et al. present an upper
bound for the frequency-flat fading case ($N=1$) with a peak power
constraint. In this case the upper bound in \cite{sethuraman2009low}
is tighter than the bounds in (\ref{e: low snr approx UB peak}), but
the difference is negligible for underspread channels ($\hTc[_0]\gg
N$). The difference in the upper bound results from the relaxation
$\log(1+x)\le x$ in (\ref{e. Gaussian upper bounding}). This
relaxation can be easily avoided, but the resulting bounds are much
more complicated for $N>1$, (in particular in the context of
underspread channels and effective coherence time) while the
difference between the bounds is negligible. Durisi et al. presented
upper and lower bounds for frequency selective fading channels, but
only for the case of a peak power constraint both in time and in
frequency. Both works also presented results on low SNR capacity
asymptote that will be discussed in the next subsection. }

\subsubsection{\CommentApril{Low SNR limit when $\hTc[_0]$ exists}}
\CommentApril{In  the low SNR limit, if the limit of the effective
coherence time, $\hTc[_0]$, exists as the SNR goes to zero (\ref{d.
Tc max}), the channel capacity is known and matches the results
presented above both with quadratic power constraint and peak power
constraint. For the quadratic power constraint \cite{Medard:02}, the
capacity at the low SNR limit} is $\alpha \Tc \frac{E[x^4]}{2}$,
which is exactly equal to the low SNR limit of the upper bound
(\ref{e: low snr approx UB quad}). Note that the low SNR limit of
the lower bound (\ref{e: low snr approx LB quad}) is lower by a
factor of $(\hTc[_0]-N)/\hTc[_0]$ than the upper bound, which is
negligible for $\hTc[_0]\gg N$. \CommentApril{(The main reason for
this difference between the upper and lower bounds is the lower
bounding transmission scheme which assumes that  each OFDM symbol
($N$ samples) is decoded using channel estimation based only on past
symbols.) }

\CommentApril{For the peak power constraint flat fading case
($N=1$), the capacity is $p_x^2 (\hTc[_0]-1) / 2$
\cite{sethuraman2009low}, which matches the results presented here
(again up to a negligible factor of $(\hTc[_0]-N)/\hTc[_0]$ as
discussed above). Note that for the peak power constraint we did not
analyze the case where the average power constraint is lower than
the peak power constraint.

For $N>1$ the same work shows that the low SNR capacity asymptote
for the delay separable underspread stationary channel is also given
by $p_x^2 (\hTc[_0]-1) / 2$. Under the assumption $\hTc[_0]\gg N$
this channel is approximately equal to the OFDM channel considered
here, and the low SNR asymptotes also match. This result also
matches the low SNR asymptotes presented in
\cite{durisi2010noncoherent} for the case of peak power constraint
only in time. It is interesting to note that for this case the peak
power constraint used in
\cite{sethuraman2009low,durisi2010noncoherent} is stricter than the
peak power constraint used here. The power constraint in this work
limits the energy of a single OFDM symbol. Translating to the time
domain the constraint is applied to the average energy of groups of
$N$ samples. On the other hand the constraint in
\cite{sethuraman2009low} applies for each time domain sample.
Interestingly, the bounds are almost identical, even though
\cite{sethuraman2009low} clearly shows that relaxing the peak power
constraint results in a higher capacity for the same average
transmission power. Comparing these two works one can conclude that
the relaxation of the peak power constraint is effective only if it
is applicable for time periods which are at least on the order of
the channel coherence time. Allowing signal ``peakiness" which must
be averaged over periods ($N$) which are significantly shorter than
the channel coherence time ($\hTc[_0]$) is not sufficient to
increase the channel capacity. }

Since the effective coherence time can be interpreted as the block
length of a block fading model that achieves the same capacity, our
results naturally match results that were derived for the block
fading model (e.g., \cite{sethuraman2005cpu,Telatar:00}).

\CommentApril{
\subsubsection{Analysis for a given channel estimation error}
Several works (e.g. \cite{Medard:00,Lapidoth:02}) consider the
effect of channel estimation error with a given variance on the
channel capacity. These works have been the basis for the lower
bounds presented here, but they miss the effect of the transmitted
signal on the ability to estimate the channel (\cite{Medard:00}
considers also an estimation from an out-of-band pilot signal). In
this sense, one can say that the QPSK lower bound presented above
(Theorem \ref{T: QPSK lower bound}) is the most straightforward part
of the work. It combines an extension of the bounds in
\cite{Medard:00,Lapidoth:02} to the OFDM model, the channel
estimation scheme of \cite{jin2005ulc,li2007successive} which allows
to estimate the channel using all (relevant) past transmitted and
received symbols, and results on channel estimation errors for
constat amplitude modulations. Perhaps the most important
contribution of Theorem \ref{T: QPSK lower bound} is the derivation
of the bound in a way that shows the role of the effective coherence
time. Note that the truncated Gaussian lower bound (Theorem \ref{T:
TG lower bound}) is more complicated as the modulation is not
constant amplitude, and the upper bounds are more complicated as
they cannot assume any estimation scheme. }

\subsubsection{High SNR limit}
As detailed above, for regular channels our analysis holds only up
to some finite \CommentApril{SNR}. The results show that the
coherent channel upper bound is achievable as long as the effective
coherence time is large enough. This is in agreement with known
results \cite{Lapidoth:03,Lapidoth:05,koch2005fna} showing that in
the very high SNR limit the coherent channel upper bound is not
achievable. Our analysis cannot predict the actual double
logarithmic behavior of the capacity, as it happens at too high
\CommentApril{SNR} where the assumption on the value of the
effective coherence time is not valid.

For irregular channels, although not proven, the effective coherence
time also typically decreases as SNR increases (although at lower
rate). Hence, our analysis  is limited in SNR even for
irregular channels.

The modeling problem at high SNR is also discussed by Durisi et al.
\cite{Durisi2009ISIT}, where they focus on the characterization of
the range of SNRs in which the channel discretization is reliable.
Note that their approach is quite different from the one taken
above. In \cite{Durisi2009ISIT}, the analysis starts from a
continuous time channel, and tries to analyze all modeling errors in
the discretization process. In the analysis above we study the
behavior of a discrete time OFDM model without assuming any
discretization imperfections. Yet, we reach the same conclusion,
even the discrete time OFDM model reveals the high SNR model limit.

\subsubsection{Dependence of coherence time on SNR}
Few works analyze the dependence between the SNR and the coherence
time. In these works the coherence time characterizes the channel
(and does not change for each channel). For example Zheng et al.
\cite{zheng2007ccl} consider a block-fading channel, where the
channel remains constant for a block of $l$ symbols, before changing
to an independent realizations. Chen and Veeravalli \cite{Chen:07}
consider a block stationary channel where the fading is constant
across a block of length $T$ and changes in a stationary manner
between blocks.

In both cases the analysis considers a set of channels, and
\CommentApril{tries} to characterize relations between the channel
coherence time and the system SNR that will result in a certain
capacity behavior. Chen and Veeravalli show (\cite{Chen:07} equation
(21)) that two systems with identical block correlation have the
same capacity if the product of their peak powers and block lengths
($T \cdot {\rm SNR}$) is identical.

Zheng et al. analyze the limit capacity of a set of channels with
increasing coherence time $l$ and decreasing SNRs. They show that
different relations between coherence time and SNR of the form
$l={\rm SNR}^\alpha$ achieve different capacity behavior between the
coherent and noncoherent scheme.

Our analysis is completely different. We consider a (single)
specific channel, and show that its effective coherence time is a
function of the system SNR.

\section{Channel capacity and bounds}\label{sec. Channel capacity and bounds}
In this section we analyze the channel capacity and give the
information-theoretical part of the theorem proofs.

In general, even if
the input symbols are independent, the output symbols are dependent
due to the channel memory. Therefore, we need to evaluate the
capacity over the entire transmission, defined as:
\begin{IEEEeqnarray}{rCl}
    \label{d. channel capacity}
    C=\lim \limits_{n \to \infty } \sup \limits_{{\rm Pr} (\FV[n]{X}{0})} \frac{1}{N (n+1)}I(\FV[n]{X}{0} ;\FV[n]{Y}{0} ).
\end{IEEEeqnarray}
Note that the ergodic requirement in Assumption \ref{assump WSSUS}
guarantees the achievability of the capacity (see for example
\cite{Lapidoth:03,sethuraman2005cpu}).

Most of the bounds derived in this section depend on the conditional
distribution of the channel given the past transmitted and received
symbols. This distribution will be analyzed in detail in the next
section. For this section, it is enough to state that given the past
symbols, the channel has a complex Gaussian distribution $\FV{H}{k}
|\FV[k-1]{X}{0} ,\FV[k-1]{Y}{0}
 \sim \CN(\tilde{\bm{\mu }}_k ,\tilde{\mathcal {\bm \Leps}}_k )$.
 Lemma \ref{L: TG bound} also uses the distribution of the channel mean:
\begin{IEEEeqnarray}{rCl}\label{e: mu distribution}
 \tilde{\bm{\mu }}_k |\FV[k-1]{X}{0} \sim \CN \left(
E[\FV{H}{k}],{\openmtr{C}}_{\FV{H}{k}}-\tilde{\mathcal {\bm
\Leps}}_k
 \right)
\end{IEEEeqnarray}

\subsection{Coherent channel upper bound}\label{subsub. definition Known channel upper bound}
The first upper bound we use is the well known channel capacity when
the channel is known to the receiver (coherent channel) with only
average power constraint:
\begin{IEEEeqnarray}{rCl}
        C &\le& \lim \limits_{n \to \infty } \frac{1}{N (n+1)} \sup \limits_{{\rm Pr} (\FV[n]{X}{0})} I(\FV[n]{X}{0};\FV[n]{Y}{0}
        , \FV[n]{H}{0})
        \nonumber \\
        &=&\lim \limits_{n \to \infty } \frac{1}{N (n+1)}  \sum\limits_{k = 0}^n \sup \limits_{{\rm Pr} (\FV{X}{k})} {I(\FV{X}{k} ;\FV{Y}{k} ,\FV{H}{k} )}.
        \nonumber \\
        &=&\frac{1}{N} \sup \limits_{{\rm Pr} (\FV{X}{0})} {I(\FV{X}{0} ;\FV{Y}{0} ,\FV{H}{0} )}.
\end{IEEEeqnarray}

Using Assumptions \ref{assump WSSUS} and \ref{assump frequency
flat}, this bound is maximized when the input signal has Gaussian
distribution with $E[ \FV{X}{k}] = {\rm 0}$, and
${\Ct}_{\FV{X}{k}}=p_x \cdot \It$ \cite{Telatar:99}. The maximal
mutual information is given by:

{\it Result 1:} The channel capacity is upper bounded by
\begin{IEEEeqnarray}{rCl}
    \label{e. known channel bound ver 2}
        C \le {\rm UB}_{\rm coh} (p_x) = E_{\FV{H}{0}} \left[ {\log \left(  1+ N p_x |{\FV{h}{0,0}}|^2
        \right)}\right].
\end{IEEEeqnarray}

\CommentApril{The bound in (\ref{e. known channel bound ver 2}) is
the capacity of the channel with a constraint only on the average
power. As both types of power constraints analyzed here must satisfy
$E[\RV{p}{k} ]\le N p_x$, this bound holds for both types of
constraints}. As in the case of the block fading model
\cite{marzetta1999cmm}, we will show that for large enough effective
coherence times and SNR this bound is tight, and the channel
capacity does not degrade due to the lack of channel knowledge.

\renewcommand{\theLemma}{\arabic{Lemma}.a}
\subsection{Low SNR upper bound}\label{subsub. definition Known channel upper bound}
The channel capacity is upper bounded by the following lemma:
\begin{Lemma}\label{L: ub low pk}
    The capacity of a channel with a peak power constraint is upper bounded by:
    \begin{IEEEeqnarray}{rCl}
    \label{e. UB peak}
        C &\le& \lim \limits_{k \to \infty } \left\{p_x
        \left[ 1 + N\left|E \left[  \FV{h}{0,0} \right]
        \right|^2 \right]
        - \frac{1}{N}\log \left( 1+ N^2 p_x \min_m \inf_{\FV[k-1]{X}{0}\in\mathcal{I}(N p_x \It) } (\tilde{\mathcal {\bm \Leps}}_k)_{m,m}
        \right)\right\},\IEEEeqnarraynumspace
\end{IEEEeqnarray}
and
    the capacity of a channel with a quadratic power constraint is upper bounded by:
    \begin{IEEEeqnarray}{rCl}
    \label{e. UB qd}
        C &\le&
        \lim \limits_{k \to \infty } \left\{
        N p_x \left|E \left[  \FV{h}{0,0}\right]\right|^2
        \phantom{        \sqrt{
        \sup \limits_{{\rm Pr} (\Pc_{k})} E \left[ \left\{ 1-N
        \min_m \inf_{\FV[k-1]{X}{0}\in\mathcal{I}(\Pc_{k-1}) } (\tilde{\mathcal {\bm \Leps}}_k)_{m,m}
        \right\}^2 \right]}
}\right.
        \nonumber \\
        &&+\left.
        \sqrt{\alpha}p_x
        \sqrt{
        \sup \limits_{{\rm Pr} (\Pc_{k-1})} E \left[ \left\{ 1-N
        \min_m \inf_{\FV[k-1]{X}{0}\in\mathcal{I}(\Pc_{k-1}) } (\tilde{\mathcal {\bm \Leps}}_k)_{m,m}
        \right\}^2 \right]}
        +\frac{1}{2}\alpha N p_x^2
        \right\},
    \end{IEEEeqnarray}
    where $\mathcal{I}(\Pc_{k})=\{ \FV[k]{X}{0}:\FV[\dagger]{X}{i}
\FV{X}{i}=\RV{p}{i} , i=0,\ldots,k\}$ is the set of all input
symbols that correspond to the power matrix $\Pc_{k}$.
\end{Lemma}

%\vspace{0.4cm}
\begin{IEEEproof}[Proof of Lemma \ref{L: ub low pk}]
We start with the inequality:
\begin{IEEEeqnarray}{rCl}
    \label{e. mututal information equality 2}
    \lim \limits_{n \to \infty } \frac{1}{N (n+1)} I(\FV[n]{X}{0} ;\FV[n]{Y}{0} )
    & = & \lim \limits_{n \to \infty } \frac{1}{N (n+1)} \sum\limits_{k = 0}^n {I(\FV{Y}{k};\FV[n]{X}{0} | \FV[k-1]{Y}{0}  )}  \nonumber \\
\nonumber \\
    &=& \lim \limits_{n \to \infty } \frac{1}{N (n+1)} \sum\limits_{k = 0}^n\left\{ h(\FV{Y}{k}|\FV[k-1]{Y}{0})-h(\FV{Y}{k}|\FV[n]{X}{0} ,\FV[k-1]{Y}{0})\right\}
    \nonumber \\
    &\le& \lim \limits_{n \to \infty } \frac{1}{N (n+1)} \sum\limits_{k = 0}^n\left\{ h(\FV{Y}{k})-h(\FV{Y}{k}|\FV[n]{X}{0} ,\FV[k-1]{Y}{0})\right\}.
\end{IEEEeqnarray}
where $h(\cdot)$ is the differential entropy. The first term in the
last line of (\ref{e. mututal information equality 2}) can be upper
bounded by the maximal entropy of a random vector with a given
covariance matrix. The resulting entropy is \cite{Cover:91}:
\begin{IEEEeqnarray}{rCl}
    \label{e. Gaussian upper bounding}
    h(\FV{Y}{k})
    \le \log \left( {\|
        {\openmtr{C}}_{\FV{Y}{k}}
        \|} \right)+N \log (\pi e)
      \le  {\rm Tr} \left(
        E [ \FV{Y}{k}  \FV[\dagger]{Y}{k}   ] - {\It}
        \right) + N \log (\pi e).
\end{IEEEeqnarray}
where the last inequality uses $\log \left( {\left\| {{\It} +
\mtr{A}} \right\|} \right) \le {\rm Tr} (\mtr{A})$ which holds for
any \CommentApril{positive semi-definite} matrix
$\mtr{A}$.\footnote{Using $\left\| \It + \mtr{A} \right\| = \prod
\limits_{i=0}^{N-1} (1+\lambda_i) $, and ${\rm Tr} (\mtr{A})=\sum
\limits_{i=0}^{N-1} \lambda_i$, where
$\lambda_0,\ldots,\lambda_{N-1}$ are the eigenvalues of the matrix
$\mtr{A}$, and the inequality \CommentApril{$\log(1+x)\le x$}.} Note
that this inequality is tight for low SNR if the transmitted signal
has zero mean. Substituting the covariance matrix:
\begin{IEEEeqnarray}{rCl}
    E \left[ \FV{Y}{k}  \FV[\dagger]{Y}{k} \right]=
N \cdot E \left[\diagM  ({\FV{X}{k}}) {\FV{H}{k}}
{\FV[\dagger]{H}{k}}\diagM ({\FV{X}{k}})^\dagger \right]+ \It,
\end{IEEEeqnarray}
we get:
\begin{IEEEeqnarray}{rCl}\label{e: first entropy result}
    h(\FV{Y}{k}) - N \log (\pi e)
    &\le& N \cdot {\rm Tr} \left( E \left[ \Sf_{\RV{X}{k} }\right] \left[ {\openmtr C}_{\FV{H}{0}} + E \left[ \FV{H}{0} \right] E \left[ \FV{H}{0} \right]^\dagger  \right] \right)
    \nonumber \\
    &=& E \left[ \FV[\dagger]{X}{k}\FV{X}{k}\right] \left[ 1 + N \left|E \left[ \FV{h}{0,0} \right]
    \right|^2 \right]
\end{IEEEeqnarray}
where the first line uses the rotation property of the trace and the
spectrum definition, (\ref{e. Spectrum}). The second line uses the
fact that $\Sf_{\RV{X}{k}}$ is a diagonal matrix, while all elements
on the diagonal of the second term are equal. We also use the
normalization ${\rm Tr}[{\openmtr{C}}_{\RV{H}{0}}]=1$, which
\CommentApril{results in} $({\openmtr C}_{\FV{H}{0}})_{m,m}=1/N$.

Turning to the second entropy in the last line of (\ref{e. mututal
information equality 2}), we note that the conditional distribution
of the output given the input is Gaussian. We also observe that:
\begin{IEEEeqnarray}{rCl}
    {\openmtr{C}}_{\FV{Y}{k}|\FV[n]{X}{0} ,\FV[k-1]{Y}{0}} =
N \cdot \diagM  ({\FV{X}{k}}) \tilde{\mathcal {\bm \Leps}}_k \diagM
({\FV{X}{k}})^\dagger + \It,
\end{IEEEeqnarray}
where $\tilde{\mathcal {\bm \Leps}}_k$ is the conditional channel
covariance matrix given the past transmitted and received symbols.
Using (\ref{d: power variables}), the entropy is written as:
\begin{IEEEeqnarray}{rCl}\label{e:conditional entropy 1}
    h(\FV{Y}{k}|\FV[n]{X}{0} ,\FV[k-1]{Y}{0}) -N \log (\pi e)
    &=&
     E_{\FV[n]{X}{0} ,\FV[k-1]{Y}{0}} \left[ {\log \left( {\left\|
        {\openmtr{C}}_{\FV{Y}{k}|\FV[n]{X}{0} ,\FV[k-1]{Y}{0}}
        \right\|} \right)} \right]
    \nonumber \\
    &=&
    E_{\Pc_{n}} \left[ E_{\FV[n]{X}{0}|\Pc_{n}} \left[ \log \left( \left\|
        N \diagM  ({\FV{X}{k}}) \tilde{\mathcal {\bm \Leps}}_k \diagM
({\FV{X}{k}})^\dagger + \It
        \right\| \right) \right] \right]
    \nonumber \\
    &=&
    E_{\Pc_{k}} \left[
    E_{\FV[k]{X}{0}|\Pc_{k}} \left[ \log \left( \left\|
        N \tilde{\mathcal {\bm \Leps}}_k \Sf_{\RV{X}{k}}+ {\It}
        \right\| \right) \right]
    \right]
    \nonumber \\
    &\ge&
        E_{\Pc_{k}} \left[
        \inf_{\FV[k]{X}{0}\in\mathcal{I}(\Pc_{k}) }
        \log \left( \left\|
        N \tilde{\mathcal {\bm \Leps}}_k \Sf_{\RV{X}{k}}+ {\It}
        \right\| \right)
        \right],
\end{IEEEeqnarray}
where we use the rotation property $\left\| {\It}+\mtr{A} \mtr{B}
\right\| =\left\| {\It}+\mtr{B} \mtr{A} \right\|$.

Now, we observe that since $\Sf_{\RV{X}{k}}$ is diagonal and
\CommentApril{nonnegative},
\begin{IEEEeqnarray}{rCl}
    {\rm Tr}\(\tilde{\mathcal {\bm \Leps}}_k \Sf_{\RV{X}{k}}\)
    \ge \min_m (\tilde{\mathcal {\bm \Leps}}_k)_{m,m} {\rm Tr}\(\Sf_{\RV{X}{k}}\).
\end{IEEEeqnarray}
Using also the inequality $\log \left\| {\It}+\mtr{A} \right\| \ge
\log \(1+{\rm Tr}(\mtr{A})\)$ which holds for any
\CommentApril{positive semi-definite} matrix
$\mtr{A}$,\footnote{Using $\sum_i \log(1+\lambda_i)\ge \log
(1+\sum_i \lambda_i)$ since $\lambda_i\ge 0$.} the term inside the
expectation in the last row of (\ref{e:conditional entropy 1}) is
lower bounded by:
\begin{IEEEeqnarray}{rCl}\label{e: temp 132}
    \inf_{\FV[k]{X}{0}\in\mathcal{I}(\Pc_{k}) } \log \left( \left\|
        N\tilde{\mathcal {\bm \Leps}}_k \Sf_{\RV{X}{k}}+ {\It}
        \right\| \right)
    &\ge& \inf_{\FV[k]{X}{0}\in\mathcal{I}(\Pc_{k}) } \log \left( 1+ N {\rm Tr} \( \tilde{\mathcal {\bm \Leps}}_k
\Sf_{\RV{X}{k}}\)
         \right)
    \nonumber \\
    &\ge& \log \left( 1+N\RV{p}{k}
    \min_m \inf_{\FV[k-1]{X}{0}\in\mathcal{I}(\Pc_{k-1}) } (\tilde{\mathcal {\bm \Leps}}_k)_{m,m}  \right) .
\IEEEeqnarraynumspace
\end{IEEEeqnarray}
Note that this lower bound is achievable using a transmission
spectrum that concentrates all of the power on the frequency bin
that has minimal estimation error.

Substituting (\ref{e: temp 132}) and (\ref{e: first entropy result})
in (\ref{e. mututal information equality 2}) \CommentApril{results
in:
\begin{IEEEeqnarray}{rCl}
        C &\le& \lim \limits_{n \to \infty } \frac{1}{N (n+1)} \sum\limits_{k = 0}^n  \sup \limits_{{\rm Pr} (\Pc_{k})}
        \Bigg\{ E \left[ \RV{p}{k} \right] \left[ 1 + N\left|E \left[  \FV{h}{0,0} \right]
        \right|^2 \right]
        \nonumber \\  && - \: E \Big[  \log \Big( 1+N\RV{p}{k}
        \min_m \inf_{\FV[k-1]{X}{0}\in\mathcal{I}(\Pc_{k-1}) } (\tilde{\mathcal {\bm \Leps}}_k)_{m,m}
        \Big)
\Big]\Bigg\}.
\end{IEEEeqnarray}
Noting also that the resulting quantity is monotonically
\CommentApril{nondecreasing} in $k$, the capacity is upper bounded
by}:
\begin{IEEEeqnarray}{rCl}
        \label{e: low SNR general upper bound}
        C &\le& \frac{1}{N} \lim \limits_{k \to \infty }  \sup \limits_{{\rm Pr} (\Pc_{k})}
        \Bigg\{ E \left[ \RV{p}{k} \right] \left[ 1 + N\left|E \left[  \FV{h}{0,0} \right]
        \right|^2 \right]
        \nonumber \\  && - \: E \Big[  \log \Big( 1+N\RV{p}{k}
        \min_m \inf_{\FV[k-1]{X}{0}\in\mathcal{I}(\Pc_{k-1}) } (\tilde{\mathcal {\bm \Leps}}_k)_{m,m}
        \Big)
\Big]\Bigg\}.
\end{IEEEeqnarray}
Subject to the peak power constraint, for every $k$ the bound is
maximized if $\RV{p}{k}=N p_x$ (as $\RV{p}{k}$ appears inside the
log in the second (negative) term). Substituting this optimal
transmission power in (\ref{e: low SNR general upper bound}) results
in (\ref{e. UB peak}) and proves the first part of the lemma.

%\begin{theorem}\label{t: ub low qd}
In order to derive the bound for the quadratic power constraint we
use the inequality $\log(1+x)\ge x-\frac{1}{2} x^2$ and simplify
(\ref{e: low SNR general upper bound}) to:
\begin{IEEEeqnarray}{rCl}
    \label{e. low UB qd general}
        C &\le& \frac{1}{N}  \lim \limits_{k \to \infty } \sup \limits_{{\rm Pr} (\Pc_{k})} \left\{
        E \left[ \RV{p}{k} \right] \left[ 1 + N\left|E \left[  \FV{h}{0,0} \right]
        \right|^2 \right]
        \phantom{ \left[ \left\{ \inf_{\FV[k-1]{X}{0}\in\mathcal{I}(\Pc_{k-1}) }
         (\tilde{\mathcal {\bm \Leps}}_k)_{m,m} \right\}^2
        \right] }\right.
        \nonumber \\
        &&- \left. E \left[ N \RV{p}{k} \min_m \inf_{\FV[k-1]{X}{0}\in\mathcal{I}(\Pc_{k-1}) } (\tilde{\mathcal {\bm \Leps}}_k)_{m,m}
        -\frac{1}{2}N^2 \RV[2]{p}{k} \left\{\min_m \inf_{\FV[k-1]{X}{0}\in\mathcal{I}(\Pc_{k-1}) }
         (\tilde{\mathcal {\bm \Leps}}_k)_{m,m} \right\}^2
        \right] \right\}. \IEEEeqnarraynumspace
\end{IEEEeqnarray}
Using $(\tilde{\mathcal {\bm \Leps}}_k)_{m,m} \le ({\openmtr
C}_{\FV{H}{0}})_{m,m}=1/N $, we rewrite (\ref{e. low UB qd general})
as:
\begin{IEEEeqnarray}{rCl}
        C &\le& \frac{1}{N} \lim \limits_{k \to \infty }   \sup \limits_{{\rm Pr} (\Pc_{k})} \left\{
        N E \left[ \RV{p}{k} \right]  \left|E \left[  \FV{h}{0,0}\right]\right|^2
\phantom{\left[
        \( \inf_{\FV[k-1]{X}{0} }(\tilde{\mathcal {\bm \Leps}}_k)_{m}\)\right] }\right.
        \nonumber \\
        &&+ \left. E \left[ \RV{p}{k}
        \(1-N\min_m \inf_{\FV[k-1]{X}{0}\in\mathcal{I}(\Pc_{k-1}) }(\tilde{\mathcal {\bm \Leps}}_k)_{m,m}\)
        +\frac{1}{2}\RV[2]{p}{k} \right] \right\},
\end{IEEEeqnarray}
and use the Cauchy-Schwartz inequality and the power constraint,
(\ref{e. frequency framed Quadratic constraint}), to prove (\ref{e.
UB qd}).
\end{IEEEproof}

The exact values of the bounds depend on the conditional channel
distribution, through $\min_m
\inf_{\FV[k-1]{X}{0}\in\mathcal{I}(\Pc_{k-1}) }(\tilde{\mathcal {\bm
\Leps}}_k)_{m,m}$. This distribution is analyzed in the next
section.

\subsection{QPSK lower bound}
For the lower bound we need to select an input distribution. The
first lower bound derived herein is based on QPSK modulation. This
bound applies for both power constraints. For the quadratic power
constraint, the bound is especially significant at the low SNR
regime, where it is crucial to use a low fourth moment. The QPSK
modulation uses constant amplitude and hence minimizes the ratio
$\frac{E[|\FV{x}{k,m}|^4]}{E^2[|\FV{x}{k,m}|^2]}$.

The input symbol is given by:
\begin{IEEEeqnarray}{rCl}\label{d: QPSK modulation}
    \FV{X}{k}=\sqrt{\beta \frac{N}{r} p_x}{\bf g}\FV{B}{k},
\end{IEEEeqnarray}
where $\FV{B}{k}=\left[  \FV{b}{k,0}\ldots,\FV{b}{k,r-1},0,\ldots,0
\right]$, \CommentApril{$r$ ($1\le r\le N$)} is the number of active
frequency bins, $\{ \FV{b}{k,m} \}$ is the sequence of iid QPSK
data, and $\FV{b}{k,m} \in \{ \pm 1, \pm j\}$ with equal
probability. On the other hand, ${\bf g}\in\{0,1\}$ is a binary
random variable, statistically independent from
$\FV{b}{k,0},\ldots\FV{b}{k,r-1}$, that determines whether the
system will transmit or not (i.e., a single variable that affects
all transmitted symbols). The transmission probability is ${\rm
Pr}({\bf g}=1)=\frac{1}{\beta}$. This distribution is convenient for
analysis, although it is not reasonable for practical systems (one
cannot consider a practical system that has a positive probability
not to transmit any symbol at any time). A practical system can
achieve the same performance by transmitting in blocks with large
gaps between the blocks, as long as the block length is
significantly larger than the channel coherence time.

\begin{Lemma}\label{L: QPSK LB}
    The capacity of a channel with a quadratic power constraint is lower bounded by:
\begin{IEEEeqnarray}{rCl}
    \label{e. QPSK qd bound}
    C \ge  \lim \limits_{k \to \infty } \max_{r\le N} \max_{1\le\beta\le \alpha}
    \frac{1}{N\beta}
    \sum_{m=0}^{r-1} E\left[ \left. C_{{\rm QPSK}}\(\frac{\beta \frac{N^2}{r} p_x|{\tilde{\bm \mu}}_{k,m}|^2}
    {1+\beta \frac{N^2}{r} p_x (\tilde{\mathcal {\bm \Leps}}_k)_{m,m}}\) \right| {\bf
    g}=1\right], \IEEEeqnarraynumspace
\end{IEEEeqnarray}
where \CommentApril{$C_{{\rm QPSK}}(\rho)$, defined in (\ref{d:
C_QPSK}),} is the achievable mutual information for QPSK
transmission over AWGN channel with SNR equal to $\rho$ (see for
example \cite{baccarelli2000ssb}, which also gives some useful
bounds). The capacity of a channel with a peak power constraint is
lower bounded by:
\begin{IEEEeqnarray}{rCl}
    \label{e. QPSK iid bound}
    C \ge  \lim \limits_{k \to \infty } \max_{r\le N}
    \frac{1}{N}
    \sum_{m=0}^{r-1} E\left[ \left. C_{{\rm QPSK}}\(\frac{\frac{N^2}{r} p_x|{\tilde{\bm \mu}}_{k,m}|^2}
    {1+\frac{N^2}{r} p_x (\tilde{\mathcal {\bm \Leps}}_k)_{m,m}}\) \right| {\bf
    g}=1\right].
\end{IEEEeqnarray}

\end{Lemma}
\begin{IEEEproof}[Proof of Lemma \ref{L: QPSK LB}]
As $\textbf{g}$ is constant throughout the transmission, we can
safely assume that after long enough time the receiver can decode
${\bf g}$ with no error. Thus, we can write:
\begin{IEEEeqnarray}{rCl}\label{e: lemma 2.a eq 1}
    \label{e. mututal information equality}
    \lim \limits_{n \to \infty } \frac{1}{ N (n+1)} I(\FV[n]{X}{0} ;\FV[n]{Y}{0} ) & = & \lim \limits_{n \to \infty } \frac{1}{N (n+1)} I(\FV[n]{X}{0} ;\FV[n]{Y}{0} | {\bf g}) \nonumber \\ &=& \frac{1}{\beta} \lim \limits_{n \to \infty }\frac{1}{N (n+1)} I(\FV[n]{B}{0} ;\FV[n]{Y}{0} | {\bf g}=1).
\end{IEEEeqnarray}

Given ${\bf g}=1$ the transmitted symbols are statistically
independent. We use the following equality:
\begin{IEEEeqnarray}{rCl}
    \label{e. mututal information equality}
    I(\FV[n]{B}{0} ;\FV[n]{Y}{0} | {\bf g}=1) & = & \sum\limits_{k = 0}^n {I(\FV{B}{k} ;\FV{Y}{k} | {\bf g}=1,\FV[k-1]{B}{0} ,\FV[k-1]{Y}{0} ,\FV[n]{Y}{k+1} )}  \nonumber \\
    &&+ \sum\limits_{k = 0}^n {I(\FV{B}{k} ;\FV[k-1]{Y}{0} | {\bf g}=1,\FV[k-1]{B}{0} ,\FV[n]{Y}{k+1}
    )} \nonumber \\
    &&+ \sum\limits_{k = 0}^n {I(\FV{B}{k} ;\FV[n]{Y}{k+1} | {\bf g}=1,\FV[k-1]{B}{0} )}
\end{IEEEeqnarray}
The second and third terms on the right hand side of this equation
correspond to the mutual information between an input symbol and the
past or future output symbols. As these terms do not use the $k$-th
symbol output, they both vanish when the input symbols are iid. The
first term can be lower bounded by:
\begin{IEEEeqnarray}{rCl}
    \label{e. data aided lower bound}
    I(\FV[n]{B}{0} ;\FV[n]{Y}{0} | {\bf g}=1) &=& \sum\limits_{k = 0}^n {I(\FV{B}{k} ;\FV{Y}{k} | {\bf g}=1,\FV[k-1]{B}{0} ,\FV[k-1]{Y}{0} ,\FV[n]{Y}{k+1} )}  \nonumber \\
    & \ge & \sum\limits_{k = 0}^n {I(\FV{B}{k} ;\FV{Y}{k} | {\bf g}=1,\FV[k-1]{B}{0} ,\FV[k-1]{Y}{0}  )}
    \nonumber \\
    & \ge & \sum\limits_{k = 0}^n \sum\limits_{m = 0}^{r-1}{I(\FV{b}{k,m} ;\FV{y}{k,m} | {\bf g}=1,\FV[k-1]{B}{0} ,\FV[k-1]{Y}{0}  )}
,
\end{IEEEeqnarray}
where the second line uses the statistical independence of
$\FV[n]{Y}{k+1}$ and $\FV{B}{k}$, and the third line uses the
statistical independence between the elements of $\FV{B}{k}$.
Substituting (\ref{e. data aided lower bound}) in (\ref{e: lemma 2.a
eq 1})\CommentApril{, and noting again that the term inside the
summation is monotonic in $k$,} results in the lower bound on the
capacity:
\begin{IEEEeqnarray}{rCl}
      \label{e. information lim}
        C \ge \lim \limits_{k \to \infty } \frac{1}{N \beta}  \sum\limits_{m = 0}^{r-1}{I(\FV{b}{k,m} ;\FV{y}{k,m} | \FV[k-1]{B}{0} ,\FV[k-1]{Y}{0}  ,{\bf
        g}=1)}.
\end{IEEEeqnarray}
Note that the past symbols affect the distribution of the current
symbol only through the conditional distribution of the channel.
Thus, in this lower bound, the decoding of correlated output symbols
is replaced by the decoding of uncorrelated symbols with channel
state information (CSI). This CSI is produced by optimal channel
estimation based on past input and output symbols. A coding scheme
that allows to decode the previous symbols and use them for channel
estimation for the next symbol is shown in
\cite{jin2005ulc,li2007successive}.

Focusing on the $m$-th frequency bin of the $k$-th symbol, given
${\bf g}=1$, the channel output is:
\begin{IEEEeqnarray}{rCl}
    \FV{y}{k,m}=\sqrt{ \frac{\beta}{r} p_x}N {\tilde{\bm \mu}}_{k,m} \FV{b}{k,m} + \FV{w}{k,m} + \FV{v}{k,m}.
\end{IEEEeqnarray}
where $\FV{w}{k,m}$ is the $m$-th element of the DFT of the Complex
Gaussian noise, and $\FV{v}{k,m}=\sqrt{ \frac{\beta}{r}
p_x}N(\FV{h}{k,m}-{\tilde{\bm \mu}}_{k,m})\FV{b}{k,m}$ is the
interference term due to channel estimation error. Since the
estimation error is a circularly symmetric Gaussian random variable
and the data has constant amplitude, the interference term is also
Gaussian and is statistically independent from the data
($\FV{b}{k,m}$). Assuming optimal channel estimation, the resulting
channel is equivalent to a \CommentApril{coherent} flat fading
channel with additive independent complex Gaussian noise samples of
variance $1+ \beta\frac{N^2}{r} p_x (\tilde{\mathcal {\bm
\Leps}}_k)_{m,m}$. Evaluating the mutual information of this channel
for QPSK modulation and multiplying it by the transmission
probability ($1/\beta$) leads to (\ref{e. QPSK qd bound}) and
(\ref{e. QPSK iid bound}), depending on the allowed range of the
parameter $\beta$.
\end{IEEEproof}

\subsection{Truncated Gaussian lower bound}\label{subsub. definition lower bound}
For high SNR, the QPSK bound will not be tight (most obviously as it
is limited to 2 bits per frequency bin). The most intuitive
alternative is a complex Gaussian input distribution. Such a
distribution seems very close to optimal (based on a large set of
simulations). However, we were unable to produce an appropriate
bound on the resulting conditional channel variance, and therefore
turned to a slightly modified distribution.

Let $\uv $ be a proper complex Gaussian distribution,
 $\uv \sim \CN(0,1)$. We denote hereafter by \textit{truncated Gaussian distribution} with parameters $\eta$ and $\xi$ the
distribution of $\uv $ given : $\eta\le|\uv |^2\le\xi$. In
mathematical terms, we say that $\zv \sim \TCN(\eta,\xi)$ if for any
function $f(\cdot)$ we have:
\begin{IEEEeqnarray}{rCl}
E\left[ f(\zv ) \right]=E\left[ f(\uv )\Big | \eta\le|\uv |^2 \le
\xi\right].
\end{IEEEeqnarray}
Note that $|\uv |^2$ has an exponential distribution with parameter
1. Define:
\begin{IEEEeqnarray}{rCl}
    \theta=\textrm{Pr}(\CommentApril{\eta\le|\uv |^2 \le
\xi})=e^{-\eta}-e^{-\xi}.
\end{IEEEeqnarray}
The second moment of the truncated Gaussian distribution is:
\begin{IEEEeqnarray}{rCl}\label{e: TG 2nd moment}
    p_z&=&\frac{1}{\theta}\int_\eta^\xi{\chi
    e^{-\chi}d\chi}
    =1+\frac{\eta e^{-\eta}-\xi e^{-\xi}}{e^{-\eta}-e^{-\xi}}.
\end{IEEEeqnarray}
The Fourth moment of the truncated Gaussian distribution is:
\begin{IEEEeqnarray}{rCl}\label{e: TG 4th moment}
    E\left[|\zv |^4\right]&=&\frac{1}{\theta}\int_\eta^\xi{\chi^2
    e^{-\chi}d\chi}
    =2+\frac{(\eta^2+2 \eta)e^{-\eta}-(\xi^2+2 \xi)e^{-\xi} }
    {e^{-\eta}-e^{-\xi}}
\end{IEEEeqnarray}

Using the truncated Gaussian distribution, we define the input
symbol as:
\begin{IEEEeqnarray}{rCl}\label{e: TG modulation}
    \FV{X}{k}=\frac{c}{\sqrt{p_z}} {\bf g}\FV{Z}{k},
\end{IEEEeqnarray}
where $c$ is chosen to satisfy the power constraint, $\bf g$ is
defined as in the QPSK lower bound, $\FV{Z}{k}=\left[
\FV{z}{k,0}\ldots,\FV{z}{k,r-1},0,\ldots,0 \right]$, $1\le r\le N$
is the number of active frequency bins, and $\{ \FV{z}{k,m} \}$ is
the sequence of iid input random variables with truncated Gaussian
distribution $\FV{z}{k,m}\sim\TCN (\eta,\xi)$.

\begin{Lemma}\label{L: TG bound}
    If $E[\FV{h}{0,m}]=0$, the capacity of a channel with a
quadratic power constraint (and $\alpha>1$) is lower bounded by:
\begin{IEEEeqnarray}{rCl}\label{e: almost final TG}
        C & \ge  &\lim \limits_{k \to \infty }
        \max_{r\le N} \sup_{\eta>0} \max_{1\le\beta\le \frac{\alpha r(1+\eta)^2}{1+r(1+\eta)^2}}
-\frac{r \eta}{N\beta} \nonumber
\\
&&     +\frac{1}{N\beta}
    \sum_{m=0}^{r-1}
     \log\(1+\frac{\beta\frac{N^2}{r} p_x\left(\frac{1}{N}- E\left[ \left. (\tilde{\mathcal
{\bm \Leps}}_k)_{m,m}\right| {\bf g}=1\right]\right)\left|\uv
\right|^2}
    {1+\beta\frac{N^2}{r} p_x E\left[ \left. (\tilde{\mathcal
{\bm \Leps}}_k)_{m,m}\right| {\bf g}=1\right]}\)
\end{IEEEeqnarray}
and the capacity of a channel with a peak power constraint is lower
bounded by:
\begin{IEEEeqnarray}{rCl}\label{e: IT peak TG bound}
        C &\ge  &\lim \limits_{k \to \infty }\max_{r\le N}
        \sup_{0<\eta<\xi}
    \frac{1}{N}
    \sum_{m=0}^{r-1}
     \log\(1+\frac{\frac{N^2 p_x p_z}{r\xi}   \left(\frac{1}{N}- E\left[ \left. (\tilde{\mathcal
{\bm \Leps}}_k)_{m,m}\right| {\bf g}=1\right]\right)\left|\uv
\right|^2}
    {1+\frac{N^2 p_x p_z}{r\xi} E\left[ \left. (\tilde{\mathcal
{\bm \Leps}}_k)_{m,m}\right| {\bf g}=1\right]}\) \nonumber
\\ &&+\frac{r}{N\beta}\log(e^{-\eta}-e^{-\xi}).\IEEEeqnarraynumspace
\end{IEEEeqnarray}
where $\uv $ is a proper complex Gaussian random variable $\(\uv
\sim\CN(0,1)\)$ and $\eta$ and $\xi$ are the thresholds used to
define the truncated Gaussian distribution.
\end{Lemma}

As in the previous bounds, this lower bound depends on the
conditional channel distribution (through the expectation of the
conditional channel covariance matrix). Note that \CommentApril{the}
covariance matrix itself also depends on the definition of the
truncated Gaussian distribution, i.e., on  the parameters $\eta$ and
$\xi$.

\textit{Proof of Lemma \ref{L: TG bound}:} See Appendix \ref{app:
Proof of Lemma 3a}.

\setcounter{Lemma}{0}
\renewcommand{\theLemma}{\arabic{Lemma}.b}
\section{Channel estimation}\label{Sec. Channel estimation}
In this section we analyze the conditional distribution of the
channel given the past transmitted and received symbols, and provide
the bounds required for the capacity evaluation.

Using Assumption \ref{assump X iid H} (no feedback) the transmitted
signal is statistically independent from the channel, and hence,
given the input symbols $\FV[k-1]{X}{0}$, the output symbols
$\FV[k-1]{Y}{0}$ are jointly Gaussian with the channel
$\FV[k]{H}{0}$. Thus, the desired distribution can be analyzed
through the theory of optimal linear estimation. The auto-covariance
of the conditioned output symbols, $\FV[k-1]{Y}{0} | \FV[k-1]{X}{0}
$, is:
\begin{IEEEeqnarray}{rCl}
    \label{e. condtional channel distribution mean}
        {\Ct}_{\FV[k-1]{Y}{0}| \FV[k-1]{X}{0}}= N \diagM(\FV[k-1]{X}{0}) {\openmtr{C}}_{\FV[k-1]{H}{0}} \diagM(\FV[k-1]{X}{0})^\dagger    +
        \It
\end{IEEEeqnarray}
and their cross covariance matrix with $\FV{H}{k}$ is:
\begin{IEEEeqnarray}{rCl}
    \label{e. channel data correlation}
        {\Ct}_{\FV[k-1]{Y}{0},\FV{H}{k} | \FV[k-1]{X}{0}}  =
        \sqrt{N} \cdot \diagM(\FV[k-1]{X}{0})
        {\openmtr{C}}_{\FV[k-1]{H}{0},\FV{H}{k}}
\end{IEEEeqnarray}
where  ${\openmtr{C}}_{\FV[k-1]{H}{0},\FV{H}{k}}$ is the sub-matrix
of ${\openmtr{C}}_{\FV[k-1]{H}{0}}$ describing the cross covariance
between current channel vector and the past channel vectors.
Therefore, $\FV{H}{k} |\FV[k-1]{X}{0} ,\FV[k-1]{Y}{0}$ has complex
Gaussian distribution $\FV{H}{k} |\FV[k-1]{X}{0} ,\FV[k-1]{Y}{0}
\sim \CN(\tilde{\bm{\mu }}_k ,\tilde{\mathcal {\bm \Leps}}_k )$,
where the conditional channel mean, $\tilde{\bm{\mu }}_k$, and the
conditional channel covariance matrix, $\tilde{\mathcal {\bm
\Leps}}_k$, are given by \cite{kay1993fss}:
\begin{IEEEeqnarray}{rCl}
    \label{e. condtional channel distribution mean}
        \tilde{\bm{\mu }}_k  &= & E[\FV{H}{k}]
        +\sqrt{N} {{\openmtr{C}}_{\FV[k-1]{H}{0},\FV{H}{k}}^\dagger}  \diagM(\FV[k-1]{X}{0})^\dagger
        \left( N \diagM(\FV[k-1]{X}{0}) {\openmtr{C}}_{\FV[k-1]{H}{0}} \diagM(\FV[k-1]{X}{0})^\dagger    + \It  \right)^{ - 1}
        \nonumber \\
        &&\cdot \left( \FV[k-1]{Y}{0} - \sqrt{N} \diagM(\FV[k-1]{X}{0}) E[\FV[k-1]{H}{0}] \right)
\end{IEEEeqnarray}
\begin{IEEEeqnarray}{rCl}
    \label{e. condtional channel distribution variance}
        \tilde{\mathcal {\bm \Leps}}_k  &=& {\openmtr{C}}_{\FV{H}{k}}  -
        N {{\openmtr{C}}_{\FV[k-1]{H}{0},\FV{H}{k}}^\dagger}  \diagM(\FV[k-1]{X}{0})^\dagger
        \left( N \diagM(\FV[k-1]{X}{0}) {\openmtr{C}}_{\FV[k-1]{H}{0}} \diagM(\FV[k-1]{X}{0})^\dagger    + \It  \right)^{ - 1}
        \nonumber \\
        &&\cdot  \diagM(\FV[k-1]{X}{0}) {{\openmtr{C}}_{\FV[k-1]{H}{0},\FV{H}{k}}}
        .
\end{IEEEeqnarray}

We note that the conditional channel mean is characterized by the
conditional channel covariance matrix, $\tilde{\mathcal {\bm
\Leps}}_k$, since we can write the distribution of the channel mean
as $\tilde{\bm{\mu }}_k |\FV[k-1]{X}{0} \sim \CN \left(
E[\FV{H}{k}],{\openmtr{C}}_{\FV{H}{k}}-\tilde{\mathcal {\bm
\Leps}}_k
 \right)$. Therefore, we will focus hereon
on the characterization of the conditional channel covariance matrix
$\tilde{\mathcal {\bm \Leps}}_k$.

In this section we characterize the bounds on $(\tilde{\mathcal {\bm
\Leps}}_k)_{m,m}$ and its moments in terms of the effective
coherence time.

\begin{Lemma}\label{L: peak estimation bound}
With a peak power constraint, the conditional channel covariance
matrix is lower bounded by:
\begin{IEEEeqnarray}{rCl}
    \label{e. peak estimation bound}
    \lim \limits_{k \to \infty }\min_m \inf_{\FV[k-1]{X}{0}\in\mathcal{I}(N p_x \It) }(\tilde{\mathcal {\bm \Leps}}_k)_{m,m}
        &\ge&\frac{1}{N}
    \frac{1}{1+\frac{p_x}{2}(\hTc(p_x)-N)}
     .
\end{IEEEeqnarray}
With a quadratic power constraint, the conditional channel
covariance matrix satisfies:
\begin{IEEEeqnarray}{rCl}
    \label{e. 2 moment estimation bound}
    \lim \limits_{k \to \infty }
            \sup \limits_{{\rm Pr} (\Pc_{k-1})} E \left[ \left\{ 1-N\min_m \inf_{\FV[k-1]{X}{0}\in\mathcal{I}(\Pc_{k-1}) }(\tilde{\mathcal {\bm \Leps}}_k)_{m,m}
        \right\}^2 \right]
    &\le&
    \alpha \(1-\frac{1}{1+\frac{p_x}{2}(\hTc(p_x)-N)}\)^2
    ,\IEEEeqnarraynumspace
\end{IEEEeqnarray}
\end{Lemma}
\begin{IEEEproof}[Proof of Lemma \ref{L: peak estimation bound}]
In Appendix \ref{app: Minimization of the estimation error} we prove
the following inequality:
\begin{IEEEeqnarray}{rCl}
    \label{e. estimation error frequency minima}
    (\tilde{\mathcal {\bm \Leps}}_k)_{m,m}
        &\ge&
    \frac{1}{N}\left[1 - {{D}_{k}^\dagger}
    \Pc_{k-1}\left( \It + {{{\Delta}_{k-1}}} \Pc_{k-1}
     \right)^{-1} {{D}_{k}}\right].
\end{IEEEeqnarray}
The proof is based on showing that the conditional covariance is
minimized using $\Sf_{\RV[k-1]{X}{0}}= \Pc_{k-1} \otimes \Js_m$,
where $\Js_m$ is an $N \times N$ matrix with only one
\CommentApril{nonzero} element, the $m$-th element on the diagonal,
which equals 1.

The first part of the lemma follows directly from (\ref{e.
estimation error frequency minima}) by noting that the conditional
covariance matrix is a nonincreasing function of each element in
$\Pc_{k-1}$. Therefore, it is minimized when all elements take their
maximal value ($\Pc_{k-1}=N p_x \It$), so that
\begin{IEEEeqnarray}{rCl}
    \lim \limits_{k \to \infty }\min_m \inf_{\FV[k-1]{X}{0}\in\mathcal{I}(N p_x \It) }(\tilde{\mathcal {\bm \Leps}}_k)_{m,m}
        &\ge&
    \frac{1}{N}\left[1 - \lim \limits_{k \to \infty } N p_x {{D}_{k}^\dagger}
    \left( \It + N p_x{{{\Delta}_{k-1}}}
     \right)^{-1} {{D}_{k}}\right]
.
\end{IEEEeqnarray}
Using the definition of the effective coherence time, (\ref{d:
effective coherence time}), and the matrix inversion lemma leads to
(\ref{e. peak estimation bound}) and completes the first part of the
proof.

For the quadratic case the bounding is slightly more complicated,
and we search for bounds on moments of the conditional covariance
matrix that will hold for any distribution of the transmission
powers, $\Pc_{k-1}$. To do so, we use the same derivation as in
(\ref{e. estimation sum separation}), and apply it to (\ref{e.
estimation error frequency minima}) with $\Pc_{k-1}=E[\Pc_{k-1}] +
\breve\Pc$. We get:
\begin{IEEEeqnarray}{rCl}\label{e: estimation error bound 1}
    (\tilde{\mathcal {\bm \Leps}}_k)_{m,m}
    &\ge&
    \frac{1}{N} - \frac{1}{N}{{D}_{k}^\dagger}
    E[\Pc_{k-1}] \left( \It + {{{\Delta}_{k-1}}} {E[\Pc_{k-1}]}
    \right)^{-1} {{D}_{k}}
        \nonumber \\
        &&- \frac{1}{N}{{D}_{k}^\dagger}
        \left( \It + {E[\Pc_{k-1}]} {{{\Delta}_{k-1}}} \right)^{ - 1}
        \breve\Pc
        \left( \It + {{{\Delta}_{k-1}}} {E[\Pc_{k-1}]}
    \right)^{ - 1}
        {{D}_{k}}.
\end{IEEEeqnarray}
Taking the expectation of the square of (\ref{e: estimation error
bound 1}) and using  $(\tilde{\mathcal {\bm \Leps}}_k)_{m,m}\le
\frac{1}{N}$, we have:
\begin{IEEEeqnarray}{rCl}
    \label{e. estimation 2 moment}
    E\left[ \left\{ 1-N(\tilde{\mathcal {\bm \Leps}}_k)_{m,m} \right\}^2 \right]
    &\le&
    \(  {{D}_{k}^\dagger}
    E[\Pc_{k-1}] \left( \It + {{{\Delta}_{k-1}}} E[\Pc_{k-1}]
    \right)^{-1} {{D}_{k}}\)^2
         + E \left[ {V^\dagger
        \breve\Pc VV^\dagger  \breve\Pc
        V}
    \right].\IEEEeqnarraynumspace
\end{IEEEeqnarray}
where $\CommentApril{V}=
        \left( \It + {{{\Delta}_{k-1}}} E[\Pc_{k-1}]
    \right)^{ - 1}
        {{D}_{k}}
        $.
Focusing on the second term in the righthand side of (\ref{e.
estimation 2 moment}) we have:
\begin{IEEEeqnarray}{rCl}
    E \left[ {V^\dagger
        \breve\Pc VV^\dagger  \breve\Pc
        V}\right]&=&
    E\left[\sum_a \sum_b v^*_a
        \breve\Pc_{a,a} v_a v^*_b \breve\Pc_{b,b}
        v_b\right]
    \nonumber \\
    & = & \sum_a \sum_b \left| v_a \right|^2 \left| v_b \right|^2
        E\left[ \breve\Pc_{a,a} \breve\Pc_{b,b}
        \right]
    \nonumber \\
    & \le & \sum_a \sum_b \left| v_a \right|^2 \left| v_b \right|^2
        \sqrt{E\left[ \breve\Pc_{a,a}^2\right]E\left[ \breve\Pc_{b,b}^2\right]}
    \nonumber \\
    & \le & (\alpha -1 )N^2 p_x^2 (V^\dagger V)^2
\end{IEEEeqnarray}
where we use $\breve\Pc_{b,b}=\RV{p}{b}-E[\RV{p}{b}]$, and hence
$E\left[\breve\Pc_{b,b}^2\right]\le(\alpha -1 )N^2 p_x^2$.
Substituting back in (\ref{e. estimation 2 moment}), we observe that
the righthand side of (\ref{e. estimation 2 moment}) is maximized
with $E[\Pc_{k-1}]=N p_x \It$ and we get:
\begin{IEEEeqnarray}{rCl}
    \label{e. 2 moment estimation baount}
    E\left[ \left\{ 1-N(\tilde{\mathcal {\bm \Leps}}_k)_{m,m} \right\}^2 \right]
    &\le&
    N^2 p_x^2\( {{D}_{k}^\dagger}
    \left( \It + N p_x {{{\Delta}_{k-1}}}
    \right)^{-1} {{D}_{k}}\)^2
        \nonumber \\
        &&+ \: (\alpha -1 )N^2 p_x^2 \({{D}_{k}^\dagger}
        \left( \It + N p_x {{{\Delta}_{k-1}}} \right)^{ - 2}
        {{D}_{k}}
    \)^2
    \nonumber \\
    &\le&
    N^2 p_x^2\( {{D}_{k}^\dagger}
    \left( \It + N p_x {{{\Delta}_{k-1}}}
    \right)^{-1} {{D}_{k}}\)^2
        \nonumber \\
        && + \: (\alpha -1 )N^2 p_x^2 \({{D}_{k}^\dagger}
        \left( \It + N p_x {{{\Delta}_{k-1}}} \right)^{ - 1}
        {{D}_{k}}
    \)^2
    \nonumber \\
    &=&
    \alpha N^2 p_x^2 \({{D}_{k}^\dagger}
        \left( \It + N p_x {{{\Delta}_{k-1}}} \right)^{ - 1}
        {{D}_{k}}
    \)^2
\end{IEEEeqnarray}
where the second inequality in (\ref{e. 2 moment estimation baount})
uses the  fact that all eigenvalues of $\It + N p_x
{{{\Delta}_{k-1}}}$ are larger or equal to 1, and therefore $\(\It +
N p_x {{{\Delta}_{k-1}}}\)^{-1}\ge \(\It + N p_x
{{{\Delta}_{k-1}}}\)^{-2}$. Using the effective coherence time
definition results in (\ref{e. 2 moment estimation bound}) and
completes the proof of the lemma.
\end{IEEEproof}

\begin{Lemma}\label{L: QPSK estimaiton error}
Using QPSK modulation, as defined in (\ref{d: QPSK modulation}), for
each active frequency bin ($0\le m\le r-1$) the diagonal element of
the conditional channel covariance matrix is lower bounded by:
\begin{IEEEeqnarray}{rCl}\label{e: QPSK estimaiton error}
\lim \limits_{k \to \infty }(\tilde{\mathcal {\bm \Leps}}_k)_{m,m}
&\le&\left\{ \begin{array}{ll}
 \frac{1}{N}\frac{1}{1+\frac{\beta}{L} \frac{p_x}{2}\left(\hTc\left(
\beta p_x\right)-N\right)} &r=N
\\
\frac{1}{N}
    \frac{1}{1+\frac{\beta}{r}\frac{p_x}{2}\left(\hTc\left(\frac{\beta}{r} p_x\right)-N\right)} & r<N
\end{array}
\right.
\end{IEEEeqnarray}
\end{Lemma}
\begin{IEEEproof}[Proof of Lemma \ref{L: QPSK estimaiton error}]
Starting with the wideband QPSK modulation ($r=N$), the transmission
spectrum given $\textbf{g}=1$ is equal to
$\Sf_{\RV[k-1]{X}{0}}=\beta p_x \It$. Substituting in the estimation
error covariance matrix, (\ref{e. condtional channel distribution
variance}) (see also (\ref{e: estimation error wirh spectrun})), we
have:
\begin{IEEEeqnarray}{rCl}
    \label{d: estimation error with expectation}
    \tilde{{\bm \Leps}}_k  = {\openmtr{C}}_{\FV{H}{k}}  -
        N \beta p_x {{\openmtr{C}}_{\FV[k-1]{H}{0},\FV{H}{k}}^\dagger}
        \left( N \beta p_x{\openmtr{C}}_{\FV[k-1]{H}{0}}    + \It  \right)^{ - 1}
        {{\openmtr{C}}_{\FV[k-1]{H}{0},\FV{H}{k}}}.
\end{IEEEeqnarray}

Writing the estimation error covariance matrix, (\ref{d: estimation
error with expectation}), in the time domain, and substituting
(\ref{d: kron channel matrix}) we have:
\begin{IEEEeqnarray}{rCl}
    \label{e: tap estimation error variance}
    {\bm \Leps}_k  = \Ft \tilde{{\bm \Leps}}_k \Ft^\dagger ={\openmtr{C}}_{\RV{H}{k}}  -
        N \beta p_x {{D}_{k}^\dagger} \otimes {\openmtr{C}}_{\RV{H}{k}}
        \left( N \beta p_x {{\Delta}_{k-1}} \otimes {\openmtr{C}}_{\RV{H}{k}} + \It  \right)^{ - 1}
        {{D}_{k}} \otimes {\openmtr{C}}_{\RV{H}{k}}.
\end{IEEEeqnarray}
Note that covariance matrices are hermitian so that
${\openmtr{C}}_{\RV{H}{k}}^\dagger={\openmtr{C}}_{\RV{H}{k}}$. In
Appendix \ref{app: Derivation of equation} we show that this is a
diagonal matrix in which the $l$-th element on the diagonal is given
by:
\begin{IEEEeqnarray}{rCl}\label{e. tap etsimation error}
    ({\bm \Leps}_k)_{l,l}  &=& ({\openmtr{C}}_{\RV{H}{k}})_{l,l}  -
        N \beta p_x  {{D}_{k}^\dagger}
        ({\openmtr{C}}_{\RV{H}{k}})_{l,l} \left(N \beta p_x {{\Delta}_{k-1}}
        ({\openmtr{C}}_{\RV{H}{k}})_{l,l}+\It\right)^{-1}{{D}_{k}}
        ({\openmtr{C}}_{\RV{H}{k}})_{l,l}
.
\end{IEEEeqnarray}
For any channel tap such that $({\openmtr{C}}_{\RV{H}{k}})_{l,l}>0$
we can use the matrix inversion lemma to write the inverse of the
estimation error as:
\begin{IEEEeqnarray}{rCl}\label{e. inverse tap etsimation error first one}
    ({{\bm \Leps}}_k)_{l,l}^{-1}  &=& ({\openmtr{C}}_{\RV{H}{k}})_{l,l}^{-1}  +
        N \beta p_x
        {{D}_{k}^\dagger}
        \cdot
        \left( N \beta p_x ({\openmtr{C}}_{\RV{H}{k}})_{l,l} \left[{{\Delta}_{k-1}} -
        {{D}_{k}} {{D}_{k}^\dagger}  \right]
        + \It
        \right)^{ - 1}
        {{D}_{k}}
\end{IEEEeqnarray}
and taking the limit as $k$ goes to infinity:
\begin{IEEEeqnarray}{rCl}
    \label{e. inverse tap etsimation error}
    \lim \limits_{k \to \infty } ({{\bm \Leps}}_k)_{l,l}^{-1}
&= & ({\openmtr{C}}_{\RV{H}{k}})_{l,l}^{-1} +\frac{\beta
p_x}{2}\left(\hTc\left(\beta p_x
({\openmtr{C}}_{\RV{H}{k}})_{l,l}\right)-N\right) \nonumber \\ &\ge&
({\openmtr{C}}_{\RV{H}{k}})_{l,l}^{-1} +\frac{\beta
p_x}{2}(\hTc\left( \beta p_x\right)-N)
        .
\end{IEEEeqnarray}

Going back to the frequency domain, the properties of the DFT
guarantee that all elements on the diagonal of $\tilde{\bm \Leps}_k$
are equal. We therefore have:
\begin{IEEEeqnarray}{rCl}\label{d. c_wb}
    \lim \limits_{k \to \infty }(\tilde{\mathcal {\bm \Leps}}_k)_{m,m}&=&\frac{1}{N}{\rm Tr}\left\{\lim \limits_{k \to \infty } {\bm \Leps}_k\right\}
        \le\frac{1}{N}\sum_{l:(\Ct_{\RV{H}{k}} )_{l,l}\ne 0}\frac{1}{(\Ct_{\RV{H}{k}} )_{l,l}^{ - 1}+\beta \frac{p_x}{2}\left(\hTc\left(\beta p_x\right)-N\right)}
.
\end{IEEEeqnarray}
Using the concavity of the function $(a+1/x)^{-1}$ for positive $a$
and ${\rm Tr}[{\openmtr{C}}_{\RV{H}{k}}]=1$ proves the first (upper)
part of (\ref{e: QPSK estimaiton error}).

For $r<N$ we lower bound the diagonal element of the conditional
channel covariance matrix by the performance of a sub-optimal
estimation scheme. This estimation scheme estimates the channel at
the $m$-th frequency bin based only on data transmitted and received
over the $m$ frequency bin. Inspecting (\ref{d: estimation error
with expectation}) and taking only the elements that correspond to
the $m$ frequency bin ($0\le m\le r-1$) the conditional channel
variance is lower bounded by:
\begin{IEEEeqnarray}{rCl}\label{e: eq 93}
    (\tilde{\bm \Leps}_k)_{m,m}  \le \frac{1}{N}  -
        \frac{1}{N} \beta\frac{N}{r} p_x D_k^\dagger
        \left( \beta\frac{N}{r} p_x\Delta_{k-1}    + \It  \right)^{ - 1}
        D_k.
\end{IEEEeqnarray}
where we took into account the normalization
$({\openmtr{C}}_{\FV{H}{k}})_{m,m}=\frac{1}{N}$ and the fact that
the transmitted power for an active bin, given $\textbf{g}=1$, is
$\beta\frac{N}{r} p_x$. Using the matrix inversion lemma and the
definition of the effective coherence time, (\ref{d: effective
coherence time}),  results in the second part of (\ref{e: QPSK
estimaiton error}) and completes the proof of the lemma.
\end{IEEEproof}
\begin{Lemma}\label{L: TG estimaiton error}
Using truncated Gaussian input distribution, as defined in (\ref{e:
TG modulation}), for each active frequency bin ($0\le m\le r-1$) the
mean of the diagonal element of the conditional channel covariance
matrix is lower bounded for quadratic power constraint by:
\begin{IEEEeqnarray}{rCl}\label{e: TG estimaiton error}
\lim \limits_{k \to \infty }E\left[(\tilde{\mathcal {\bm
\Leps}}_k)_{m,m} \right]&\le&\left\{ \begin{array}{ll}
 \frac{1}{N}\frac{1}{1+\frac{\beta p_x}{(1+\eta)\log(1+\frac{1}{\eta})}\frac{1}{2 L} \left(\hTc\left(
 \frac{\beta p_x}{(1+\eta)\log(1+\frac{1}{\eta})} \right)-N\right)} &r=N
\\
 \frac{1}{N}\frac{1}{1+\frac{\beta p_x}{(1+\eta)\log(1+\frac{1}{\eta})}\frac{1}{2 r} \left(\hTc\left(
 \frac{\beta p_x}{r(1+\eta)\log(1+\frac{1}{\eta})} \right)-N\right)} & r<N
\end{array}
\right.
\end{IEEEeqnarray}
and for peak power constraint by:
\begin{IEEEeqnarray}{rCl}\label{e: TG estimaiton error peak}
\lim \limits_{k \to \infty }E\left[(\tilde{\mathcal {\bm
\Leps}}_k)_{m,m} \right]&\le&\left\{ \begin{array}{ll}
 \frac{1}{N}\frac{1}{1+\frac{p_x }{\xi}
    \frac{1-e^{\eta-\xi}}{\log(1+\frac{1}{\eta})}\frac{1}{2L}\left(\hTc\left(
 \frac{p_x }{\xi}
    \frac{1-e^{\eta-\xi}}{\log(1+\frac{1}{\eta})}\right)-N\right)} &r=N
\\
 \frac{1}{N}\frac{1}{1+\frac{p_x }{\xi}
    \frac{1-e^{\eta-\xi}}{\log(1+\frac{1}{\eta})}\frac{1}{2r}\left(\hTc\left(
 \frac{p_x }{r\xi}
    \frac{1-e^{\eta-\xi}}{\log(1+\frac{1}{\eta})}\right)-N\right)} & r<N
\end{array}
\right.
\end{IEEEeqnarray}
\end{Lemma}
\begin{IEEEproof}[Proof of Lemma \ref{L: TG estimaiton error}]
Starting from the wideband truncated Gaussian distribution ($r=N$), the
power transmitted in all frequency bins is strictly positive. We
therefore can write the conditional channel covariance matrix
(\ref{e. condtional channel distribution variance}) (see also
(\ref{e: estimation error wirh spectrun})), as:
\begin{IEEEeqnarray}{rCl}\label{e: estimation error wirh inverse spectrun}
        \tilde{\mathcal {\bm \Leps}}_k =f_1\left(\frac{1}{N}\Sf_{\RV[k-1]{X}{0}}^{-1}\right)= {\openmtr{C}}_{\FV{H}{k}}  - {\openmtr C}_{\FV[k-1]{H}{0}\FV{H}{k}}^\dagger
        \left( {\openmtr C}_{\FV[k-1]{H}{0}} + \frac{1}{N}\Sf_{\RV[k-1]{X}{0}}^{-1} \right)^{ - 1}
        {\openmtr C}_{\FV[k-1]{H}{0}\FV{H}{k}}.
\end{IEEEeqnarray}

Next we show that $f_1(\CommentApril{\mtr{A}})$ is a concave
function of each of the elements in the strictly positive diagonal
matrix $\mtr{A}$. Denote by $\Iv_i$ the vector which is all zeros
except for a 1 in the $i$-th element. The derivative of
$f_1(\mtr{A})$ with respect to $(\mtr{A})_{i,i}$ is given by:
\begin{IEEEeqnarray}{rCl}\label{e: espilon derive 1}
\frac{\partial f_1(\mtr{A})}{\partial  (\mtr{A})_{i,i}  }=
    {\openmtr C}_{\FV[k-1]{H}{0}\FV{H}{k}}^\dagger
        \left( {\openmtr C}_{\FV[k-1]{H}{0}} + \mtr{A}\right)^{ - 1} {\Iv}_i
        {\Iv}_i^\dagger \left( {\openmtr C}_{\FV[k-1]{H}{0}} + \mtr{A}\right)^{ - 1}{\openmtr C}_{\FV[k-1]{H}{0}\FV{H}{k}}
\end{IEEEeqnarray}
which is a \CommentApril{positive semi-definite} matrix. The second
derivative is given by:
\begin{IEEEeqnarray}{rCl}
\frac{\partial^2 f_1(\mtr{A})}{\partial  (\mtr{A})_{i,i}^2  }=  -2
{\openmtr C}_{\FV[k-1]{H}{0}\FV{H}{k}}^\dagger
        \left( {\openmtr C}_{\FV[k-1]{H}{0}} + \mtr{A}\right)^{ - 1} {\Iv}_i
        {\Iv}_i^\dagger \left( {\openmtr C}_{\FV[k-1]{H}{0}} + \mtr{A}\right)^{ - 1} {\Iv}_i
        {\Iv}_i^\dagger \left( {\openmtr C}_{\FV[k-1]{H}{0}} + \mtr{A}\right)^{ - 1}{\openmtr
        C}_{\FV[k-1]{H}{0}\FV{H}{k}} \IEEEeqnarraynumspace
\end{IEEEeqnarray}
which is a \CommentApril{negative semi-definite} matrix, and
guaranties that the function $f_1(\mtr{A})$ is a concave function of
$(\mtr{A})_{i,i}$.

In order to avoid the differentiation with respect to the whole
matrix $\mtr{A}$ we considered only the derivatives with respect to
each single element in it. But, as we showed that the function
$f_1(\mtr{A})$ is a concave function of each of the (diagonal)
elements in $\mtr{A}$ regardless of the others elements, we can now
apply the Jensen's inequality serially on one matrix element at a
time. Going over all matrix elements we get:
\begin{IEEEeqnarray}{rCl}\label{e: f estimation error wirh inverse spectrun}
        E\left[\tilde{\mathcal {\bm \Leps}}_k\right] =E\left[f_1\left(\frac{1}{N}\Sf_{\RV[k-1]{X}{0}}^{-1}\right)\right]\le {\openmtr{C}}_{\FV{H}{k}}  - {\openmtr C}_{\FV[k-1]{H}{0}\FV{H}{k}}^\dagger
        \left( {\openmtr C}_{\FV[k-1]{H}{0}} + \frac{1}{N}E\left[\Sf_{\RV[k-1]{X}{0}}^{-1}\right] \right)^{ - 1}
        {\openmtr C}_{\FV[k-1]{H}{0}\FV{H}{k}}. \IEEEeqnarraynumspace
\end{IEEEeqnarray}
Taking into account the definition of the transmitted signal,
(\ref{e: TG modulation}), the expectation of the inverse of the
transmitted spectrum, given $\textbf{g}=1$, is:
\begin{IEEEeqnarray}{rCl}\label{e: TG error expectation}
    E\left[\Sf_{\RV[k-1]{X}{0}}^{-1}\right]&=&\It\cdot\frac{p_z}{c^2}
    \frac{1}{\theta }\int_\eta^\xi{\chi^{-1}
    e^{-\chi}d\chi}
    <\It\cdot\frac{p_z}{c^2}
    \frac{E_1(\eta)}{e^{-\eta}-e^{-\xi}}
    <\It\cdot\frac{p_z}{c^2}
    \frac{e^{-\eta}\log(1+\frac{1}{\eta})}{e^{-\eta}-e^{-\xi}}
\end{IEEEeqnarray}
where $E_1()$ is the exponential integral function, and the second
inequality used $E_1(a)<e^{-a}\log(1+\frac{1}{a})$
\cite{abramowitz1965handbook}.

Substituting (\ref{e: TG error expectation}) in (\ref{e: f
estimation error wirh inverse spectrun}) \CommentApril{results in}:
\begin{IEEEeqnarray}{rCl}\label{e: TG expec bound 3}
        E\left[\tilde{\mathcal {\bm \Leps}}_k\right] \le {\openmtr{C}}_{\FV{H}{k}}  - N \zeta {\openmtr C}_{\FV[k-1]{H}{0}\FV{H}{k}}^\dagger
        \left( N \zeta {\openmtr C}_{\FV[k-1]{H}{0}} + \It\right)^{ - 1}
        {\openmtr C}_{\FV[k-1]{H}{0}\FV{H}{k}} \IEEEeqnarraynumspace
\end{IEEEeqnarray}
where for quadratic power constraint, substituting (\ref{e: TG 2nd
moment}), (\ref{e: c vlaue quad}), $r=N$ and $\xi=\infty$ we have:
\begin{IEEEeqnarray}{rCl}\label{e: zeta quad}
    \zeta&=&\beta \frac{N}{r}p_x\frac{1}{(1+\eta)\log(1+\frac{1}{\eta})
    }.
\end{IEEEeqnarray}
For the peak power constraint, substituting (\ref{e: c vlaue peak})
and $\beta=1$
\begin{IEEEeqnarray}{rCl}\label{e: zeta peak}
        \zeta&=&\frac{N p_x }{r\xi}
    \frac{e^{-\eta}-e^{-\xi}}{e^{-\eta}\log(1+\frac{1}{\eta})}.
\end{IEEEeqnarray}

Comparing  (\ref{e: TG expec bound 3}) with (\ref{d: estimation
error with expectation}) we note that the only difference is the
change of the factor $\beta p_x$ with the factor $\zeta$. Hence,
repeating the derivation of Lemma \ref{L: QPSK estimaiton error}
(Equations (\ref{e: tap estimation error variance}) to (\ref{d.
c_wb})) with the replaced factor \CommentApril{results in}:
\begin{IEEEeqnarray}{rCl}\label{e: one of lemma 3b equations}
\lim \limits_{k \to \infty }E\left[(\tilde{\mathcal {\bm
\Leps}}_k)_{m,m} \right]&\le&
 \frac{1}{N}\frac{1}{1+\frac{\zeta}{2 L} \left(\hTc\left(
\zeta \right)-N\right)}.
\end{IEEEeqnarray}
Substituting the proper values of $\zeta$ in (\ref{e: one of lemma
3b equations}) lead to the upper parts of (\ref{e: TG estimaiton
error}) and (\ref{e: TG estimaiton error peak}) and completes the
first part of the lemma.

Next for $r<N$. As in the QPSK case, we turn to the suboptimal
single frequency bin estimation. In this case, the estimation error
in the $m$-th frequency bin ($0\le m\le r-1$) is given by:
\begin{IEEEeqnarray}{rCl}\label{e: bin estimation error wirh inverse spectrun}
        \left(\tilde{\mathcal {\bm \Leps}}_k\right)_{m,m} \le \frac{1}{N}  - \frac{1}{N^2}D_{k}^\dagger
        \left( \frac{1}{N}\Delta_{k-1} + \frac{1}{N}{\cal A}_m \right)^{ - 1}
        D_{k}
\end{IEEEeqnarray}
where the diagonal matrix ${\cal A}_m$ is generated by taking from
the matrix $\Sf_{\RV[k-1]{X}{0}}^{-1}$ only the terms corresponding
to the $m$-th element of each OFDM symbol.

Comparing (\ref{e: bin estimation error wirh inverse spectrun}) with
(\ref{e: estimation error wirh inverse spectrun}), they have exactly
the same structure, and hence we can repeat the first part of the
proof (Equations (\ref{e: espilon derive 1}) to (\ref{e: TG expec
bound 3})), adjusting only the corresponding matrices. The resulting
bound is:
\begin{IEEEeqnarray}{rCl}\label{e: narrow TG expec bound 3}
        E\left[\tilde{\mathcal {\bm \Leps}}_k\right] \le \frac{1}{N}  -  \frac{1}{N}\zeta D_k^\dagger
        \left(  \zeta \Delta_{k-1} + \It\right)^{ - 1}
        D_k.
\end{IEEEeqnarray}
with the values of $\zeta$ given by (\ref{e: zeta quad}) and
(\ref{e: zeta peak}).

Using the matrix inversion lemma \CommentApril{$ \left( {A + UCV}
\right)^{ - 1}  = A^{ - 1}  - A^{ - 1} U\left( {C^{ - 1}  + VA^{ -
1} U} \right)^{ - 1} VA^{ - 1} $} and the definition of the
effective coherence time, (\ref{d: effective coherence time}),
results in the lower parts of (\ref{e: TG estimaiton error}) and
(\ref{e: TG estimaiton error peak}) and completes the proof of the
lemma.
\end{IEEEproof}

\section{Conclusions}\label{sec: summary}
In this paper we derived bounds on the capacity of the noncoherent
stationary underspread complex Gaussian OFDM-WSSUS channel with a
peak power constraint or a quadratic power constraint. The bounds
are characterized only by the system signal-to-noise ratio (SNR)
$p_x$ and by a newly defined effective coherence time $\hTc(p_x)$,
which measures the capability to estimate the channel and is a
nonincreasing function of the system SNR. The bounds show that:
\begin{itemize}
\item{The coherent channel capacity is achievable if $p_x \hTc(p_x)\gg 1 $ and $\hTc(p_x)\gg N$.}
\item{For the peak power constraint, if the channel has zero mean and $p_x \ll 1/\hTc(p_x)$ the capacity is approximately $\frac{1}{2}p_x^2 \hTc(p_x)$.}
\item{For the quadratic power constraint if the channel has zero mean and $p_x \ll 1/\hTc(p_x)$ the capacity ranges between $\frac{1}{2}p_x^2 \hTc(p_x)$ and
$\frac{\alpha}{2}p_x^2 \hTc(p_x)$ (typically the effective coherence
time changes slowly with SNR and the higher bound better
characterizes the capacity). If the limit $\hTc[_0]=\lim_{p_x
\rightarrow 0} \hTc(p_x)$ exists, the capacity in the very low SNR limit is $\frac{\alpha}{2}p_x^2 \hTc(p_x)$.}
\item{As long as the effective coherence time is large enough ($\hTc(p_x)\gg
N$) the channel capacity is achievable using a receiver that
performs channel estimation based on past received and transmitted
symbols, and decodes based on minimum Euclidean distance.}
\end{itemize}

The paper presented an initial study of the effective coherence time
and plotted it for the auto-regressive AR1 channel. Owing to its
relevance in the characterization of the channel capacity, it is
important to better understand its properties. Future work is
required to characterize the effective coherence time of commonly
used channels, and to study  the relationship between the effective
coherence time and channel capacity in more complicated channel
structures.

%\useRomanappendicesfalse
\begin{appendices}
\renewcommand{\theequation}{\Alph{section}.\arabic{equation}}
\setcounter{equation}{0}
\section{Proof of Lemma \ref{L: TG bound}}\label{app: Proof of Lemma 3a}
As stated in the proof of Lemma \ref{L: QPSK LB}, we can safely
assume that ${\bf g}$ can be decoded with no error. We reuse the
derivation of Lemma \ref{L: QPSK LB} up to Equation (\ref{e.
information lim}), and focus on the mutual information of a single
frequency bin:
\begin{IEEEeqnarray*}{rCl}
        C \ge \lim \limits_{k \to \infty } \frac{1}{N \beta}  \sum\limits_{m = 0}^{r-1}{I(\FV{z}{k,m} ;\FV{y}{k,m} | \FV[k-1]{Z}{0} ,\FV[k-1]{Y}{0}  ,{\bf
        g}=1)}.
\end{IEEEeqnarray*}

In order to lower bound the mutual information we use the
generalized mutual information\CommentApril{ (GMI, see for example
\cite{Lapidoth:02})}. The derivation of the GMI can give a lower
bound on the mutual information evaluated by:
\begin{IEEEeqnarray}{rCl}\label{E: GMI lower bound}
I(\zv ;\yv )\ge E\left[ \log \frac{e^{-sd(\zv ,\yv
)}}{E\left[e^{-sd(\zv ',\yv )}|\yv \right]}\right]
\end{IEEEeqnarray}
where $s$ is any positive constant, $d(\zv ,\yv )$ is any distance
metric that defines the operation of the receiver (which chooses the
codeword that minimizes the distance to the received signal), and
$\zv '$ is a random variable with the same distribution of $\zv $
but statistically independent of $\zv $ and $\yv
$\CommentDeleteApril{ (see for example \cite{Lapidoth:02})}.

We next evaluate such lower bound based on \CommentApril{the
}transmission of a truncated Gaussian distributed symbol over a flat
fading Gaussian channel. Consider the channel:
\begin{IEEEeqnarray}{rCl}
    \yv =\textbf{h}\zv +\textbf{v}
\end{IEEEeqnarray}
where $\zv \sim \TCN (\eta,\xi)$ is the input symbol with truncated
Gaussian distribution, $\textbf{h}$ is the Gaussian channel with
$\textbf{h}\sim\CN(\mu,\epsilon)$ and $\textbf{v}$ is an additive
white proper complex Gaussian noise with $\textbf{v}\sim\CN(0,1)$.
We assume a conventional coherent receiver, i.e., the distance
metric is:
\begin{IEEEeqnarray}{rCl}
d(z,y)=\left|y-\mu z\right|^2.
\end{IEEEeqnarray}

The more difficult part in the evaluation of (\ref{E: GMI lower
bound}) is typically the evaluation of the expectation in the
denominator. Luckily, we can use the fact that the input
distribution is close to the Gaussian distribution. For a proper
complex Gaussian random variable $\uv \sim\CN(0,1)$ we have:
\begin{IEEEeqnarray}{rCl}
E\left[e^{-\left|\nu-a \uv
\right|^2}\right]=\frac{1}{1+|a|^2}e^{-\frac{|\nu|^2}{1+|a|^2}}.
\end{IEEEeqnarray}
As the expression inside the expectation is always positive, we can
lower bound this expectation using the truncated Gaussian
distribution by:
\begin{IEEEeqnarray}{rCl}\label{e: cond Gauss exp}
\frac{1}{1+|a|^2}e^{-\frac{|\nu|^2}{1+|a|^2}}\ge\theta
E\left[\left.e^{-\left|\nu-a \uv \right|^2}\right|\eta\le|\uv
|^2\le\xi\right].
\end{IEEEeqnarray}
where $\theta=e^{-\eta}-e^{-\xi}$ is the probability that
$\eta\le|\uv |^2\le\xi$. Substituting $\nu=\sqrt{s}\yv $ and
$a=\sqrt{s}\mu$, we have for a truncated Gaussian distributed $\zv
'$:
\begin{IEEEeqnarray}{rCl}\label{e: temp A.6}
E\left[e^{-sd(\zv ',\yv )}\Big|\yv \right]\le
\frac{1}{e^{-\eta}-e^{-\xi}}\frac{1}{1+s|\mu|^2}
e^{-\frac{s\left|\yv \right|^2}{1+s|\mu|^2}}
\end{IEEEeqnarray}
Substituting (\ref{e: temp A.6}) in (\ref{E: GMI lower bound}), the
resulting bound is:
\begin{IEEEeqnarray}{rCl}\label{e: GMI intermediate bound}
I(\zv ;\yv )&\ge&
\log(e^{-\eta}-e^{-\xi})+\log\left(1+s|\mu|^2\right)+E\left[
-s\left|\yv -\mu\zv \right|^2+\frac{s|\yv |^2}{1+s|\mu|^2}\right]
\nonumber \\
&=&\log(e^{-\eta}-e^{-\xi}) +\log\left(1+s|\mu|^2\right) -s(1+p_z
\epsilon)+\frac{s(|\mu|^2 p_z +1+p_z \epsilon)}{1+s|\mu|^2}.
\end{IEEEeqnarray}
We arbitrarily choose $s$ to zero the last 2 terms in (\ref{e: GMI
intermediate bound}), i.e.:
\begin{IEEEeqnarray}{rCl}
s=\frac{p_z}{1+p_z\epsilon},
\end{IEEEeqnarray}
which \CommentApril{results in}:
\begin{IEEEeqnarray}{rCl}
I(\zv ;\yv )&\ge& \log(e^{-\eta}-e^{-\xi})+\log\left(1+\frac{p_z
|\mu|^2}{1+p_z \epsilon}\right).
\end{IEEEeqnarray}

To use this bound for the actual channel at hand and the modulation
described by (\ref{e: TG modulation}), we substitute
$\mu=c\tilde{\bm \mu}_{k,m}\sqrt{N/p_z}$ and $\epsilon=N c^2
(\tilde{\mathcal {\bm \Leps}}_k)_{m,m}/p_z$, which
\CommentApril{results in} the bound:
\begin{IEEEeqnarray}{rCl}\label{e: TG LB}
C &\ge &\lim \limits_{k \to \infty }
    \frac{1}{N\beta}
    \sum_{m=0}^{r-1}
E\left[ \left.\log\(1+\frac{N c^2   |\tilde{\bm \mu}_{k,m}|^2}
    {1+N c^2 (\tilde{\mathcal {\bm \Leps}}_k)_{m,m}}\) \right| {\bf
    g}=1\right]+\frac{r}{N\beta}\log(e^{-\eta}-e^{-\xi})
.\IEEEeqnarraynumspace
\end{IEEEeqnarray}

Next, we show that (\ref{e: TG LB}) is convex with respect to
$(\tilde{\mathcal {\bm \Leps}}_k)_{m,m}$, and hence can be lower
bounded using the expectation of $(\tilde{\mathcal {\bm
\Leps}}_k)_{m,m}$. To do so we use the distribution of $\tilde{\bm
\mu}_{k,m}$, (\ref{e: mu distribution}). At this point we also use
$E[\FV{h}{k,m}]=0$, as was stated in the body of the lemma. We
introduce a new proper complex Gaussian random variable, ${\bf u}
\sim {\cal CN}(0,1)$, statistically independent of $(\tilde{\mathcal
{\bm \Leps}}_k)_{m,m}$, and note that the distribution of
${\tilde{\bm \mu}}_{k,m}$ is identical to the distribution of $\uv
\sqrt{\frac{1}{N}-(\tilde{\mathcal {\bm \Leps}}_k)_{m,m} } $. Using
the random variable $\uv $ we can rewrite (\ref{e: TG LB}) as:
\begin{IEEEeqnarray}{rCl}
\label{tot_exp_bound}
        C &\ge &\lim \limits_{k \to \infty }
    \frac{1}{N\beta}
    \sum_{m=0}^{r-1}
E\left[ \left.\log\(1+\frac{N c^2 \left(\frac{1}{N}-(\tilde{\mathcal
{\bm \Leps}}_k)_{m,m}\right)\left|\uv  \right|^2}
    {1+N c^2 (\tilde{\mathcal {\bm \Leps}}_k)_{m,m}}\) \right| {\bf
    g}=1\right]+\frac{r}{N\beta}\log(e^{-\eta}-e^{-\xi})
    \nonumber \\
    &=&\frac{r}{N\beta}\log(e^{-\eta}-e^{-\xi})+ \lim \limits_{k \to \infty }
    \frac{1}{N\beta}
    \sum_{m=0}^{r-1}
    \nonumber \\
    && E\left[ \left. E\left[ \left.\log\(1+\frac{N c^2
\left(\frac{1}{N}-(\tilde{\mathcal {\bm
\Leps}}_k)_{m,m}\right)\left|\uv  \right|^2}
    {1+N c^2 (\tilde{\mathcal {\bm \Leps}}_k)_{m,m}}\) \right| (\tilde{\mathcal {\bm \Leps}}_k)_{m,m}\right] \right| {\bf
    g}=1\right].
    \IEEEeqnarraynumspace
\end{IEEEeqnarray}
where in the second equation we used the law of total expectations,
and the inner expectation is taken only with respect to $\uv $.

We define the function:
\begin{IEEEeqnarray}{rCl}\label{d: f for TG LB}
    f(\epsilon)= E\left[ \log\(1+\frac{N c^2  \left(\frac{1}{N}- \epsilon\right)\left|\uv \right|^2}
    {1+N c^2  \epsilon}\) \right]
\end{IEEEeqnarray}
and rewrite (\ref{tot_exp_bound}) as:
\begin{IEEEeqnarray}{rCl}\label{e: mean f C}
        C \ge  &\lim \limits_{k \to \infty } &
    \frac{1}{N\beta}
    \sum_{m=0}^{r-1}
    E\left[ \left. f\left((\tilde{\mathcal {\bm \Leps}}_k)_{m,m}\right) \right| {\bf
    g}=1\right]+\frac{r}{N\beta}\log(e^{-\eta}-e^{-\xi}).
\end{IEEEeqnarray}
We next show that $f(\epsilon)$ is convex for
$0\le\epsilon\le\frac{1}{N}$ by proving that its second derivative
with respect to $\epsilon$ is \CommentApril{nonnegative}. The first
derivative is given by:
\begin{IEEEeqnarray}{rCl}
    \frac{d f(\epsilon)}{d\epsilon}&=&
 E\left[ \frac{-N c^2  |\uv |^2
    \left(1+N c^2 \epsilon\right)-N c^2 \left(N c^2   \left(\frac{1}{N}- \epsilon\right)|\uv |^2\right)
    }{\left(1+N c^2
\epsilon+N c^2  \left(\frac{1}{N}- \epsilon\right)|\uv
|^2\right)\left(1+N c^2 \epsilon\right)}\right]
\nonumber \\
&=& N c^2   E\left[ \frac{-(1+c^2 )|\uv |^2
    }{\left(1+N c^2
\epsilon+N c^2  \left(\frac{1}{N}- \epsilon\right)|\uv
|^2\right)\left(1+N c^2 \epsilon\right)}\right].
\end{IEEEeqnarray}
The second derivative is:
\begin{IEEEeqnarray}{rCl}\label{e:f_second_derivative}
    \frac{d^2 f(\epsilon)}{d\epsilon^2}
&=&N^2 c^4   E\left[ \frac{(1+c^2)|\uv |^2  \left(1 -  |\uv
|^2\right)\left(1+N c^2 \epsilon\right)
    }{\left(1+N c^2
\epsilon+N c^2   \left(\frac{1}{N}- \epsilon\right)|\uv
|^2\right)^2\left(1+N c^2
\epsilon\right)^2}\right] \nonumber \\
&&+ N^2 c^4    E\left[ \frac{(1+c^2 )|\uv |^2 \left(1+N c^2
\epsilon+N c^2 \left(\frac{1}{N}- \epsilon\right)|\uv |^2\right)
    }{\left(1+N c^2
\epsilon+N c^2   \left(\frac{1}{N}- \epsilon\right)|\uv
|^2\right)^2\left(1+N c^2
\epsilon\right)^2}\right]\IEEEeqnarraynumspace
\end{IEEEeqnarray}
Inspecting the expectation in the second line of
(\ref{e:f_second_derivative}), the term inside the expectation is
always positive. Using $\epsilon<\frac{1}{N}$ we can lower bound
this expectation by removing the term $N c^2  \left(\frac{1}{N}-
\epsilon\right)|\uv |^2$ from the numerator. The resulting
expectation is quite similar to the expectation in the first line of
(\ref{e:f_second_derivative}). Combining the two expectation results
in:
\begin{IEEEeqnarray}{rCl}\label{e: f_second_derivative_bound}
    \frac{d^2 f(\epsilon)}{d\epsilon^2}
&\ge& N^2 c^4   E\left[ \frac{(1+c^2)|\uv |^2 \left(2 - |\uv
|^2\right)
    }{\left(1+N c^2
\epsilon+N c^2   \left(\frac{1}{N}- \epsilon\right)|\uv
|^2\right)^2\left(1+N c^2 \epsilon\right)}\right].
\end{IEEEeqnarray}

We next use the inequality $(a-x)/(b+x)\ge(a-x)/(b+a)$ which holds
for $a,x>-b$. Slightly rewriting this inequality we have
$(2-x)/(c+dx)\ge(2-x)/(c+2d)$, and using it in (\ref{e:
f_second_derivative_bound}) we have:
\begin{IEEEeqnarray}{rCl}\label{e: f_second_derivative_bound 2}
    \frac{d^2 f(\epsilon)}{d\epsilon^2}
&\ge& \frac{N^2 c^4  (1+c^2)
    }{\left(1+N c^2
\epsilon+N c^2\left(\frac{1}{N}- \epsilon\right)\cdot
2\right)^2\left(1+\beta\frac{N^2}{r} p_x \epsilon\right)}E\left[|\uv
|^2  \left(2 -  |\uv |^2\right)\right].
\end{IEEEeqnarray}
To complete this part we note that the multiplicative term in
(\ref{e: f_second_derivative_bound 2}) is positive, and the
expectation is easily evaluated using $E[|\uv |^2]=1$ and $E[|\uv
|^4]=2$. Thus, the second derivative is always nonnegative:
\begin{IEEEeqnarray}{rCl}\label{e: f_second_derivative_positive}
    \frac{d^2 f(\epsilon)}{d\epsilon^2}
&\ge& 0.
\end{IEEEeqnarray}

Equation (\ref{e: f_second_derivative_positive}) shows that the
function $f(\epsilon)$ is convex. Using the Jensen's inequality we
have:
\begin{IEEEeqnarray}{rCl}\label{e: jansen TG}
E\left[ \left. f\left((\tilde{\mathcal {\bm \Leps}}_k)_{m,m}\right)
\right| {\bf g}=1\right] \ge f\left(E\left[ \left. (\tilde{\mathcal
{\bm \Leps}}_k)_{m,m}\right| {\bf g}=1\right]\right).
\end{IEEEeqnarray}
Substituting (\ref{e: jansen TG}) and (\ref{d: f for TG LB}) in
(\ref{e: mean f C}) we get:
\begin{IEEEeqnarray}{rCl}\label{e: almost final TG temp}
        C \ge  &\lim \limits_{k \to \infty } &
    \frac{1}{N\beta}
    \sum_{m=0}^{r-1}
     \log\(1+\frac{N c^2   \left(\frac{1}{N}- E\left[ \left. (\tilde{\mathcal
{\bm \Leps}}_k)_{m,m}\right| {\bf g}=1\right]\right)\left|\uv
\right|^2}
    {1+N c^2 E\left[ \left. (\tilde{\mathcal
{\bm \Leps}}_k)_{m,m}\right| {\bf
g}=1\right]}\)+\frac{r}{N\beta}\log(e^{-\eta}-e^{-\xi}).\IEEEeqnarraynumspace
\end{IEEEeqnarray}

For the last stage of the proof we need to consider the different
power constraints. For the quadratic constraint we set $\xi=\infty$
and $c$ in (\ref{e: TG modulation}) to:
\begin{IEEEeqnarray}{rCl}\label{e: c vlaue quad}
c=\sqrt{\beta \frac{N}{r} p_x}
\end{IEEEeqnarray}
using (\ref{e: TG modulation}), (\ref{e: TG 2nd moment}) and
(\ref{e: TG 4th moment}) we have:
\begin{IEEEeqnarray}{rCl}
E[\FV[\dagger]{X}{k}\FV{X}{k}]=N p_x
\end{IEEEeqnarray}
\begin{IEEEeqnarray}{rCl}
E[(\FV[\dagger]{X}{k}\FV{X}{k})^2]&=&  \frac{\beta N^2 p_x^2}{r^2}
\frac{r(r-1) (1+\eta)^2 +r(2+\eta^2+2 \eta)}{(1+\eta)^2}  = \beta
N^2 p_x^2 \left[1+ \frac{1}{r(1+\eta)^2} \right]
.\IEEEeqnarraynumspace
\end{IEEEeqnarray}
Adding to (\ref{e: almost final TG temp}) the requirement to satisfy
the quadratic power constraint results in (\ref{e: almost final TG})
and proves the first part of the lemma.

For the peak power constraint we need to satisfy:
\begin{IEEEeqnarray}{rCl}
    \frac{c^2}{p_z} r \xi \le N p_x
\end{IEEEeqnarray}
We thus set $\beta=1$ and
\begin{IEEEeqnarray}{rCl}\label{e: c vlaue peak}
    c^2=\frac{N p_x p_z}{r\xi},
\end{IEEEeqnarray}
which leads to (\ref{e: IT peak TG bound}) and completes the proof
of the lemma.

\setcounter{equation}{0}
\section{Minimization of the conditional covariance: Proof of Equation (\ref{e. estimation error frequency minima})}\label{app: Minimization of the estimation error}
We search for the minimal estimation error in the $m$-th frequency
bin given the transmitted power in each symbol. We use the following
inequality which holds for the square matrices $\openmtr D$,
$\openmtr E$ and $\openmtr F$ if $\openmtr F$ and ${\openmtr
D}+{\openmtr E}$ are \CommentApril{positive semi-definite}:
\begin{IEEEeqnarray}{rCl}
\label{e: matrix inequality} ({\openmtr D}+{\openmtr E})\({\openmtr
F}({\openmtr D}+{\openmtr E})+\It\)^{-1}\le {\openmtr D}\({\openmtr
F}{\openmtr D}+\It\)^{-1} +\({\openmtr D}{\openmtr F}+\It\)^{-1}
{\openmtr E}\({\openmtr F}{\openmtr D}+\It\)^{-1} .
\end{IEEEeqnarray}
where $\mtr{A}\ge \mtr{B}$ means $\mtr{A}-\mtr{B}$ is
\CommentApril{positive semi-definite}. To prove the inequality we
write:
\begin{IEEEeqnarray}{rCl}
({\openmtr D}+{\openmtr E})\({\openmtr F}({\openmtr D}+{\openmtr
E})+\It\)^{-1}&=& ({\openmtr D}+{\openmtr E})\({\openmtr F}{\openmtr
D}+\It\)^{-1} \nonumber \\&& - \: ({\openmtr D}+{\openmtr
E})\({\openmtr F}({\openmtr D}+{\openmtr E})+\It\)^{-1} {\openmtr
F}{\openmtr E}\({\openmtr F}{\openmtr D}+\It\)^{-1} \nonumber
\\&=& {\openmtr D}\({\openmtr F}{\openmtr D}+\It\)^{-1} \nonumber \\&& + \left[ \It
- ({\openmtr D}+{\openmtr E}){\openmtr F}\(({\openmtr D}+{\openmtr
E}){\openmtr F}+\It\)^{-1}\right] {\openmtr E}\({\openmtr
F}{\openmtr D}+\It\)^{-1} \nonumber
\\&=& {\openmtr D}\({\openmtr F}{\openmtr D}+\It\)^{-1} \nonumber \\&& +\(({\openmtr
D}+{\openmtr E}){\openmtr F}+\It\)^{-1} {\openmtr E}\({\openmtr
F}{\openmtr D}+\It\)^{-1} \nonumber \\&=& {\openmtr D}\({\openmtr
F}{\openmtr D}+\It\)^{-1} \nonumber \\&& +\({\openmtr D}{\openmtr
F}+\It\)^{-1} {\openmtr E}\({\openmtr F}{\openmtr D}+\It\)^{-1}
\nonumber
\\&& -\({\openmtr D}{\openmtr F}+\It\)^{-1}{\openmtr EF}\(({\openmtr D}+{\openmtr E}){\openmtr
F}+\It\)^{-1} {\openmtr E}\({\openmtr F}{\openmtr D}+\It\)^{-1}
 ,
\end{IEEEeqnarray}
where the first and fourth equalities use the identity
$(\mtr{A}+\mtr{B})^{-1}={\openmtr A}^{-1}-({\openmtr A}+{\openmtr
B})^{-1}{\openmtr BA}^{-1}$ and its transpose respectively, which
hold for any \CommentApril{nonsingular} matrix ${\openmtr A}$, as
long as $({\openmtr A}+{\openmtr B})^{-1}$ exists. The second
equality uses the identity $(\It +{\openmtr BA})^{-1}{\openmtr
B}={\openmtr B}(\It+{\openmtr AB})^{-1}$, which is obvious if the
matrix $\openmtr B$ is nonsingular, but holds for any
\CommentApril{positive semi-definite} matrix ${\openmtr B}$ as long
as the inverse of $(\It +{\openmtr BA})$ exists. The inequality in
(\ref{e: matrix inequality}) results from the fact that the matrix
${\openmtr F}\(({\openmtr D}+{\openmtr E}){\openmtr F}+\It\)^{-1}$
is \CommentApril{positive semi-definite}.

We use the identity ${\openmtr B}^\dagger(\It +{\openmtr
BAB}^{\dagger})^{-1}{\openmtr B}={\openmtr B}^{\dagger}{\openmtr
B}(\It+\mtr{A}{\openmtr B}^{\dagger}{\openmtr B})^{-1}$, which holds
as long as the inverse exists, and rewrite the estimation error
covariance (\ref{e. condtional channel distribution variance}) as:
\begin{IEEEeqnarray}{rCl}\label{e: estimation error wirh spectrun}
        &\tilde{\mathcal {\bm \Leps}}_k  &= {\openmtr{C}}_{\FV{H}{k}}  - N {\openmtr C}_{\FV[k-1]{H}{0}\FV{H}{k}}^\dagger
        \Sf_{\RV[k-1]{X}{0}}
        \left( N{\openmtr C}_{\FV[k-1]{H}{0}}\Sf_{\RV[k-1]{X}{0}} + \It  \right)^{ - 1}
        {\openmtr C}_{\FV[k-1]{H}{0}\FV{H}{k}}.
\end{IEEEeqnarray}
Setting $\Sf_{\RV[k-1]{X}{0}}=\Jc_m+\Jc_{\overline{m}}$ where
$\Jc_m$ and $\Jc_{\overline{m}}$ are both diagonal, and using
Inequality (\ref{e: matrix inequality}) \CommentApril{results in}:
{\setlength\arraycolsep{1pt}
\begin{IEEEeqnarray}{rCl}
    \label{e. estimation sum separation}
        \tilde{\mathcal {\bm \Leps}}_k  &=& {\openmtr{C}}_{\FV{H}{k}}  - N {\openmtr C}_{\FV[k-1]{H}{0}\FV{H}{k}}^\dagger
        (\Jc_m+\Jc_{\overline{m}})
        \left( N{\openmtr C}_{\FV[k-1]{H}{0}}(\Jc_m+\Jc_{\overline{m}}) + \It  \right)^{ - 1}
        {\openmtr C}_{\FV[k-1]{H}{0}\FV{H}{k}}
        \nonumber \\
        &\ge& {\openmtr{C}}_{\FV{H}{k}}  - N {\openmtr
        C}_{\FV[k-1]{H}{0}\FV{H}{k}}^\dagger
        \Jc_m
        \left( N{\openmtr C}_{\FV[k-1]{H}{0}}\Jc_m + \It  \right)^{ - 1}
        {\openmtr C}_{\FV[k-1]{H}{0}\FV{H}{k}}
        \nonumber \\
        && - \: N {\openmtr
        C}_{\FV[k-1]{H}{0}\FV{H}{k}}^\dagger
        \left( N\Jc_m{\openmtr C}_{\FV[k-1]{H}{0}} + \It  \right)^{ - 1}\Jc_{\overline{m}}
        \left( N{\openmtr C}_{\FV[k-1]{H}{0}}\Jc_m + \It  \right)^{ - 1}
        {\openmtr C}_{\FV[k-1]{H}{0}\FV{H}{k}}
\end{IEEEeqnarray}}

Without loss of generality we will assume in this appendix that
$m=0$. Now assume that $\Jc_0 = \Pc_{k-1} \otimes \Js_0$ and
$\Js_0=\diagM([1,0,\ldots,0])$. Also recall that ${\openmtr
C}_{\FV[k-1]{H}{0}}={{{\Delta}_{k-1}}} \otimes {\openmtr
C}_{\FV{H}{0}}$. Using the identity $(\mtr{A}\otimes\mtr{B})\cdot
(\mtr{C}\otimes\mtr{D})=\mtr{AC}\otimes\mtr{BD}$, the matrix inverse
in the second line of (\ref{e. estimation sum separation}) can be
written as:
\begin{IEEEeqnarray}{rCl}\label{e: matrix for inversion}
    \left(N{\openmtr C}_{\FV[k-1]{H}{0}} \Jc_0 + {\It}\right)^{ - 1}
    &=&
    \left( N\({{{\Delta}_{k-1}}} \Pc_{k-1}\)
    \otimes \({\openmtr C}_{\FV{H}{0}} \Js_0\) + {\It}\right)^{ - 1}.
\end{IEEEeqnarray}
Noting that $({\openmtr C}_{\FV{H}{0}} \Js_0)^2=({\openmtr
C}_{\FV{H}{0}})_{0,0}({\openmtr C}_{\FV{H}{0}} \Js_0)$, we use the
identity:
\begin{IEEEeqnarray}{rCl}\label{e: kron inverse identity}
\(\It + \mtr{A}\otimes{\openmtr B}\)^{-1}=\It-\(\(\It+\phi {\openmtr
A}\)^{-1}\mtr{A}\)\otimes{\openmtr B}
\end{IEEEeqnarray}
which holds if the inverse exists and ${\openmtr B}^2=\phi {\openmtr
B}$, and can be easily verified by direct multiplication. Using
Identity (\ref{e: kron inverse identity}), the inverse in (\ref{e:
matrix for inversion}) can be written as:
\begin{equation}\label{e: matrix 1 inverse}
    \left(N{\openmtr C}_{\FV[k-1]{H}{0}} \Jc_0 + {\It}\right)^{ - 1}
    =
    {\It} - \left\{
    N\( {\It} + N{{{\Delta}_{k-1}}}\Pc_{k-1}  ({\openmtr C}_{\FV{H}{0}})_{0,0} \) ^{-1} {{{\Delta}_{k-1}}} \Pc_{k-1} \right\}
    \otimes \({\openmtr C}_{\FV{H}{0}} \Js_0\).
\end{equation}
Substituting  also ${\openmtr C}_{\FV[k-1]{H}{0}\FV{H}{k}}
={{D}_{k}} \otimes {\openmtr C}_{\FV{H}{0}}$, we can write the
multiplication:
\begin{IEEEeqnarray}{rCl}
    \IEEEeqnarraymulticol{3}{l}{ \left( N{\openmtr C}_{\FV[k-1]{H}{0}}\Jc_0 + {\It}\right)^{ -1}
        {\openmtr C}_{\FV[k-1]{H}{0}\FV{H}{k}} }
    \nonumber \\  \quad &  = &
    {{D}_{k}} \otimes {\openmtr C}_{\FV{H}{0}}
    - N\left\{
    \( {\It} + N{{{\Delta}_{k-1}}} \Pc_{k-1} ({\openmtr C}_{\FV{H}{0}})_{0,0} \) ^{-1} {{{\Delta}_{k-1}}} \Pc_{k-1} {{D}_{k}}\right\}
    \otimes \({\openmtr C}_{\FV{H}{0}}\Js_0 {\openmtr C}_{\FV{H}{0}}\).\IEEEeqnarraynumspace
\end{IEEEeqnarray}
As we test the estimation error in the first frequency bin, we need
to evaluate the first column of the matrix $\left( N{\openmtr
C}_{\FV[k-1]{H}{0}}\Jc_0 + {\It}\right)^{ -1}
 {\openmtr C}_{\FV[k-1]{H}{0}\FV{H}{k}}$. This column
is given by:
\begin{IEEEeqnarray}{rCl}
\IEEEeqnarraymulticol{3}{l}{    v_{aN+b} }
    \nonumber \\
    &=&
    ( {{D}_{k}} )_a ({\openmtr C}_{\FV{H}{0}})_{b,0}
    - N\left(
    \( {\It} + N{{{\Delta}_{k-1}}} \Pc_{k-1} ({\openmtr C}_{\FV{H}{0}})_{0,0} \) ^{-1} {{{\Delta}_{k-1}}} \Pc_{k-1}
    {{D}_{k}}\right)_a
    ({\openmtr C}_{\FV{H}{0}})_{0,0} ({\openmtr C}_{\FV{H}{0}})_{b,0}
    \nonumber \\
    &=& g(a) ({\openmtr C}_{\FV{H}{0}})_{b,0}
\end{IEEEeqnarray}
Now the %LM%
top left element of the matrix described by the %LM%
last line of (\ref{e. estimation sum separation}) can be written as:
\begin{IEEEeqnarray}{rCl}
    \IEEEeqnarraymulticol{3}{l}{ -N\sum_{a=0}^{k-1}\sum_{b=0}^{N-1} |v_{aN+b}|^2 ( \Jc_{\overline{0}})_{aN+b,aN+b} }
    \nonumber \\
    \quad \quad &=& -N\sum_{a=0}^{k-1} |g(a)|^2 \sum_{b=0}^{N-1} \left|({\openmtr C}_{\FV{H}{0}})_{b,0}\right|^2 ( \Jc_{\overline{0}})_{aN+b,aN+b}
    \nonumber \\
    &=& -N\sum_{a=0}^{k-1} |g(a)|^2 \left[ \left|({\openmtr C}_{\FV{H}{0}})_{0,0}\right|^2 ( \Jc_{\overline{0}})_{aN,aN} + \sum_{b=1}^{N-1} \left|({\openmtr C}_{\FV{H}{0}})_{b,0}\right|^2 ( \Jc_{\overline{0}})_{aN+b,aN+b} \right]
    \nonumber \\
    &\ge& 0
\end{IEEEeqnarray}
where we use the fact that
$\Sf_{\RV[k-1]{X}{0}}=\Jc_{0}+\Jc_{\overline{0}}$ and ${\rm
Tr}(\Sf_{\RV{X}{a}})\le \RV{p}{a}$, and therefore $(
\Jc_{\overline{0}})_{aN,aN}\le 0$ and $\sum_{b=0}^{N-1} (
\Jc_{\overline{0}})_{aN+b,aN+b} = 0$. We also use the
\CommentApril{positive semi-definiteness} of the matrix $ {\openmtr
C}_{\FV{H}{0}}$ so that $|({\openmtr C}_{\FV{H}{0}})_{0,0}|^2 \ge
|({\openmtr C}_{\FV{H}{0}})_{b,0}|^2$.

We conclude that the bound on the estimation error for the first
frequency bin is achieved when $\Jc_{\overline{0}}=0$. Using
$\Js_0{\openmtr C}_{\FV{H}{0}}\Js_0=({\openmtr
C}_{\FV{H}{0}})_{0,0}\Js_0$ and (\ref{e: matrix 1 inverse}) we
evaluate the matrix:
\begin{IEEEeqnarray}{rCl}\label{e: temp B.11}
    \Jc_0 \cdot \left(N{\openmtr C}_{\FV[k-1]{H}{0}} \Jc_0 + {\It}\right)^{ - 1}
    &=& \Pc_{k-1} \otimes \Js_0
    \nonumber \\ &&
    - \left\{
    \Pc_{k-1} N\( {\It} + N{{{\Delta}_{k-1}}}\Pc_{k-1}  ({\openmtr C}_{\FV{H}{0}})_{0,0} \) ^{-1} {{{\Delta}_{k-1}}} \Pc_{k-1} ({\openmtr C}_{\FV{H}{0}})_{0,0}\right\}
    \otimes \Js_0
    \nonumber \\
    &=&
    \left\{\Pc_{k-1}
    \( {\It} + N{{{\Delta}_{k-1}}}\Pc_{k-1}  ({\openmtr C}_{\FV{H}{0}})_{0,0} \) ^{-1} \right\}
    \otimes \Js_0.
\end{IEEEeqnarray}
Substituting (\ref{e: temp B.11}) and $\Jc_{\overline{0}}=0$ in
(\ref{e: estimation error wirh spectrun}), the bound is given by:
\begin{IEEEeqnarray}{rCl}
        (\tilde{\mathcal {\bm \Leps}}_k)_{0,0}
        &\ge&
    ( {\openmtr C}_{\FV{H}{0}} )_{0,0}
-  N \( \left( {{D}_{k}^\dagger} \otimes {\openmtr C}_{\FV{H}{0}}
    \right)\Jc_{0}\left( N{\openmtr C}_{\FV[k-1]{H}{0}}\Sf_{\RV[k-1]{X}{0}} + \It  \right)^{ - 1}
     \left( {{D}_{k}} \otimes {\openmtr C}_{\FV{H}{0}}
    \right)\)_{0,0}
    \nonumber \\
    &=&
    ( {\openmtr C}_{\FV{H}{0}} )_{0,0}
-  N  {{D}_{k}^\dagger}  \Pc_{k-1}
    \( {\It} + N{{{\Delta}_{k-1}}}\Pc_{k-1}  ({\openmtr C}_{\FV{H}{0}})_{0,0} \) ^{-1} {{D}_{k}}
        \cdot\(  {\openmtr C}_{\FV{H}{0}} \Js_0 {\openmtr C}_{\FV{H}{0}}\)_{0,0}
    \nonumber \\
    &=&
    ( {\openmtr C}_{\FV{H}{0}} )_{0,0}
    -N( {\openmtr C}_{\FV{H}{0}} )_{0,0}^2 {{D}_{k}^\dagger}
    \Pc_{k-1}
    \( {\It} + N{{{\Delta}_{k-1}}}\Pc_{k-1}  ({\openmtr C}_{\FV{H}{0}})_{0,0} \) ^{-1} {{D}_{k}}.
\end{IEEEeqnarray}
Using our normalization $( {\openmtr C}_{\FV{H}{0}} )_{0,0}=1/N$,
and we get (\ref{e. estimation error frequency minima}).

\setcounter{equation}{0}
\section{Derivation of Equation (\ref{e. tap etsimation error})}\label{app: Derivation of equation}
In this appendix we present the derivation of Equation (\ref{e. tap
etsimation error}) from Equation (\ref{e: tap estimation error
variance}). The easiest way to do so is to define a matrix modulo
concatenation, and use its properties to simplify the derivation.
Note that the matrix modulo concatenation can be seen as a
permutation of a block diagonal matrix concatenation. Therefore the
two concatenations share the same properties. Yet, for the problems
at hand, the modulo concatenation is more convenient as the matrices
we handle have the desired form.

For the matrices of identical size $\mtr{A}_0,
\mtr{A}_1,{\ldots},\mtr{A}_{\CommentApril{v}-1}$ we define the
modulo concatenation as $\mtr{A}={\bm \backslash} \mtr{A}_0,
\mtr{A}_1,{\ldots},\mtr{A}_{v-1} {\bm \backslash}$ so that its
elements satisfy:
\begin{IEEEeqnarray}{rCl}
\left(\mtr{A}\right)_{k,l}=\left\{
\begin{array}{ll}
0 & k \ {\rm mod} \ v \ne l \ {\rm mod} \  v \\
\left( \mtr{A}_{k \ {\rm mod} \  v} \right)_{{\lfloor \frac{k}{v}\rfloor},{\lfloor \frac{l}{v}\rfloor}} & k \ {\rm mod} \  v = l \ {\rm mod} \  v \\
\end{array} \right. ,
\end{IEEEeqnarray}
i.e., the elements of the concatenated matrices are placed on the
diagonals of the blocks of the matrix $\mtr{A}$. If the size of the
matrices is $r \times N$ then the size of their diagonal
concatenation is $v r \times v N$.

The modulo concatenation has the following useful properties:
\begin{enumerate}
\item{If $\mtr{A}={\bm \backslash} \mtr{A}_0, \mtr{A}_1,{\ldots},\mtr{A}_{v-1} {\bm \backslash}$ and  $\mtr{B}={\bm \backslash}
\mtr{B}_0, \mtr{B}_1,{\ldots},\mtr{B}_{v-1} {\bm \backslash}$ are
the modulo concatenation of $r \times N$ matrices, then
$\mtr{A}+\mtr{B}={\bm \backslash} \mtr{A}_0+ \mtr{B}_0, \mtr{A}_1
+\mtr{B}_1,{\ldots},\mtr{A}_{v-1}+ \mtr{B}_{v-1} {\bm \backslash}$.}
\item{If $\mtr{A}={\bm \backslash} \mtr{A}_0, \mtr{A}_1,{\ldots},\mtr{A}_{v-1} {\bm \backslash}$ is
the modulo concatenation of $r \times N$ matrices, and $\mtr{B}={\bm
\backslash} \mtr{B}_0, \mtr{B}_1,{\ldots},\mtr{B}_{v-1} {\bm
\backslash}$ is the
modulo concatenation of $N \times U$ matrices, then $AB=\\
{\bm \backslash} \mtr{A}_0 \mtr{B}_0, \mtr{A}_1
\mtr{B}_1,{\ldots},\mtr{A}_{v-1} \mtr{B}_{v-1} {\bm \backslash}$.}
\item{$\It={\bm \backslash} \It, \It,{\ldots},\It {\bm \backslash}$.}
\item{If the inverse exists then $\mtr{A}^{-1}={\bm \backslash} \mtr{A}_0^{-1}, \mtr{A}_1^{-1},{\ldots},\mtr{A}_{v-1}^{-1}
{\bm \backslash}$.}
\item{If $D=\diagM([d_0,d_1,\ldots,d_{v-1}])$ is a diagonal matrix, then
$\mtr{A} \otimes D={\bm \backslash} \mtr{A} d_{0}, \mtr{A}
d_{1},{\ldots},\mtr{A} d_{v-1} {\bm \backslash} $.}
\end{enumerate}
Note that properties 1,3 and 5 follow directly from the definition
of modulo concatenation, and property 4 follows from 2 and 3.
Property 2 can be proved by permutating the matrices into block
diagonal matrices. But, for the completeness of the treatment we
give here a short direct proof:

The $k,l$ element of the multiplication in property 2 is:
\begin{IEEEeqnarray}{rCl}
\left( \mtr{A} \mtr{B} \right)_{k,l}&=&\sum_{w=0}^{N v-1}
\mtr{A}_{k,w} \mtr{B}_{w,l}
\nonumber \\
&=&\sum_{w=0}^{N-1} \mtr{A}_{k,w v+(k\ {\rm mod} \  v)} \mtr{B}_{w
v+(k\ {\rm mod} \  v),l}
\end{IEEEeqnarray}
which is 0 if $k\ {\rm mod} \  v\ne l\ {\rm mod} \  v$, if $k\ {\rm
mod} \  v= l\ {\rm mod} \  v$ the $k,l$ element is:
\begin{IEEEeqnarray}{rCl}
\left( \mtr{A} \mtr{B} \right)_{k,l}&=& \sum_{w=0}^{v-1} \left(
\mtr{A}_{k\ {\rm mod} \  v}\right)_{{\lfloor \frac{k}{v}\rfloor},w}
\left( \mtr{B}_{k\ {\rm mod} \ v}\right)_{w,{\lfloor
\frac{l}{v}\rfloor}}=\left(\mtr{A}_{k\ {\rm mod} \  v} \mtr{B}_{k\
{\rm mod} \  v}\right)_{{\lfloor \frac{k}{v}\rfloor},{\lfloor
\frac{l}{v}\rfloor}}
\end{IEEEeqnarray}
which completes the proof.

Using the properties of the modulo concatenation and noting that the
matrix ${\openmtr{C}}_{\RV{H}{k}}$ is diagonal, we can now rewrite
Equation (\ref{e: tap estimation error variance}) as:
\begin{IEEEeqnarray}{rCl}
    \hat{\Leps}_k  &=& {\openmtr{C}}_{\RV{H}{k}}  -
        N \beta p_x {{D}_{k}^\dagger} \otimes {\openmtr{C}}_{\RV{H}{k}}
        \left( N \beta p_x {{\Delta}_{k-1}} \otimes {\openmtr{C}}_{\RV{H}{k}} + \It  \right)^{ - 1}
        {{D}_{k}} \otimes {\openmtr{C}}_{\RV{H}{k}}
        \nonumber \\
         &=& {\openmtr{C}}_{\RV{H}{k}}  -
        N \beta p_x {\bm \backslash} {{D}_{k}^\dagger}
        ({\openmtr{C}}_{\RV{H}{k}})_{0,0},\ldots,{{D}_{k}^\dagger}
        ({\openmtr{C}}_{\RV{H}{k}})_{N-1,N-1}{\bm \backslash}
        \nonumber \\
        &&\cdot {\bm \backslash} N \beta p_x {{\Delta}_{k-1}}
        ({\openmtr{C}}_{\RV{H}{k}})_{0,0}+\It,\ldots,N \beta p_x {{\Delta}_{k-1}}
        ({\openmtr{C}}_{\RV{H}{k}})_{N-1,N-1}+\It{\bm \backslash}^{-1}
        \nonumber \\
        &&\cdot  {\bm \backslash} {{D}_{k}}
        ({\openmtr{C}}_{\RV{H}{k}})_{0,0},\ldots,{{D}_{k}}
        ({\openmtr{C}}_{\RV{H}{k}})_{N-1,N-1}{\bm \backslash}
        \nonumber \\
         &=& {\openmtr{C}}_{\RV{H}{k}}  -
        N \beta p_x {\bm \backslash} {{D}_{k}^\dagger}
        ({\openmtr{C}}_{\RV{H}{k}})_{0,0} \left(N \beta p_x {{\Delta}_{k-1}}
        ({\openmtr{C}}_{\RV{H}{k}})_{0,0}+\It\right)^{-1}{{D}_{k}}
        ({\openmtr{C}}_{\RV{H}{k}})_{0,0},\ldots
        \nonumber \\
        &&,{{D}_{k}^\dagger}
        ({\openmtr{C}}_{\RV{H}{k}})_{N-1,N-1} \left(N \beta p_x {{\Delta}_{k-1}}
        ({\openmtr{C}}_{\RV{H}{k}})_{N-1,N-1}+\It\right)^{-1}{{D}_{k}}
        ({\openmtr{C}}_{\RV{H}{k}})_{N-1,N-1}{\bm {\bm \backslash}}
.
\end{IEEEeqnarray}
which leads to (\ref{e. tap etsimation error}).

\section{Alternative truncated Gaussian lower bound}\label{app: aleternative TG bound}
\CommentApril{In this appendix we derive an alternative bound
that can replace the one in Theorem \ref{T: TG lower bound} without the need
for the zero mean assumption. This bound is given by:}

The capacity of a channel with a quadratic power constraint is lower
bounded by:
\begin{equation}\label{ae: TG bound maximization quad}
        C \ge \sup_{\eta>0}\max_{r\le N} \max_{1\le\beta\le \frac{\alpha r(1+\eta)^2}{1+r(1+\eta)^2}} {\rm LB}_{\rm TG}^{({\rm qd})}
        (p_x,r,\beta,\eta)-\frac{r}{N\beta}\eta
\end{equation}
and the capacity of a channel with a peak power constraint is lower
bounded by:
\begin{equation}\label{ae: TG bound maximization peak}
        C \ge \sup_{\eta>0}\sup_{\xi>\eta}\max_{r\le N} {\rm LB}_{\rm TG}^{({\rm pk})}
        (p_x,r,\eta,\xi)-\frac{r}{N}\log(e^{-\eta}-e^{-\xi})
\end{equation}
where
\begin{equation}\label{ae: fianl LB TG}
{\rm LB}_{\rm TG}^{({\rm qd})}(p_x,r,\beta,\eta)=\left\{
\begin{array}{ll}
        \frac{r}{N\beta}
     \log\(1+\frac{\beta\frac{N^2}{r} p_x\left| E \left[  \FV{h}{0,0}
\right]+\frac{\uv}{\sqrt{N}}\sqrt{1 - \frac{1}{1+\frac{\beta p_x
\eta}{1+\eta}\frac{1}{2 r} \left(\hTc\left(
 \frac{\beta p_x \eta}{r(1+\eta)} \right)-N\right)}  } \right|^2}
    {1+\frac{\beta\frac{N}{r} p_x}{1+\frac{\beta p_x \eta}{1+\eta}\frac{1}{2 r} \left(\hTc\left(
 \frac{\beta p_x \eta}{r(1+\eta)} \right)-N\right)} }\)
& r<N \\
    \frac{1}{\beta}
     \log\(1+\frac{\beta N p_x\left| E \left[  \FV{h}{0,0}
\right]+\frac{\uv}{\sqrt{N}} \sqrt{1 - \frac{1}{1+\frac{\beta p_x
\eta}{1+\eta}\frac{1}{2 L} \left(\hTc\left(
 \frac{\beta p_x \eta}{1+\eta} \right)-N\right)} } \right|^2}
    {1+\frac{\beta p_x }{1+\frac{\beta p_x \eta}{1+\eta}\frac{1}{2 L} \left(\hTc\left(
 \frac{\beta p_x \eta}{1+\eta} \right)-N\right)}}\)  & r=N
\end{array} \right.
    , \IEEEeqnarraynumspace
    \end{equation}
\begin{equation}\label{ae: fianl LB TG peak}
{\rm LB}_{\rm TG}^{({\rm pk})}(p_x,r,\eta,\xi)=\left\{
\begin{array}{ll}
    \frac{r}{N}
     \log\(1+\frac{\frac{N^2 p_x p_z}{r\xi}   \left| E \left[  \FV{h}{0,0}
\right]+\frac{\uv}{\sqrt{N}} \sqrt{1-\frac{1}{1+\frac{p_x
\eta}{\xi}\frac{1}{2r}\left(\hTc\left(
 \frac{p_x \eta}{r \xi}\right)-N\right)} } \right|^2}
    {1+\frac{\frac{p_x }{\xi}\left(1+\frac{\eta e^{-\eta}-\xi e^{-\xi}}{e^{-\eta}-e^{-\xi}}\right) }{1+\frac{p_x \eta}{\xi}\frac{1}{2r}\left(\hTc\left(
 \frac{p_x \eta}{r \xi}\right)-N\right)}}\)
& r<N \\
     \log\(1+\frac{\frac{N p_x p_z}{\xi}   \left| E \left[  \FV{h}{0,0}
\right]+\frac{\uv}{\sqrt{N}} \sqrt{1-\frac{1}{1+\frac{p_x \eta}{\xi}
    \frac{1}{2L}\left(\hTc\left(
 \frac{p_x \eta}{\xi}
    \right)-N\right)} } \right|^2}
    {1+\frac{\frac{p_x }{\xi}\left(1+\frac{\eta e^{-\eta}-\xi e^{-\xi}}{e^{-\eta}-e^{-\xi}}\right) }{1+\frac{p_x \eta}{\xi}
    \frac{1}{2L}\left(\hTc\left(
 \frac{p_x \eta}{\xi}
    \right)-N\right)}}\)
 & r=N
\end{array} \right.
    , \IEEEeqnarraynumspace
    \end{equation}
and ${\bf u} \sim {\cal CN}(0,1)$ is a proper complex Gaussian
random variable.

$ $

The proof of this bound follows the same lines as the proof of
Theorem \ref{T: TG lower bound} with one major shortcut. We use the
fact that the estimation error variance is a decreasing function of
each of the transmitted symbols' power. We thus upper bound it by
the estimation error variance that is achieved when all previous
symbol used the minimal allowed transmission power. Taking into
account the transmitted signal structure, (\ref{e: TG modulation}),
the bound results by replacing  the term $\zeta$ by $\frac{c^2
\eta}{p_z}$ in (\ref{e: TG expec bound 3}) and (\ref{e: narrow TG
expec bound 3}).

As this bound does not use the convexity argument,
(\ref{tot_exp_bound}) - (\ref{e: f_second_derivative_positive}), it
does not require the zero mean assumption and holds for any
expectation of the channel.

\end{appendices}
\bibliographystyle{IEEEtran}
\bibliography{uwb_new}

\begin{thebibliography}{10}
\providecommand{\url}[1]{#1}
\csname url@samestyle\endcsname
\providecommand{\newblock}{\relax}
\providecommand{\bibinfo}[2]{#2}
\providecommand{\BIBentrySTDinterwordspacing}{\spaceskip=0pt\relax}
\providecommand{\BIBentryALTinterwordstretchfactor}{4}
\providecommand{\BIBentryALTinterwordspacing}{\spaceskip=\fontdimen2\font plus
\BIBentryALTinterwordstretchfactor\fontdimen3\font minus
  \fontdimen4\font\relax}
\providecommand{\BIBforeignlanguage}[2]{{%
\expandafter\ifx\csname l@#1\endcsname\relax
\typeout{** WARNING: IEEEtran.bst: No hyphenation pattern has been}%
\typeout{** loaded for the language `#1'. Using the pattern for}%
\typeout{** the default language instead.}%
\else
\language=\csname l@#1\endcsname
\fi
#2}}
\providecommand{\BIBdecl}{\relax}
\BIBdecl

\bibitem{tse2005fwc}
D.~Tse and P.~Viswanath, \emph{{Fundamentals of Wireless Communication}}.\hskip
  1em plus 0.5em minus 0.4em\relax Cambridge University Press, 2005.

\bibitem{kozek1998nonorthogonal}
W.~Kozek and A.~Molisch, ``{Nonorthogonal pulseshapes for multicarrier
  communications in doubly dispersive channels},'' \emph{IEEE Journal on
  Selected Areas in Communications}, vol.~16, no.~8, pp. 1579--1589, 1998.

\bibitem{durisi2010noncoherent}
G.~Durisi, U.~Schuster, H.~B\"olcskei, and S.~Shamai, ``{Noncoherent capacity
  of underspread fading channels},'' \emph{IEEE Trans. on Information Theory},
  vol.~56, no.~1, pp. 367--395, 2010.

\bibitem{Hlawatsch2011}
F.~Hlawatsch and G.~Matz, \emph{{Wireless Communications Over Rapidly
  Time-Varying Channels}}.\hskip 1em plus 0.5em minus 0.4em\relax Academic
  Press, 2011.

\bibitem{Medard:02}
M.~Medard and R.~G. Gallager, ``Bandwith {S}caling for {F}ading {M}ultipath
  {C}hannels,'' \emph{IEEE Trans. on Information Theory}, vol.~48, no.~4, pp.
  840--852, Apr. 2002.

\bibitem{sethuraman2005cpu}
V.~Sethuraman and B.~Hajek, ``{Capacity Per Unit Energy of Fading Channels With
  a Peak Constraint},'' \emph{IEEE Trans. on Information Theory}, vol.~51,
  no.~9, pp. 3102--3120, 2005.

\bibitem{sethuraman2009low}
V.~Sethuraman, L.~Wang, B.~Hajek, and A.~Lapidoth, ``{Low-SNR capacity of
  noncoherent fading channels},'' \emph{IEEE Trans. on Information Theory},
  vol.~55, no.~4, pp. 1555--1574, 2009.

\bibitem{Telatar:00}
I.~E. Telatar and D.~N.~C. Tse, ``Capacity and {M}utual {I}nformation of
  {W}ideband {M}ultipath {F}ading {C}hannels,'' \emph{IEEE Trans. on
  Information Theory}, vol.~46, no.~4, pp. 1384--1400, July 2000.

\bibitem{porrat2007channel}
D.~Porrat, D.~Tse, and S.~Nacu, ``{Channel Uncertainty in Ultra-Wideband
  Communication Systems},'' \emph{IEEE Trans. on Information Theory}, vol.~53,
  no.~1, pp. 194--208, 2007.

\bibitem{Doob:53}
J.~Doob, \emph{{Stochastic processes}}.\hskip 1em plus 0.5em minus 0.4em\relax
  Wiley, New York, 1953.

\bibitem{Liang:04}
Y.~Liang and V.~V. Veeravalli, ``Capacity of {N}oncoherent {T}ime-selective
  {R}ayleigh-fading {C}hannels,'' \emph{IEEE Trans. on Information Theory},
  vol. IT-50, no.~12, pp. 3095--3110, Dec. 2004.

\bibitem{Lapidoth:03}
A.~Lapidoth and S.~M. Moser, ``Capacity {B}ounds {V}ia {D}uality {W}ith
  {A}pplications to {M}ultiple-{A}ntenna {S}ystems on {F}lat-{F}ading
  {C}hannels,'' \emph{IEEE Trans. on Information Theory}, vol.~49, no.~10, pp.
  2426--2443, Oct. 2003.

\bibitem{Lapidoth:05}
A.~Lapidoth, ``On the {A}symptotic {C}apacity of {S}tationary {G}aussian
  {F}ading {C}hannels,'' \emph{IEEE Trans. on Information Theory}, vol.~51,
  no.~2, pp. 437--446, Feb. 2005.

\bibitem{koch2005fna}
T.~Koch and A.~Lapidoth, ``{The fading number and degrees of freedom in
  non-coherent MIMO fading channels: a peace pipe},'' in \emph{Proceedings of
  the International Symposium on Information Theory, ISIT}, 2005, pp. 661--665.

\bibitem{Proakis:95}
J.~G. Proakis, \emph{Digital Communications}.\hskip 1em plus 0.5em minus
  0.4em\relax McGraw-Hill, 1995.

\bibitem{bolckei2002cob}
H.~B\"olcskei, D.~Gesbert, and A.~Paulraj, ``{On the capacity of OFDM-based
  spatial multiplexing systems},'' \emph{IEEE Trans. on Communications},
  vol.~50, no.~2, pp. 225--234, 2002.

\bibitem{schafhuber2005maa}
D.~Schafhuber and G.~Matz, ``{MMSE and adaptive prediction of time-varying
  channels for OFDM systems},'' \emph{IEEE Trans. on Wireless Communications},
  vol.~4, no.~2, pp. 593--602, 2005.

\bibitem{Medard:00}
M.~Medard, ``The {E}ffect upon {C}hannel {C}apacity in {W}ireless
  {C}ommunications of {I}mperfect {K}nowledge of the {C}hannel,'' \emph{IEEE
  Trans. on Information Theory}, vol.~46, no.~3, pp. 933--946, May 2000.

\bibitem{neeser1993pcr}
F.~Neeser and J.~Massey, ``{Proper complex random processes with applications
  to information theory},'' \emph{IEEE Trans. on Information Theory}, vol.~39,
  no.~4, pp. 1293--1302, 1993.

\bibitem{baccarelli2000ssb}
E.~Baccarelli and A.~Fasano, ``{Some simple bounds on the symmetric capacity
  and outage probability for QAM wireless channels with Rice and Nakagami
  fadings},'' \emph{IEEE Journal on Selected Areas in Communications}, vol.~18,
  no.~3, pp. 361--368, 2000.

\bibitem{marzetta1999cmm}
T.~Marzetta and B.~Hochwald, ``{Capacity of a mobile multiple-antenna
  communication link in Rayleigh flat fading},'' \emph{IEEE Trans. on
  Information Theory}, vol.~45, no.~1, pp. 139--157, 1999.

\bibitem{Bergel:10}
I.~Bergel and S.~Benedetto, ``{The effective coherence time of common channel
  models},'' in \emph{IEEE 11th Workshop on Signal Processing Advances in
  Wireless Communications. SPAWC}, 2010.

\bibitem{clarke1968statistical}
R.~Clarke, ``{A statistical theory of mobile-radio reception},'' \emph{Bell
  Syst. Tech. J}, vol.~47, no.~6, pp. 957--1000, 1968.

\bibitem{Lapidoth:02}
A.~Lapidoth and S.~Shamai, ``Fading {C}hannels: {H}ow {P}erfect {N}eed
  ``{P}erfect {S}ide {I}nformation'' {B}e?'' \emph{IEEE Trans. on Information
  Theory}, vol.~48, no.~5, pp. 1118--1134, May 2002.

\bibitem{jin2005ulc}
X.~Jin, T.~Li, O.~Collins, and T.~Fuja, ``{The universality of LDPC codes on
  correlated fading channels with decision feedback based receiver},''
  \emph{IEEE Global Telecommunications Conference, GLOBECOM 2005}, vol.~3,
  2005.

\bibitem{li2007successive}
T.~Li and O.~Collins, ``{A successive decoding strategy for channels with
  memory},'' \emph{IEEE Trans. on Information Theory}, vol.~53, no.~2, pp.
  628--646, 2007.

\bibitem{Durisi2009ISIT}
G.~Durisi, V.~Morgenshtern, and H.~B\"olcskei, ``{On the Sensitivity of
  Noncoherent Capacity to the Channel Model},'' in \emph{Proceedings of the
  International Symposium on Information Theory, ISIT}, 2009, pp. 2174--2178.

\bibitem{zheng2007ccl}
L.~Zheng, D.~Tse, and M.~Medard, ``{Channel Coherence in the Low-SNR Regime},''
  \emph{IEEE Trans. on Information Theory}, 2007.

\bibitem{Chen:07}
J.~Chen and V.~Veeravalli, ``{Capacity Results for Block-Stationary Gaussian
  Fading Channels With a Peak Power Constraint},'' \emph{IEEE Trans. on
  Information Theory}, vol.~53, no.~12, pp. 4498--4520, 2007.

\bibitem{Telatar:99}
I.~E. Telatar, ``Capacity of {M}ulti-{A}ntenna {G}aussian {C}hannels,''
  \emph{European Trans. on Telecommunications}, vol.~10, pp. 585--595, Nov.
  1999.

\bibitem{Cover:91}
T.~M. Cover and J.~A. Thomas, \emph{Elements of Information Theory}.\hskip 1em
  plus 0.5em minus 0.4em\relax John Wiley \& Sons, Inc., 1991.

\bibitem{kay1993fss}
S.~Kay, \emph{{Fundamentals of statistical signal processing: estimation
  theory}}.\hskip 1em plus 0.5em minus 0.4em\relax Prentice Hall PTR, 1993.

\bibitem{abramowitz1965handbook}
M.~Abramowitz and I.~Stegun, \emph{{Handbook of mathematical functions: with
  formulas, graphs, and mathematical tables}}.\hskip 1em plus 0.5em minus
  0.4em\relax Courier Dover Publications, 1965.

\end{thebibliography}
\newpage


\end{document}